\documentclass[aps,prd,twocolumn,showpacs,superscriptaddress, preprintnumbers]{revtex4} 

\usepackage{graphicx}
\usepackage{multirow}
\usepackage{amsmath}
\usepackage{feyn}

\usepackage{color}

\font\FermiSmallfont=cmssq8 scaled 1200

\def\LANLppthead#1{
\null 
\begin{center}\vskip -1.0truein{\hbox to 7.5truein {
\hfill
\vbox to 1in {\vfill \FermiSmallfont
              \hbox{#1}
              \vfill}
}}\vskip-0.0truein\end{center}}

\begin{document}

\title{Neutrino Quantum Kinetics}
\preprint{LA-UR-13-27035}

\author{Alexey Vlasenko}
\affiliation{Department of Physics, University of California, San Diego, La Jolla, CA 92093-0319, USA}
\affiliation{Neutrino Engineering Institute, New Mexico Consortium, Los Alamos, NM 87545, USA}
\author{George M. Fuller}
\affiliation{Department of Physics, University of California, San Diego, La Jolla, CA 92093-0319, USA}
\affiliation{Neutrino Engineering Institute, New Mexico Consortium, Los Alamos, NM 87545, USA}
\author{Vincenzo Cirigliano}
\affiliation{Theoretical Division, Los Alamos National Laboratory, Los Alamos, NM 87545, USA}
\affiliation{Neutrino Engineering Institute, New Mexico Consortium, Los Alamos, NM 87545, USA}

\begin{abstract}
We present a formulation of the quantum kinetic equations (QKEs) which govern the evolution of neutrino flavor at high density and temperature.  Here, 
the structure of the QKEs is derived from the ground up, using fundamental neutrino interactions and quantum field theory.  We show that the resulting QKEs describe coherent flavor evolution with an effective mass when inelastic scattering is negligible. The QKEs also contain a collision term. This term can reduce to the collision term in the Boltzmann equation when scattering is dominant and the neutrino effective masses and density matrices become diagonal in the interaction basis. We also find that the QKE's include equations of motion for a new dynamical quantity related to neutrino spin. This quantity decouples from the equations of motion for the density matrices at low densities or in isotropic conditions. However, the spin equations of motion allow for the possibility of coherent transformation between neutrinos and antineutrinos at high densities and in the presence of anisotropy. Although the requisite conditions for this exist in the core collapse supernova and compact object merger environments, it is likely that only a self consistent incorporation of the QKEs in a sufficiently realistic model could establish whether or not significant neutrino-antineutrino conversion occurs.
\end{abstract}


\maketitle

\section{Introduction}

In this paper we  address
the difficult problem of how neutrino flavor evolves in a general medium. The stakes are high because neutrino weak interactions with matter, dictated in part by the neutrino flavor states, may lie at the heart of our understanding of neutrino-affected astrophysical environments, and these can be important sites for the origin of the elements.

This paper represents a first step towards the derivation of 
practicable generalized  kinetic equations, useful in actual simulations of neutrino propagation in anisotropic media,  
in any density regime.   Here we set up the formalism, 
identify the degrees of freedom needed to describe the neutrino ensemble (these include both flavor and spin),  
and derive the correct structure of the quantum kinetic equations (QKEs), including  coherent evolution and a collision term accounting for inelastic scattering. 
Our final results, summarized in Eq.~(\ref{eq:165-1}), are somewhat formal, since  self-energies entering into the collision term on the right-hand side are
not  fully calculated.  Nonetheless, all the medium-induced  potentials appearing on the left-hand-side of 
Eq.~(\ref{eq:165-1}) are computed in Section VI.A,  so this paper  provides a complete description of coherent
spin and flavor evolution in the absence of collisions.     
We will complete our program in a future paper, devoted to a detailed analysis of the collision term.

In this work, we have sought a well-posed prescription for treating general neutrino flavor evolution, one which can describe how neutrinos propagate and possibly change their flavors in environments ranging from low density regimes, where quantum mechanical phases are important and the evolution is Schr\"odinger-like, to very high temperature or very high matter density environments where phases are unimportant and the propagation/evolution is governed by the Boltzmann equation, and to all conditions between these limits. As a result, interaction-induced de-coherence,
an historically thorny issue in relativistic and nonrelativistic quantum systems \cite{Schwinger:1961uq,Harris:1981qy,Harris:1982fk,Stodolsky:1987lr,Manohar:1987ys,Habib:1996yq,Cooper:1997rt,Berges:2002vn,Berges:2003fj,Berges:2005lr,Muller:2006kx,Giraud:2010lr}, must be addressed directly and self-consistently.

The approach we take differs from previous treatments. Those studies examined neutrino or general fermion flavor conversion in both the active-active channel \cite{Raffelt:1993fj,Sigl:1993fr,Raffelt:1993kx,McKellar:1994uq,Sawyer:2005yg,Strack:2005fk,Cardall:2008lr,Herranen:2008qy,Herranen:2009fk,Gava:2009yq,Volpe:2013lr} and in the active-sterile channel \cite{Enqvist:1991yq,Barbieri:1991fj,Dodelson:1994rt,Shi:1996vn,Foot:1997rt,Bell:1999fk,Dolgov:2000fr,Volkas:2000uq,Abazajian:2001lr,Dolgov:2002ve,Kusenko:2005qy,Boyanovsky:2007fk,Boyanovsky:2007kx,Boyanovsky:2007lr,Kishimoto:2008pd,Kusenko:2009lr}, with a number of different approaches. Here we follow the general prescription used in Ref.s~\cite{Cirigliano:2010lr,Cirigliano:2011lr} for bosons, but adapted and extended appropriately for fermions. In this development, we start from the most fundamental considerations of quantum field theory, and then build QKEs which describe neutrino flavor evolution.

In hot and dense environments in astrophysics, like those associated with the early universe, core collapse supernovae, and compact object mergers, neutrinos may carry a significant fraction of the energy and entropy. The way these particles interact with and communicate with the medium is through the weak interaction. As a consequence, ascertaining the flavor states (weak interaction states) of the neutrino fields is these environments can be a key part of understanding, for example, how neutrinos set the neutron-to-proton ratio \cite{Qian93} and deposit energy in supernovae \cite{Bethe:1980zr,Bethe:1985lr,Fuller:1992eu,Dasgupta:2011uq}, or whether neutrinos decouple in mass or in flavor states in the very early universe \cite{Fuller:2009lr,Dodelson:2009qy}.

A feature of both the early universe and core collapse supernovae is that neutrinos propagate from very hot, high energy density regions or epochs, where transport mean free paths could be short compared to neutrino flavor oscillation lengths, to environments where the opposite is true. (We know that collective neutrino oscillations can readily occur in the latter regime, as reviewed in Ref.~\cite{Duan:2010fr} and references therein, and can be sensitive to small-scale density inhomogeneities \cite{Sawyer:1990lr,Loreti:1994qy,Loreti:1995fk,Kneller:2008rt,Kneller:2010ys} and the angular distribution of neutrino flux \cite{Raffelt:2013qy,Mirizzi:2013uq,Mirizzi:2013fj}.) 

Between these extremes, a poorly understood and complicated interplay of coherent neutrino flavor oscillations and scattering-induced de-coherence can govern how flavor develops. Partly because of this complication, modelers of supernova neutrino propagation with energy and flavor evolution have relied on a clear separation of regimes: Boltzmann equation treatments inside the proto-neutron star, and in the vicinity of the chemical and thermal equilibrium decoupling zone (neutrino sphere); and a coherent treatment in which only forward-scattering is considered in the low density environment sufficiently far above the neutron star. 

However, at some level these regimes cannot be separated. Indeed, recent work \cite{Cherry:2012lr} shows that in some supernova envelope models, well above the neutrino sphere, neutrinos which suffer direction-changing scattering, though comprising only a seemingly negligible fraction ({\it e.g.,} one in a thousand) of all neutrinos coming from the neutron star, nevertheless may make significant contributions to the potentials which govern flavor transformation. Though this neutrino \lq\lq halo\rq\rq\ effect has been argued \cite{Sarikas:2012fk,Mirizzi:2012qy} to make little difference in flavor evolution during the supernova accretion phase, in the one completely self-consistent calculation \cite{Cherry:2013lr} that has been done to date it produces a significant modification in collective neutrino oscillations and the expected signal for an O-Ne-Mg core collapse neutronization burst.

These studies point out that understanding neutrino flavor evolution in some supernova and compact object merger environments ultimately may require following the interplay of nuclear composition, three-dimensional radiation hydrodynamics, and the QKEs for neutrino flavor. From a computational astrophysics modeling standpoint, the essential complication of the QKEs over conventional Boltzmann neutrino transport schemes is the necessity of following high frequency quantum flavor oscillations along with scattering. The QKEs we derive in this paper are no exception. And though our QKEs can have the expected physically intuitive limits of being Schr\"odinger-like at low density and Boltzmann-like in scattering-dominated regions, they also have features that are new and surprising, and which were not revealed by more {\it ad hoc} treatments. 

Chief among these is the possibility of neutrino spin coherence. Since that, in principle, could mediate transformation between neutrinos and antineutrinos, it could be of importance in understanding compact object physics and nucleosynthesis as outlined above. The asymmetry between $\nu_e$ and $\bar\nu_e$ flowing from compact object environments can be, for example, a key arbiter of neutrino energy deposition and neutrino-heated nucleosynthesis. However, as will be evident in our subsequent exposition, implementing our QKEs in realistic simulations of astrophysical environments may require a radical alteration of the current approaches, and possibly a leap in computing capabilities. 

In what follows we give some background on two-component spinor notation and introduce our model for Majorana neutrinos in Section II.  We also describe how to extend our treatment to Dirac neutrinos.  We present the approach for deriving equations of motion for neutrino correlation functions from quantum field theory in Section III.  In Section IV we relate these correlation functions to physical quantities, such as neutrino densities and coherence terms, and present a scheme for perturbative expansion of the equations of motion.  We then derive the kinetic equations for neutrino densities and coherence terms in Section V, and calculate the potentials that describe neutrino interactions with matter in Section VI.  
In Section VII, we present a discussion of some properties of the quantum kinetic equations, identifying the limits in which we obtain Schr\"odinger-like flavor evolution and Boltzmann-like kinetics.  Also, in Section VII we identify some potential novel phenomena that are absent in the approximate treatments, including the possibility of coherent conversion between neutrinos and anti-neutrinos.
In Section VIII  we compare our work to existing approaches to neutrino QKEs and in Section IX we present our conclusions.

\section{Preliminaries}
\subsection{Two-Component Spinor Notation}
In this paper, we will primarily use two-component spinor notation, common in the supersymmetry literature and explained in detail in Ref.~\cite{Martin:2011v6}, an arXiv-published monograph by Stephen P. Martin, and Ref.~\cite{Dreiner:2010lr}.  A key reason for this choice of notation is that the two-component language is the most natural one for describing ultra-relativistic Majorana neutrinos. Moreover, this notation allows us to neatly separate components of physical quantities in a way that corresponds to their different physical meaning.  In this section, we briefly review two-component spinor notation and the relation to four-component spinor notation.

The Lorentz group, ${\rm{SO}}\left(3,1\right)$, is equivalent to ${\rm{SU}}\left(2\right)_L\times{\rm{SU}}\left(2\right)_R$.  Left-handed two-component spinors are objects that transform in the $\left(2,1\right)$  representation of the Lorentz group ${\rm{SU}}\left(2\right)_L\times{\rm{SU}}\left(2\right)_R$, while right-handed two-component spinors transform in the $\left(1,2\right)$ representation.  By convention, left-handed spinors are labeled by undotted two-component indices, $\alpha, \beta$, etc, while right-handed spinors are labeled by dotted indices, $\dot{\alpha},\dot{\beta}$, etc.  The presence or absence of a dot on a spinor index simply indicates which ${\rm{SU}}\left(2\right)$ factor is associated with the index.

Hermitian conjugation interchanges ${\rm{SU}}\left(2\right)_L$ and ${\rm{SU}}\left(2\right)_R$, so the Hermitian conjugate of a left-handed spinor is a right-handed spinor:  $\psi^{\dagger\dot{\alpha}}\equiv\left(\psi^\alpha\right)^\dagger$.  We adopt the convention that left-handed spinors (those with undotted indices) are always written without the dagger symbol, while right-handed spinors are always written with the dagger.

Four-component spinors are objects that transform in the $\left(2,1\right)+\left(1,2\right)$ representation of the Lorentz group.   A four-component Dirac spinor consists of two independent two-component spinors, and can be written as $\Psi_D=\left(\chi_\alpha,\xi^{\dagger\dot{\alpha}}\right)$.  A four-component Majorana spinor consists of a two-component spinor and its Hermitian conjugate:  $\Psi_M=\left(\psi_\alpha,\psi^{\dagger\dot{\alpha}}\right)$.  

Note that a Dirac spinor has the same physical content as two Majorana spinors, and therefore Dirac spinors can always be represented as pairs of Majorana spinors.  We will always do so; for example, we represent the charged leptons, which are Dirac spinors, as pairs of Majorana spinors (the lepton and the anti-lepton).  In this paper, the statement that a pair of Majorana spinors forms a Dirac spinor should be taken to mean that the Lagrangian has a $U\left(1\right)$ symmetry under which the two Majorana fields carry opposite charge.  This symmetry constrains the mass term to be proportional to a product of the two oppositely charged fields.

Two-component spinor indices can be raised or lowered with the antisymmetric symbol $\epsilon^{\alpha\beta}$ or $\epsilon^{\dot{\alpha}\dot{\beta}}$, both variants defined by $\epsilon^{12}=-\epsilon^{21}=1$ and $\epsilon_{21}=-\epsilon_{12}=1$.  A raised and a lowered index can be contracted (summed over), provided the indices are either both dotted or both undotted.  Due to the antisymmetric nature of $\epsilon^{\alpha\beta}$, $\psi_\alpha\chi^\alpha=-\psi^\alpha\chi_\alpha$, and similarly for the dotted indices.  

By convention, contracted undotted indices are always written with the first index raised, {\it e.g.,} $\psi^\alpha\chi_\alpha$, while contractions on dotted indices are written with the first index lowered, {\it e.g.,} $\psi^\dagger_{\dot{\alpha}}\chi^{\dagger\dot{\alpha}}$.  This allows us to adopt an index-free notation for contraction of spinor indices:  $\psi\chi\equiv\psi^\alpha\chi_\alpha$ and $\psi^\dagger\chi^\dagger\equiv\psi^\dagger_{\dot{\alpha}}\chi^{\dagger\dot{\alpha}}$.

In this paper, we will primarily deal with spinor bilinears.  These quantities can either carry two undotted indices, two dotted indices, or one of each.  All spinor bilinears can be written in terms of Lorentz tensors and Lorentz invariant spinor matrices:
\begin{eqnarray}
\label{eq:1}
\Gamma_{\alpha\dot{\alpha}} & = & \Gamma^L_\mu \sigma^\mu_{\alpha\dot{\alpha}}
\nonumber\\
\Gamma^{\dot{\alpha}\alpha} & = & \Gamma^R_\mu \bar{\sigma}^{\mu \dot{\alpha}\alpha}
\nonumber\\
\Gamma_\alpha^{\ \beta} & = & \Gamma^L\delta_\alpha^{\ \beta}+\frac{1}{2}i\Gamma^L_{\mu\nu}\left(S_L^{\mu\nu}\right)_\alpha^{\ \beta}
\nonumber\\
\Gamma^{\dot{\alpha}}_{\ \dot{\beta}} & = & \Gamma^R\delta^{\dot{\alpha}}_{\ \dot{\beta}}+\frac{1}{2}i\Gamma^R_{\mu\nu}\left(S_R^{\mu\nu}\right)^{\dot{\alpha}}_{\ \dot{\beta}}
\end{eqnarray}
where $\mu$ and $\nu$ are conventional spacetime indices, {\it i.e.,} assuming values $0$, $1$, $2$, or $3$.  

The labels $L$ and $R$ on the various components of $\Gamma$ are used to indicate which spinor bilinear the component belongs to.  The basis spinor matrices are given by

\begin{eqnarray}
\label{eq:2}
\sigma^\mu & = & \left(1,\vec{\sigma}\right)
\nonumber\\
\bar{\sigma}^\mu & = & \left(1,-\vec{\sigma}\right)
\nonumber\\
\left(S_L^{\mu\nu}\right)_\alpha^{\ \beta} & = & -\frac{1}{4}i\left(\sigma_{\alpha\dot{\alpha}}^\mu\bar{\sigma}^{\nu\dot{\alpha}\beta}-\sigma_{\alpha\dot{\alpha}}^\nu\bar{\sigma}^{\mu\dot{\alpha}\beta}\right)
\nonumber\\
\left(S_R^{\mu\nu}\right)^{\dot{\alpha}}_{\ \dot{\beta}} & = & \frac{1}{4}i\left(\bar{\sigma}^{\mu\dot{\alpha}\alpha}\sigma^\nu_{\alpha\dot{\beta}}-\bar{\sigma}^{\nu\dot{\alpha}\alpha}\sigma^\mu_{\alpha\dot{\beta}}\right)
\end{eqnarray}

The signs in the definitions of $S_L$ and $S_R$ are a matter of convention.

The spinor matrices $\sigma^\mu$ and $\bar{\sigma}^\mu$ satisfy the following relations:
\begin{eqnarray}
\label{eq:3}
\sigma^\mu_{\alpha\dot{\alpha}}\bar{\sigma}^{\nu\dot{\alpha}\beta}+\sigma^\nu_{\alpha\dot{\alpha}}\bar{\sigma}^{\mu\dot{\alpha}\beta}=2g^{\mu\nu}\delta_{\alpha}^{\ \beta}
\nonumber\\
\bar{\sigma}^{\mu\dot{\alpha}\alpha}\sigma^\nu_{\alpha\dot{\beta}}+\bar{\sigma}^{\nu\dot{\alpha}\alpha}\sigma^\mu_{\alpha\dot{\beta}}=2g^{\mu\nu}\delta^{\dot{\alpha}}_{\ \dot{\beta}}
\end{eqnarray}
where $g^{\mu\nu}$ is the usual spacetime (inverse) metric

It can be shown that the antisymmetric tensor quantities $\left(S_L^{\mu\nu}\right)$ and $\left(S_R^{\mu\nu}\right)$ are anti-self-dual and self-dual, respectively; that is, $S_L^{\mu\nu}=-i\left(S_L^{\mu\nu}\right)^\star$ and $S_R^{\mu\nu}=i\left(S_R^{\mu\nu}\right)^\star$, where $\left(T^{\mu\nu}\right)^\star\equiv\frac{1}{2}\epsilon^{\mu\nu\rho\sigma}T_{\rho\sigma}$.  Anti-self-dual and self-dual antisymmetric tensors transform in separate irreducible representations of the Lorentz group, specifically in $\left(3,1\right)$ and $\left(1,3\right)$, respectively.  Since $\Gamma^L_{\mu\nu}$ can be expressed using the basis of $S_L^{\mu\nu}$ matrices, it is an anti-self-dual tensor, while $\Gamma^R_{\mu\nu}$ is a self-dual tensor.

We can use index-free notation to denote products of spin matrices, using the conventions given above for contracting dotted and undotted indices, and in addition assuming that contractions are performed in the usual order of matrix multiplication.  For example,
\begin{eqnarray}
\label{eq:4}
\sigma^\mu\bar{\sigma}^\nu\sigma^\rho=\left(\sigma^\mu\bar{\sigma}^\nu\sigma^\rho\right)_{\alpha\dot{\alpha}}=\sigma^\mu_{\alpha\dot{\beta}}\bar{\sigma}^{\nu\dot{\beta}\beta}\sigma^\rho_{\beta\dot{\alpha}}
\end{eqnarray}

Products of $\sigma$ or $\bar{\sigma}$ matrices can always be written in terms of the basis matrices $\delta, \sigma, \bar{\sigma}, S_L$ and $S_R$.  The products of three $\sigma$ or $\bar{\sigma}$ matrices are
\begin{eqnarray}
\label{eq:5}
\sigma^\mu\bar{\sigma}^\nu\sigma^\rho=g^{\mu\nu}\sigma^\rho-g^{\mu\rho}\sigma^\nu+g^{\nu\rho}\sigma^\mu+i\epsilon^{\mu\nu\rho}_{\ \ \ \sigma}\sigma^\sigma
\nonumber\\
\bar{\sigma}^\mu\sigma^\nu\bar{\sigma}^\rho=g^{\mu\nu}\bar{\sigma}^\rho-g^{\mu\rho}\bar{\sigma}^\nu+g^{\nu\rho}\bar{\sigma}^\mu-i\epsilon^{\mu\nu\rho}_{\ \ \ \sigma}\bar{\sigma}^\sigma
\end{eqnarray}

Products of four or more $\sigma$ matrices can be systematically reduced to expressions involving only the basis matrices, by repeated use of equations (\ref{eq:3}), (\ref{eq:5}), and the definitions of $S_L^{\mu\nu}$ and $S_R^{\mu\nu}$.

We will often use 4-component spinor bilinears which combine all four types of two-component spinor bilinears into a single $4\times 4$ matrix:
\begin{eqnarray}
\label{eq:6}
\Gamma\equiv\left(\begin{array}{cc}\Gamma_\alpha^{\ \beta} & \Gamma_{\alpha\dot{\beta}} \\
\Gamma^{\dot{\alpha}\beta} & \Gamma^{\dot{\alpha}}_{\ \dot{\beta}}\end{array}\right)
\end{eqnarray}

With the spinor indices arranged as in equation $(6)$, we can write contractions of 4-component spinor bilinears in an index-free way. That is, if $\Gamma$ and $\Delta$ are $4\times 4$ spin matrices having the form of equation $(6)$, so is the product $\Gamma\Delta$, where it is understood that $\Gamma$ and $\Delta$ are contracted together in the usual manner of matrix multiplication.

In this paper we have adopted a commonly used representation of 4-component spinor matrices $\gamma^\mu$ and $\gamma^5$ where
\begin{eqnarray}
\label{eq:7}
\gamma^\mu=\left(\begin{array}{cc}0 & \sigma^\mu \\ \bar{\sigma}^\mu & 0\end{array}\right)\ \ \ \ \ 
\gamma^5 = \left(\begin{array}{cc}-1 & 0 \\ 0 & 1\end{array}\right)
\end{eqnarray}

Choice of a particular representation of these matrices provides a dictionary by which expressions in 2-component spinor notation can be translated to standard 4-component spinor notation, and {\it vice versa}.

\subsection{The Model}
In what follows we will consider Standard Model neutrinos with small Majorana masses.  We will work in the low-energy limit, where the energy of the particles is much smaller than the $W$ and $Z$ boson masses, so that the $W$ and $Z$ bosons are not dynamical.  In this paper we will not consider the interactions of neutrinos with nucleons and nuclei; these interactions in certain limits and environments can be similar to the interactions of neutrinos with charged leptons. The ultimate forms of the QKEs we develop are crafted to allow straightforward incorporation of these interactions when necessary for realistic calculations.  As a consequence, for simplicity we will restrict our development to the lepton sector.

After breaking electroweak symmetry, the Standard Model Lagrangian in the lepton sector is:
\begin{eqnarray}
\label{eq:8}
i\psi^\dagger_I \bar{\sigma}_\mu \partial^\mu \psi_I+ie^\dagger_I\bar{\sigma}_\mu \partial^\mu e_I+i\bar{e}^\dagger_I\bar{\sigma}_\mu \partial^\mu\bar{e}_I
\nonumber\\
-\frac{1}{2}m_{IJ}\psi_I\psi_J-m^e_{IJ} e_I \bar{e}_J
\nonumber\\
+e^\dagger_I \frac{g\left(2\sin^2\theta_W-1\right)\bar{\sigma}_\mu Z^\mu}{2\cos\theta_W}e_I
+\bar{e}_I^\dagger\frac{g\sin^2\theta_W\bar{\sigma}_\mu Z^\mu}{\cos\theta_W}\bar{e}_I
\nonumber\\
+\psi_I^\dagger\frac{g\bar{\sigma}_\mu Z^\mu}{2\cos\theta_W}\psi_I
+\psi_I^\dagger\frac{g\bar{\sigma}_\mu W^{+\mu}}{\sqrt{2}}e_I+e^\dagger_I\frac{g\bar{\sigma}_\mu W^{-\mu}}{\sqrt{2}}\psi_I
\nonumber\\
+e^\dagger_I g_e\bar{\sigma}_\mu A^\mu e_I-\bar{e}^\dagger_I g_e \bar{\sigma}_\mu A^\mu\bar{e}_I
\nonumber\\
-M_W^2 W^+_\mu W^{-\mu}-\frac{1}{2}M_Z^2 Z_\mu Z^\mu
\nonumber\\
+\textrm{ gauge boson kinetic terms}+{\rm h.c.}\ \ \ \ \ \ \ \ 
\end{eqnarray}
Here, $\psi_I$ is the neutrino field, where $I$ is the flavor index.   In this notation $e_I$ and $\bar{e}_I$ are the charged lepton fields, where the former describes left-handed electrons (muons, tauons) and right-handed positrons, and the latter is its Dirac counterpart, describing right-handed electrons and left-handed positrons.  $A^\mu$ is the photon field, $Z^\mu$ and $W^{\pm\mu}$ are the weak boson fields.  $M_W$ and $M_Z$ are the $W$ and $Z$ boson masses.  $g_e$ is the electromagnetic coupling constant (electron charge), $g$ is the weak coupling constant, and $\theta_W$ is the Weinberg angle.  $m_{IJ}$ is the Majorana mass matrix for neutrinos, and $m^e_{IJ}$ is the Dirac mass matrix for charged fermions.  In the flavor basis, $m^e_{IJ} = {\rm{diag}}\left(m_e, m_\mu, m_\tau\right)$, where $m_e$ is the electron mass, $m_\mu$ is the muon mass, and $m_\tau$ is the tauon mass.  For Majorana neutrinos, $m_{IJ}=m_{JI}$.

\subsection{Feynman Rules}

To compute various quantities that arise in the quantum kinetic equations, we will need the Feynman rules that are derived from the Lagrangian.  In deriving the Feynman rules, we make several assumptions.  First, we assume that the energy of the neutrinos and charged leptons is much smaller than the $W$ and $Z$ boson masses, and thus the $W$ and $Z$ bosons are not dynamical and we can neglect their kinetic terms.  Second, in this low-energy regime, the electromagnetic interaction is much stronger than the weak interaction, and the distributions of charged particles thermalize on a much shorter timescale than the neutrino distributions.  Therefore we will follow the dynamics of neutrinos associated with the weak interaction, and make the assumption, valid for the astrophysical regimes of interest to us, that the effect of the electromagnetic interaction is simply to ensure that the plasma (charged leptons, described by the fields $e_I$ and $\bar{e}_I$, and photons, described by the field $A^\mu$) can be adequately represented as thermal distributions of particles.

The Feynman rules for the weak interaction vertices are
\begin{eqnarray}
\label{eq:9}
\Diagram{\vertexlabel^{\nu^{\alpha J}} \\ fdV & g \vertexlabel_{Z_\mu} \\ \vertexlabel_{\nu^{\dot{\alpha}I}} fuA\\}\ &=& \frac{-ig}{2\cos\theta_W}\delta^{IJ}\bar{\sigma}_\mu^{\dot{\alpha}\alpha}\ {\rm{or}}\ \ 
\frac{ig}{2\cos\theta_W}\delta^{IJ}\sigma^\mu_{\alpha\dot{\alpha}}
\nonumber\\\nonumber\\\nonumber\\
\Diagram{\vertexlabel^{e^{\alpha J}} \\ fdV & g \vertexlabel_{Z_\mu} \\ \vertexlabel_{e^{\dot{\alpha}I}} fuA\\}\ &=& -ig\frac{\sin^2\theta_W-\frac{1}{2}}{\cos\theta_W}\delta^{IJ}\bar{\sigma}_\mu^{\dot{\alpha}\alpha}
\nonumber\\
&{\rm or}&\ \ 
ig\frac{\sin^2\theta_W-\frac{1}{2}}{\cos\theta_W}\delta^{IJ}\sigma^\mu_{\alpha\dot{\alpha}}
\nonumber\\\nonumber\\\nonumber\\
\Diagram{\vertexlabel^{\bar{e}^{\alpha J}} \\ fdV & g \vertexlabel_{Z_\mu} \\ \vertexlabel_{\bar{e}^{\dot{\alpha}I}} fuA\\}\ &=& 
-ig\frac{\sin^2\theta_W}{\cos\theta_W}\delta^{IJ}\bar{\sigma}_\mu^{\dot{\alpha}\alpha}
\nonumber\\
&{\rm or}&\ \ 
ig\frac{\sin^2\theta_W}{\cos\theta_W}\delta^{IJ}\sigma^\mu_{\alpha\dot{\alpha}}
\nonumber\\\nonumber\\\nonumber\\
\Diagram{\vertexlabel^{\nu^{\alpha J}} \\ fdV & g \vertexlabel_{W_\mu} \\ \vertexlabel_{e^{\dot{\alpha}I}} fuA\\}&=& 
\Diagram{\vertexlabel^{e^{\alpha J}} \\ fdV & g \vertexlabel_{W_\mu} \\ \vertexlabel_{\nu^{\dot{\alpha}I}} fuA\\} = \frac{-ig}{\sqrt{2}}\delta^{IJ}\bar{\sigma}_\mu^{\dot{\alpha}\alpha} \ {\rm{or}} \ 
\frac{ig}{\sqrt{2}}\delta^{IJ}\sigma^\mu_{\alpha\dot{\alpha}}
\end{eqnarray}
Whether the $\bar{\sigma}$ or the $\sigma$ version of the vertex is used depends on the two-component index structure of the diagram.  The requirement that spinor indices be contracted in the usual order of matrix multiplication unambiguously determines which form of the vertex appears in the expression.

Next, we write down the Feynman rules for the propagators.  In this paper we will be calculating quantities derived from the 2PI (two-particle irreducible) effective action.   In this formalism, fermion lines represent the full expressions for neutrino and charged lepton two-point functions; these two-point functions are, in general, dynamical quantities that depend on particle densities and interactions. They are not just the vacuum propagators.   In position space, we will write the general form of the neutrino two-point functions as
\begin{eqnarray}
\label{eq:10}
\feyn{\vertexlabel_{\dot{\alpha},I,x} !{fA}{\ \ \ \ \ \ \nu} fA \vertexlabel_{\alpha, J, y}}\ = G_{\nu, IJ}^{\dot{\alpha}\alpha}\left(x,y\right)
\ \ 
\feyn{\vertexlabel_{\alpha,I,x} !{fV}{\ \ \ \ \ \ \nu} fV \vertexlabel_{\dot{\alpha}, J, y}}\ = G_{\nu, IJ}^{\alpha\dot{\alpha}}\left(x,y\right)
\nonumber\\\nonumber\\
\feyn{\vertexlabel_{\dot{\alpha},I,x} !{fA}{\ \ \ \ \ \ \nu} fV \vertexlabel_{\dot{\beta}, J, y}}\ = G_{\nu, IJ}^{\dot{\alpha}\dot{\beta}}\left(x,y\right)
\ \ 
\feyn{\vertexlabel_{\alpha,I,x} !{fV}{\ \ \ \ \ \ \nu} fA \vertexlabel_{\beta, J, y}}\ = G_{\nu, IJ}^{\alpha\beta}\left(x,y\right)
\end{eqnarray}

The two-point functions are defined as time-ordered expectation values of spinor field bilinears.  Thus, for example, $G_{\nu, IJ}^{\alpha\dot{\alpha}}\left(x,y\right) = \left<{\rm T}_P\left(\psi^\alpha_I\left(x\right)\psi^{\dagger\dot{\alpha}}_J\left(y\right)\right)\right>$, and similarly for the other components of $G$.  Here, ${\rm T}_P$ is the time ordering operator along a specific path.  As we explain below, we will use the closed time path (CTP) contour.  Since we are dealing with out of equilibrium, non-vacuum states described by a nontrivial density operator, the brackets, $<>$, denote an ensemble average rather than a vacuum expectation value.

Note that in two-component spinor notation the arrows on fermion propagators do not denote the flow of momentum or any conserved current, but rather simply indicate whether the two-component spinor index associated with the arrow is dotted or undotted. This is illustrated in the above equations for the two-point functions. For example, it can be seen that \lq\lq clashing arrows,\rq\rq\ where the arrows point toward each other, correspond to two point functions with right-handed spinor indices, while diverging arrows go with left-handed spinor indices, {\it etc.} 

As described below, the two-point function contains both the vacuum propagator and the particle density matrix. The density matrix encodes the particle occupation numbers and additional degrees of freedom describing flavor and possibly spin (handedness) coherence.  We will treat the neutrino two-point function as a fully dynamical entity, the time development of which allows us to solve for the time evolution of the neutrino occupation numbers. 

Similarly, the general Feynman rules for the charged lepton two-point functions are:
\begin{eqnarray}
\label{eq:11}
\feyn{\vertexlabel_{\dot{\alpha},I,x} !{fA}e !{fA}e \vertexlabel_{\alpha, J, y}}\ = G_{e, IJ}^{\dot{\alpha}\alpha}\left(x,y\right)
\ \ 
\feyn{\vertexlabel_{\alpha,I,x} !{fV}e !{fV}e \vertexlabel_{\dot{\alpha}, J, y}}\ = G_{e, IJ}^{\alpha\dot{\alpha}}\left(x,y\right)
\nonumber\\\nonumber\\
\feyn{\vertexlabel_{\dot{\alpha},I,x} !{fA}{\bar{e}} !{fA}{\bar{e}} \vertexlabel_{\alpha, J, y}}\ = G_{\bar{e}, IJ}^{\dot{\alpha}\alpha}\left(x,y\right)
\ \ 
\feyn{\vertexlabel_{\alpha,I,x} !{fV}{\bar{e}} !{fV}{\bar{e}} \vertexlabel_{\dot{\alpha}, J, y}}\ = G_{\bar{e}, IJ}^{\alpha\dot{\alpha}}\left(x,y\right)
\nonumber\\\nonumber\\
\feyn{\vertexlabel_{\dot{\alpha},I,x} !{fA}e !{fV}{\bar{e}} \vertexlabel_{\dot{\beta}, J, y}}\ = G_{e\bar{e}, IJ}^{\dot{\alpha}\dot{\beta}}\left(x,y\right)
\ 
\feyn{\vertexlabel_{\alpha,I,x} !{fV}e !{fA}{\bar{e}} \vertexlabel_{\beta, J, y}}\ = G_{e\bar{e}, IJ}^{\alpha\beta}\left(x,y\right)
\end{eqnarray}

In this development we will assume that the charged lepton distributions are thermal. With this assumption, the form of the charged lepton two-point function will depend only on the charged lepton temperature, chemical potential, and mass.

Note that since charged leptons are Dirac particles, the arrow-clashing propagator for charged leptons always connects the charged lepton field with its Dirac counterpart.  On the other hand, for Majorana neutrinos, the arrow-clashing propagator connects the field to itself.

In the low-energy limit the electroweak bosons are not dynamical, and their position space Feynman rules are simply given by
\begin{eqnarray}
\label{eq:12}
\feyn{\vertexlabel_{\mu,x} !{g}Z \vertexlabel_{\nu, y}}\ = \frac{ig^{\mu\nu}}{M_Z^2}\delta^4\left(x-y\right)\ \ \ \ \feyn{\vertexlabel_{\mu,x} !{g}W \vertexlabel_{\nu,y}}\ = \frac{ig^{\mu\nu}}{M_W^2}\delta^4\left(x-y\right)
\end{eqnarray}
Here, we have used the Feynman gauge, but other choices of gauge give physically equivalent expressions.\\

We will often express combinations of coupling constants and electroweak boson masses that appear in the Feynman diagrams in terms of the Fermi constant
\begin{eqnarray}
\label{eq:13}
G_F\equiv\frac{g^2}{4\sqrt{2}M_W^2}
\end{eqnarray}
and use
\begin{eqnarray}
\label{eq:14}
\cos\theta_W=\frac{M_W}{M_Z}
\end{eqnarray}

It is sometimes convenient to denote the combination of all components of a two-point function or vertex by omitting the arrows.  This is equivalent to using the four-component spinor notation. For example, we can write

\begin{eqnarray}
\label{eq:15}
\feyn{\vertexlabel_{I,x} !{f}\nu !{f}\nu \vertexlabel_{J, y}}\ = G_{\nu, IJ}\left(x,y\right),
\end{eqnarray}
where
\begin{eqnarray}
\label{eq:16}
G_{\nu, IJ}\equiv\left(\begin{array}{cc}{\left(G_{\nu,IJ}\right)_{\alpha}}^{\beta} & {\left(G_{\nu,IJ}\right)}_{{\alpha}{\dot{\beta}}} \\ \left(G_{\nu,IJ}\right)^{\dot{\alpha} {\beta}} & \left(G_{\nu,IJ}\right)^{\dot{\alpha}}_{\ \dot{\beta}}\end{array}\right).
\end{eqnarray}
The use of diagrams without arrows is simply shorthand notation which implies a sum of every possible combination of arrow directions that gives a nonzero contribution to the amplitude.

\section{Equations of Motion for the Two-Point Function}
\subsection{2PI Effective Action and the Two-Point Function}
The equations of motion for neutrino two-point functions can be derived from the two-particle irreducible (2PI) effective action.  The complete, general procedure is presented in Ref.s~\cite{Berges:2004qy,Berges:2005lr}.  Here, we outline the key steps in this derivation as they apply to the dynamics of neutrinos.  

The 2PI effective action is a functional of the two-point function $G=G^{ab}_{IJ}\left(x,y\right)$, corresponding to equation $(16)$, where $a$ and $b$ are four-component spinor indices (for example, $a=\left(\alpha,\dot{\alpha}\right)$), $I$ and $J$ are flavor indices, and $x$ and $y$ are position four-vectors.  The 2PI effective action consists of Feynman diagrams with no external lines that are two-particle irreducible, that is, cannot be disconnected by cutting two fermion lines (we do not consider cutting weak boson lines, since the weak bosons are not dynamical in our formalism, and can be reduced to 4-fermion vertices).  We separate the $2PI$ effective action into a 1-loop piece (a single fermion loop, the only contribution to $\Gamma^{2PI}$ in free field theory), and the rest:

\begin{eqnarray}
\label{eq:17}
\Gamma^{2PI}\left[G\right] = \Gamma^{2PI}_1\left[G\right] + \Gamma^{2PI}_2\left[G\right].
\end{eqnarray}
In this equation $\Gamma_1$ is the one-loop expression, and $\Gamma_2$ is the sum of all higher-loop contributions.  The diagrams are drawn and calculated, in position space, as usual, except that the general form for the two-point functions is used instead of the tree-level propagator, thereby incorporating effects from nonzero particle density and corrections to the propagator stemming from interactions.  We use the general result from quantum field theory:
\begin{eqnarray}
\label{eq:18}
\Gamma^{2PI}_1=-i\left({\rm Tr} \ln G^{-1}+{\rm Tr}\ G_0^{-1}G\right)
\end{eqnarray}
where $G_0^{-1}$ is the tree-level inverse propagator, and $G$ is the complete dynamical two-point correlation function.  Here, we are suppressing spin and flavor indices, but the quantities in this expression are $4\times 4$ matrices in spin space and $3\times 3$ matrices in flavor space, with an explicit form given by the expression in equation (\ref{eq:16}).  Products and traces of such quantities in our equations imply contraction of both spinor and flavor indices in the usual order of matrix multiplication.

We can now find the equations of motion for $G$ by setting $\frac{\delta\Gamma^{2PI}\left[G\right]}{\delta G}=0$.  This gives the following expression:
\begin{eqnarray}
\label{eq:19}
G^{-1}\left(x,y\right)=G_0^{-1}\left(x,y\right)-\Sigma\left[x,y; G\right],
\end{eqnarray}
where we define
\begin{eqnarray}
\label{eq:20}
\Sigma\left[x,y; G\right]\equiv -i\frac{\delta\Gamma_2^{2PI}\left[G\right]}{\delta G\left(y,x\right)}.
\end{eqnarray}

Since $\Gamma_2^{2PI}$ is the sum of two-loop and higher order 2PI diagrams with no external lines, $\Sigma$ is proportional to the sum of one-loop and higher order 1PI diagrams with two external neutrino lines. Consequently, $\Sigma$ corresponds to the neutrino proper self-energy.  For the purposes of this paper, we will calculate $\Sigma$ to 2-loop order; the corresponding Feynman diagrams and calculations will be given in a subsequent section.

We can eliminate the dependence of equation (\ref{eq:19}) on $G^{-1}$ by acting from the right with $G$, to obtain
\begin{eqnarray}
\label{eq:21}
\left(i\not\partial^x-M\right)G\left(x,y\right)-i\int d^4z \Sigma\left(x,z\right)G\left(z,y\right)
\nonumber\\
={\rm \bf 1}\ i\delta^4\left(x-y\right)
\end{eqnarray}
where ${\not\partial}^x$ and $M$ are spin $\times$ flavor matrices given by ${\not\partial}^x = \left(\begin{array}{cc}0 & \sigma_{\alpha\dot{\alpha}}^\mu\partial^x_\mu \\ \bar{\sigma}^{\mu\dot{\alpha}\alpha}\partial^x_\mu & 0\end{array}\right)\delta_{IJ}$ and $M = \left(\begin{array}{cc}\delta_\alpha^{\ \beta}\ m_{IJ} & 0 \\ 0 & \delta^{\dot{\alpha}}_{\ \dot{\beta}}\left(m_{IJ}\right)^\dagger\end{array}\right)$.  Here ${\rm \bf 1}$ is the spin $\times$ flavor unit matrix, given by ${\rm \bf 1}=\left(\begin{array}{cc}\delta_\alpha^{\ \beta} & 0 \\ 0 & \delta^{\dot{\alpha}}_{\ \dot{\beta}}\end{array}\right)\delta_{IJ}$.

\subsection{Spectral and Statistical Functions}

We can use the dynamics of the two-point function $G$ to describe the evolution of neutrino distributions, starting with arbitrary non-equilibrium initial conditions, by employing the closed time path (CTP) formalism \cite{Berges:2004qy}.  In the CTP formalism, the time ordering in the path integral is taken along a closed real-time contour, starting from the point at which initial conditions are given, to the point in time of interest in the calculation, and then back to the initial point.  The two-point correlation function $G$ is time ordered on the CTP contour:  $G\left( x, y\right)=\left<{\rm T}_{CTP}{\left( \Psi\left(x\right)\bar\Psi\left(y\right)\right)}\right>$, where ${\rm T}_{CTP}$ is an operator that imposes time ordering with respect to the CTP contour, and $\Psi$ is a Majorana spinor given by $\Psi = \left(\psi_\alpha,\psi^{\dagger\dot{\alpha}}\right)$ and $\bar{\Psi}=\left(\psi^\alpha,\psi^\dagger_{\dot{\alpha}}\right)$. 

The time ordering can be made explicit by decomposing $G$ into the following components:
\begin{eqnarray}
\label{eq:22}
G\left(x,y\right)=F\left(x,y\right)-\frac{1}{2}i\rho\left(x,y\right){\rm{sign}}_{CTP}\left(x^0-y^0\right)
\end{eqnarray}
where ${\rm{sign}}_{CTP}$ is a function of the ordering of $x$ and $y$ along the time path, taking on a value of $1$ or $-1$, depending on whether $y$ precedes or follows $x$ on the CTP contour.  For fermions, $F$ and $\rho$ are defined as follows:
\begin{eqnarray}
\label{eq:23}
F\left(x,y\right)=\frac{1}{2}\left<\left[\Psi\left(x\right),\bar{\Psi}\left(y\right)\right]\right>
\end{eqnarray}
\begin{eqnarray}
\label{eq:24}
\rho\left(x,y\right)=i\left<\left\{\Psi\left(x\right),\bar{\Psi}\left(y\right)\right\}\right>.
\end{eqnarray}
In the above expressions, $\rho$ is the spectral function, and carries information on the particle states that can appear in the theory; it is related to the usual vacuum propagator.  $F$ is the statistical function, and encodes the occupation numbers of these states.  Since we wish to solve for the evolution of neutrino occupation numbers, we will primarily be interested in the dynamics of the statistical function $F$.

Similarly, we decompose the neutrino self-energy $\Sigma$ into a local piece, plus spectral and statistical components:
\begin{eqnarray}
\label{eq:25}
\Sigma\left(x,y\right) = -i\Sigma\left(x\right)\delta^4_{\rm CTP}\left(x-y\right)
\nonumber\\
+\Pi_F\left(x,y\right)-\frac{1}{2}i\Pi_\rho\left(x,y\right){\rm{sign}}_{\rm CTP}\left(x^0-y^0\right).
\end{eqnarray}
We will show how to compute these components later, but for now, we note that for our model, the local term $\Sigma\left(x\right)$ contains contributions from 1-loop diagrams, while the spectral and statistical terms contain only contributions from 2-loop and higher diagrams.  Thus, the $\Pi_\rho\left(x,y\right)$ and $\Pi_F\left(x,y\right)$ terms carry higher powers of the coupling constant than does $\Sigma\left(x\right)$

Using equations (\ref{eq:22}) and (\ref{eq:25}) in (\ref{eq:21}) gives the following equation for the statistical function:
\begin{eqnarray}
\label{eq:26}
\left(i{\not\partial}^x-M-\Sigma\left(x\right)\right) F\left(x,y\right)=\nonumber\\
\int_{0}^{x^0}dz^0\int d^3z\ \Pi_{\rho}\left(x,z\right)F\left(z,y\right)\nonumber\\
-\int_{0}^{y^0}dz^0\int d^3z\ \Pi_{F}\left(x,z\right)\rho\left(z,y\right).
\end{eqnarray}

In addition, there is another form of the equation for $F$, which is obtained by acting on equation (\ref{eq:19}) from the left with $G$, then separating into spectral and statistical components. This gives
\begin{eqnarray}
\label{eq:27}
F\left(x,y\right)\left(-i\overleftarrow{\not\partial}^y-M-\Sigma\left(y\right)\right)=\nonumber\\
\int_{0}^{y^0}dz^0\int d^3z\ F\left(x,z\right)\Pi_\rho\left(z,y\right)\nonumber\\
-\int_{0}^{x^0}dz^0\int d^3z\ \rho\left(x,z\right)\Pi_F\left(z,y\right)
\end{eqnarray}

There are similar equations for the spectral function.  However, for the purpose of this paper, we will not need these equations.  The reason is that the spectral function does not depend on the occupation numbers of particles, but rather only on the mass and the interaction strength.  For particles with a small mass and experiencing only weak interactions, $\rho$ will deviate only slightly from its massless, free-field value.  In equations (\ref{eq:26}) and (\ref{eq:27}), $\rho$ only enters in conjunction with $\Pi_F$, which is already at two-loop order.  Because we are only computing quantities to this order, any corrections to the spectral function due to the neutrino mass or interactions will give terms in the equation that are beyond the order of our expansion.  Thus, we can simply use the massless, free-field expression for the spectral function, which will be derived below.

\section{Wigner Transform and Separation of Scales}
\subsection{The Wigner Transform}
Equations (\ref{eq:26}) and (\ref{eq:27}) give the complete dynamics of the neutrinos, approximate only insofar as we are expanding $\Sigma$ to 2-loop order, and decoupling the dynamics of the spectral function from those of the statistical function by dropping higher-order terms on the right-hand side.  However, solving these equations in their current form is impractical.  First, the connection of the object $F\left(x,y\right)$ to actual neutrino occupation numbers is somewhat complicated, so the physical meaning of these equations is difficult to elucidate.  Second, the two-point function undergoes rapid oscillations, on the scale of the neutrino de Broglie wavelength, with respect to the relative coordinate $r=x-y$.  On the other hand, for weakly coupled particles, such as neutrinos, physically meaningful quantities change much more slowly, and vary as a function of the average coordinate, $X=\frac{1}{2}\left(x+y\right)$.  Resolving the rapid oscillations associated with the neutrino de Broglie wavelength is clearly undesirable from a computational standpoint.

We derive more useful expressions from (\ref{eq:26}) and (\ref{eq:27}) by performing a Wigner transform and then expanding in small parameters.  In this, we follow the procedure of Ref.~\cite{Cirigliano:2010lr}. (Applications of some of these techniques in the context of electroweak baryogenesis are presented in Ref.s~\cite{Prokopec:2004lr,Prokopec:2004fk,Konstandin:2005qy,Konstandin:2006uq,Cirigliano:2010yq,Cirigliano:2011lr}.)

To perform the Wigner transform, we change to the relative coordinate $r$ and the average coordinate $X$. Note that eventually, after the change of coordinates, we will simply name the average coordinate $x$; it should be clear from context whether $x$ refers to the average coordinate or to one of the two spacetime arguments of a two-point function.  We then Fourier transform with respect to the relative coordinate. The Wigner transform of the statistical function $F\left(x,y\right)$ is then:

\begin{eqnarray}
\label{eq:28}
F\left(X,k\right)\equiv
\int d^4r\ e^{ik\cdot r} F\left(X+\frac{1}{2}r,X-\frac{1}{2}r\right)
\end{eqnarray}

and similarly for other functions of $\left(x,y\right)$.

\subsection{Spectral and Statistical Functions for Free, Massless Fermions}
Before we Wigner transform equations (\ref{eq:26}) and (\ref{eq:27}), we derive the expressions for the spectral and statistical functions in terms of the particle densities, neglecting neutrino mass and interactions but allowing for nonzero neutrino densities.  Neutrino masses and interactions will result in slight changes to these expressions; we will later calculate these changes perturbatively.  As we will see, the Wigner transformed functions have a straightforward physical interpretation.  In particular, the Wigner transformed statistical function, $F\left(X,k\right)$, contains components proportional to neutrino and antineutrino density matrices, $f_{IJ}\left(X,k\right)$ and $\bar{f}_{IJ}\left(X,k\right)$, while the spectral function in free field theory contains no dynamical components, and therefore simply encodes the possible particle states.  For anisotropic particle distributions, $F\left(X,k\right)$ can contain an additional dynamical quantity, which can be interpreted as describing coherence between left-handed and right-handed fermion states.

We begin with the statistical function.  In terms of the 4-component Majorana spinor fields, this is given by
\begin{eqnarray}
\label{eq:29}
F_{IJ}\left(X,k\right) =
\nonumber\\
 \frac{1}{2}  \int d^4r\ e^{ik\cdot r}\left< \left[ \Psi_I \left(X+\frac{1}{2}r\right),\bar{\Psi}_J\left(X-\frac{1}{2}r\right)\right]\right>
\end{eqnarray}

For convenience of notation, we will evaluate this expression at $X=0$, and later generalize the results to any position $X$:
\begin{eqnarray}
\label{eq:30}
F_{IJ}\left(0,k\right)=\frac{1}{2}\int d^4r\ e^{ik\cdot r}\left<\left[\Psi_I\left(\frac{r}{2}\right),\bar{\Psi}_J\left(-\frac{r}{2}\right)\right]\right>
\end{eqnarray}

We will calculate the various components of $F$ in two-component spinor notation, in which the Majorana spinors are given by $\Psi_I=\left(\psi_{I,\alpha},\psi_I^{\dagger\dot{\alpha}}\right)$ and $\bar{\Psi}_J=\left(\psi_J^\beta,\psi^\dagger_{J,\dot{\beta}}\right)$.  First, we calculate
\begin{eqnarray}
\label{eq:31}
F_{IJ,\alpha\dot{\beta}}\left(0,k\right)
\nonumber\\
=\frac{1}{2}\int d^4 r\ e^{ik\cdot r}\left<\left[\psi_{I,\alpha}\left(\frac{r}{2}\right),\psi^\dagger_{J,\dot{\beta}}\left(-\frac{r}{2}\right)\right]\right>
\end{eqnarray}

The two-component spinor field $\psi_{I,\alpha}$ is given by
\begin{eqnarray}
\label{eq:32}
\psi_{I,\alpha}\left(x\right)=
\nonumber\\
\int \tilde{dq}\left(b_I\left(\vec{q}\right)u_\alpha\left(\vec{q}\right)e^{-iq\cdot x}+d^\dagger_I\left(\vec{q}\right)v_{\alpha}\left(\vec{q}\right)e^{iq\cdot x}\right)
\end{eqnarray}

In manifestly Lorentz invariant notation, $\tilde{dq}=\frac{d^4q}{\left(2\pi\right)^4}2\pi\delta\left(q^2\right)\theta\left(q^0\right)$.  $b_I\left(\vec{q}\right)$ is an operator that annihilates a left-handed neutrino of flavor $I$ and momentum $\vec{q}$, and $d^\dagger_I\left(\vec{q}\right)$ is an operator that creates a right-handed anti-neutrino of flavor $I$ and momentum $\vec{q}$.  Note that for Majorana neutrinos, particles and antiparticles simply correspond to opposite spin states; as a result, we could instead have used the spin-dependent operators $b_s$, where $s=\pm$.  In our notation, $b = b_-$ and $d = b_+$.  The creation and annihilation operators satisfy the anticommutation relations:
\begin{eqnarray}
\label{eq:33}
\left\{b_I\left(\vec{q}_1\right),b^\dagger_J\left(\vec{q}_2\right)\right\}=\left(2\pi\right)^3\delta^3\left(\vec{q}_1-\vec{q}_2\right)2E_q\delta_{IJ}
\nonumber\\
\left\{d_I\left(\vec{q}_1\right),d^\dagger_J\left(\vec{q}_2\right)\right\}=\left(2\pi\right)^3\delta^3\left(\vec{q}_1-\vec{q}_2\right)2E_q\delta_{IJ}
\end{eqnarray}
All other anticommutators are zero.

$u^\alpha\left(\vec{q}\right)$ and $v^\alpha\left(\vec{q}\right)$ are two-component spinors that satisfy
\begin{eqnarray}
\label{eq:34}
q_\mu\bar{\sigma}^{\mu\dot{\alpha}\alpha}u_\alpha\left(\vec{q}\right)=0
\nonumber\\
q_\mu\bar{\sigma}^{\mu\dot{\alpha}\alpha}v_\alpha\left(-\vec{q}\right)=0
\end{eqnarray}
where $q_\mu \equiv \left(q_0, \vec{q}\right)$, with the timelike component taken to be positive definite.  $u$ and $v$ are normalized as follows:
\begin{eqnarray}
\label{eq:35}
u_\alpha\left(\vec{q}\right)u^\dagger_{\dot{\beta}}\left(\vec{q}\right)=q_\mu\sigma^\mu_{\alpha\dot{\beta}}
\nonumber\\
v_\alpha\left(-\vec{q}\right)v^\dagger_{\dot{\beta}}\left(-\vec{q}\right)=-q_\mu\sigma^\mu_{\alpha\dot{\beta}}
\end{eqnarray}
Substituting equation $(32)$ into equation $(31)$ gives an expression with four terms:
\begin{eqnarray}
\label{eq:36}
F_{IJ,\alpha\dot{\beta}}\left(0,k\right)=\frac{1}{2}\int d^4r \int \tilde{dq}_1\tilde{dq}_2
\nonumber\\
\left<\left[b_I\left(\vec{q}_1\right), d_J\left(\vec{q}_2\right)\right]\right>u_\alpha\left(\vec{q}_1\right)v^\dagger_{\dot{\beta}}\left(\vec{q}_2\right)e^{i\left(k-\frac{q_1-q_2}{2}\right)\cdot r}
\nonumber\\
+\left<\left[b_I\left(\vec{q}_1\right), b^\dagger_J\left(\vec{q}_2\right)\right]\right>u_{\alpha}\left(\vec{q}_1\right)u^\dagger_{\dot{\beta}}\left(\vec{q}_2\right)e^{i\left(k-\frac{q_1+q_2}{2}\right)\cdot r}
\nonumber\\
+\left<\left[d^\dagger_I\left(\vec{q}_1\right),d_J\left(\vec{q}_2\right)\right]\right>v_\alpha\left(\vec{q}_1\right)v^\dagger_{\dot{\beta}}\left(\vec{q}_2\right)e^{i\left(k+\frac{q_1+q_2}{2}\right)\cdot r}
\nonumber\\
+\left<\left[d^\dagger_I\left(\vec{q}_1\right),b^\dagger_J\left(\vec{q}_2\right)\right]\right>v_\alpha\left(\vec{q}_1\right)u^\dagger_{\dot{\beta}}\left(\vec{q}_2\right)e^{i\left(k+\frac{q_1-q_2}{2}\right)\cdot r}
\end{eqnarray}

The commutators of creation and annihilation operators are clearly related to the particle number operator, and consequently depend on the neutrino distributions.  We make the assumption that the neutrino distributions are approximately homogenous and time-invariant on the scale of the de Broglie wavelength, so that the integral over $r$ can be formally taken to infinity while still assuming that the expectation values of the commutators do not vary over the integration range.  In the astrophysical venues we target for application of our QKEs there are unlikely to be any density fluctuations on scales comparable with the neutrino de Broglie wavelength ($\sim$10 fm).

With the assumption of approximate time invariance, the first and last terms in equation (\ref{eq:36}) do not contribute to the integral, since a pair of creation operators or a pair of annihilation operators acting on a state will always change its energy.  Since a time invariant state is an energy eigenstate, the action of the pair of operators will always give a state that is orthogonal to the original, and as a result the expectation value vanishes.  Note that this result does not hold true for states describing neutrino distributions that vary on a scale comparable to the de Broglie frequency; here, we assume that there is no such rapid variation.

Similarly, we can use the assumption of approximate homogeneity to show that the remaining terms, involving a creation operator and an annihilation operator, must be proportional to $\delta^3\left(\vec{q}_1-\vec{q}_2\right)$, since the expectation value will be zero unless the operators create and annihilate a particle with the same momentum. All of this allows us to write the commutators of the creation and annihilation operators as
\begin{eqnarray}
\label{eq:37}
\left<\left[b_I\left(\vec{q}_1\right),b^\dagger_J\left(\vec{q}_2\right)\right]\right>  = 
\nonumber\\
\left<\left\{b_I\left(\vec{q}_1\right)b^\dagger_J\left(\vec{q}_2\right)\right\}\right>-2\left<b^\dagger_J\left(\vec{q}_2\right)b_I\left(\vec{q}_1\right)\right>
\nonumber\\
 =  \left(2\pi\right)^3\delta^3\left(\vec{q}_1-\vec{q}_2\right)2E_q\left(\delta_{IJ}-2f_{IJ}\left(\vec{q_1}\right)\right).
\end{eqnarray}
\noindent
Here $f_{IJ}\left(\vec{q}_1\right)$ is the density matrix for neutrinos.  For $I = J$, $f_{II}\left(\vec{q}_1\right)$ simply corresponds to the expectation value of the number operator for flavor $I$, and gives the occupation number of neutrinos of flavor $I$ and momentum $\vec{q}_1$.  For $I\not= J$, $f_{IJ}$ corresponds to coherence between neutrinos of different flavors.

Similarly,
\begin{eqnarray}
\label{eq:38}
\left<\left[d^\dagger_I\left(\vec{q}_1\right),d_J\left(\vec{q}_2\right)\right]\right>=
\nonumber\\
-\left(2\pi\right)^3\delta^3\left(\vec{q}_1-\vec{q}_2\right)2E_q\left(\delta_{IJ}-2\bar{f}_{IJ}\left(\vec{q}_1\right)\right)
\end{eqnarray}
\noindent
where $\bar{f}_{IJ}\left(\vec{q}_1\right)$ is the density matrix for anti-neutrinos.

From this point on in our exposition we will use $x$ to mean the average coordinate $X$ in Wigner transformed quantities. Using equations (\ref{eq:37}) and (\ref{eq:38}), to perform the integrals in equation (\ref{eq:36}), simplifying the spinor bilinears by using equation (\ref{eq:35}), and generalizing from $x = 0$ to any position gives
\begin{eqnarray}
\label{eq:39}
F_{\alpha\dot{\beta}}\left(x,k\right)=
2\pi\delta\left(k^2\right)k_\mu\sigma^\mu_{\alpha\dot{\beta}}
\times\nonumber\\
\left(\frac{1}{2}-\theta\left(k^0\right)f\left(x,\vec{k}\right)-\theta\left(-k^0\right)\bar{f}\left(x,-\vec{k}\right)\right),
\end{eqnarray}
where we have suppressed flavor indices on $f_{IJ}$ and $\bar{f}_{IJ}$.  Similarly,
\begin{eqnarray}
\label{eq:40}
F^{\dot{\alpha}\beta}\left(x,k\right)=2\pi\delta\left(k^2\right)k_\mu\bar{\sigma}^{\mu\dot{\alpha}\beta}
\times\nonumber\\
\left(\frac{1}{2}-\theta\left(k^0\right)\bar{f}^T\left(x,\vec{k}\right)-\theta\left(-k^0\right)f^T\left(x,-\vec{k}\right)\right).
\end{eqnarray}

Note that $F^{\dot{\alpha}\beta}\left(k\right)$ is related to $F_{\alpha\dot{\beta}}^T\left(-k\right)$, where the transpose is over flavor indices.\\

We next calculate $F_\alpha^{\ \beta}$.  This is given by the expression
\begin{eqnarray}
\label{eq:41}
F_{IJ,\alpha}^{\ \ \ \ \beta}\left(0,k\right)=\frac{1}{2}\int d^4r \int\tilde{dq}_1\tilde{dq}_2
\nonumber\\
\left<\left[b_I\left(\vec{q}_1\right),d^\dagger_J\left(\vec{q}_2\right)\right]\right>u_\alpha\left(\vec{q}_1\right)v^\beta\left(\vec{q}_2\right)e^{i\left(k-\frac{q_1+q_2}{2}\right)\cdot r}
\nonumber\\
+\left<\left[d^\dagger_I\left(\vec{q}_1\right),b_J\left(\vec{q}_2\right)\right]\right>v_\alpha\left(\vec{q}_1\right)u^\beta\left(\vec{q}_2\right)e^{i\left(k+\frac{q_1+q_2}{2}\right)\cdot r}
\end{eqnarray}
where we have omitted vanishing terms.  Since the anticommutators of $b$ and $d^\dagger$ vanish, we can write the commutators as
\begin{eqnarray}
\label{eq:42}
\left<\left[b_I\left(\vec{q}_1\right),d^\dagger_J\left(\vec{q}_2\right)\right]\right>=-2\left<d_J^\dagger\left(\vec{q}_1\right)b_I\left(\vec{q}_2\right)\right>
\nonumber\\
=-\left(2\pi\right)^3\delta^3\left(\vec{q}_1-\vec{q}_2\right)2E_q\left(2\phi_{IJ}\left(\vec{q}_1\right)\right)
\end{eqnarray}
The matrix $\phi_{IJ}$ is a correlation function between neutrino and anti-neutrino creation and annihilation operators, and so describes coherence between neutrino and anti-neutrino states.  We will see that this object vanishes with the assumption of isotropy (as expected from conservation of angular momentum), but may, in general, be present in an anisotropic environment.

We simplify the spinor bilinears in equation (\ref{eq:41}) by using
\begin{eqnarray}
\label{eq:43}
u_\alpha\left(\vec{q}\right)v^\beta\left(\vec{q}\right)=
v_\alpha\left(-\vec{q}\right)u^\beta\left(-\vec{q}\right)
\nonumber\\
=\frac{1}{2}iq^{[\mu}\left(x^1-ix^2\right)^{\nu]}\left(S^L_{\mu\nu}\right)_\alpha^{\ \beta}
\end{eqnarray}
Here, $x^1$ and $x^2$ are spacelike unit vectors orthogonal to the direction of the momentum and to each other.  Equation (\ref{eq:43}) may be directly verified by choosing a coordinate system in which $q^\mu = \left(q, 0, 0, q\right)$, $x^{1,\mu}=\left(0,1,0,0\right)$ and $x^{2,\mu}=\left(0,0,1,0\right)$, then solving equation (\ref{eq:34}) for the spinors $u$ and $v$, imposing the normalization conditions (\ref{eq:35}), explicitly calculating the spinor bilinears and comparing to the expressions for $\left(S_L^{\mu\nu}\right)_\alpha^{\ \beta}$.  Note that the pre-factor $q^{[\mu}\left(x^1+ix^2\right)^{\nu]}$ is chosen to be anti-self-dual.  We choose a pre-factor of this form because the contraction with $S^L_{\mu\nu}$ projects out the self-dual component, so any self-dual component in the pre-factor would not contribute to equation (\ref{eq:43}).\\

Using equations (\ref{eq:42}) and (\ref{eq:43}) and performing the integrals in (\ref{eq:41}) gives
\begin{eqnarray}
\label{eq:44}
F_\alpha^{\ \beta}\left(x,k\right) &=& -2\pi\delta\left(k^2\right)\frac{1}{2}ik^{[\mu}\left(\hat{x}^1-i\hat{x}^2\right)^{\nu]}\left(S^L_{\mu\nu}\right)_\alpha^{\ \beta}
\nonumber\\
&\times& \left(\theta\left(k^0\right)\phi\left(\vec{k}\right)+\theta\left(-k^0\right)\phi^T\left(-\vec{k}\right)\right)
\end{eqnarray}
Similarly,
\begin{eqnarray}
\label{eq:45}
F^{\dot{\alpha}}_{\ \dot{\beta}}\left(x,k\right) &=& -2\pi\delta\left(k^2\right)\frac{1}{2}ik^{[\mu}\left(\hat{x}^1+i\hat{x}^2\right)^{\nu]}\left(S^R_{\mu\nu}\right)^{\dot{\alpha}}_{\ \dot{\beta}}
\nonumber\\
&\times& \left(\theta\left(k^0\right)\phi^\dagger\left(\vec{k}\right)+\theta\left(-k^0\right)\phi^\star\left(-\vec{k}\right)\right)
\end{eqnarray}

We now turn to the spectral function.  Unlike the statistical function, in free field theory the spectral function is completely determined by the anticommutation relations between creation and annihilation operators.  Thus, the only nonzero components of the spectral function are
\begin{eqnarray}
\label{eq:46}
\rho_{\alpha\dot{\beta}, IJ}\left(x,k\right)=2i\pi\delta\left(k^2\right){\rm{sign}}\left(k^0\right)k_\mu\sigma^\mu_{\alpha\dot{\beta}}\delta_{IJ}
\end{eqnarray}
\begin{eqnarray}
\label{eq:47}
\rho^{\dot{\alpha}\beta}_{IJ}\left(x,k\right)=2i\pi\delta\left(k^2\right){\rm{sign}}\left(k^0\right)k_\mu\bar{\sigma}^{\mu\dot{\alpha}\beta}\delta_{IJ}
\end{eqnarray}

\subsection{Wigner-Transformed Equations of Motion for the Statistical Function}
Having determined the physical content of the statistical function, we return to the Wigner transform of equations (\ref{eq:26}) and (\ref{eq:27}).  The full Wigner transformed expressions contain gradient expansions, which are infinite series of derivatives with respect to $x$ and $k$.  We truncate these infinite series by expanding in a small parameter $\epsilon$.

In our expansion, we make use of the fact that, in the regime we are considering, neutrino masses and interaction potentials are small compared to the neutrino energy.  Also, we expect the variation of physical quantities with respect to the average coordinate $x$ to be slow compared to the inverse neutrino de Broglie frequency.  These considerations lead us to introduce the following power counting: 
\begin{eqnarray}
\label{eq:48}
\frac{\partial_x, M, \Sigma}{E}=O\left(\epsilon\right)
\ \ \ \ \ \ \frac{\Pi_\rho,\Pi_F}{E}=O\left(\epsilon^2\right)
\end{eqnarray}
where $E$ is the neutrino energy.  The contributions to self-energy $\Pi_\rho$ and $\Pi_F$ are $O\left(\epsilon^2\right)$ because they appear only at two-loop order in the Feynman diagram expansion, while $\Sigma$ appears at one-loop order.

This power counting includes the standard gradient expansion (see, for example, Ref.s \cite{Calzetta:1988qy,Cirigliano:2010lr,Cirigliano:2011lr}).  However, our approach is specialized to the ultra-relativistic neutrinos that are relevant for supernova and compact object merger environments.  Moreover, since this work involves neutrinos having energies far below the electroweak scale, the interactions are always weak.

We keep terms to $O\left(\epsilon^2\right)$, since this allows us to include terms involving $\Pi_\rho$ and $\Pi_F$, which describe inelastic and non-forward scattering of neutrinos.  To $O\left(\epsilon^2\right)$, the Wigner transformed equations for $F$ are
\begin{eqnarray}
\label{eq:49}
\left(\frac{1}{2}i{\not\partial}+{\not k}\right)F\left(x,k\right)-\left(M+\Sigma\left(x\right)\right)F\left(k,x\right)
\nonumber\\
+\frac{1}{2}i\left(\partial_x^\mu\Sigma\left(x\right)\right)\left(\partial_\mu^k F\left(x,k\right)\right)=
\nonumber\\
-\frac{1}{2}i\left(\Pi^+\left(x,k\right)G^-\left(x,k\right)-\Pi^-\left(x,k\right)G^+\left(x,k\right)\right)
\end{eqnarray}
\noindent
and its Hermitian conjugate.  Here, $\partial_\mu^k \equiv \frac{\partial}{\partial k^\mu}$.  We have made the right-hand side of the equation more compact by introducing the notation

\begin{eqnarray}
\label{eq:50}
G^\pm &\equiv& -\frac{1}{2}i\rho \pm F
\nonumber\\
\Pi^\pm &\equiv& -\frac{1}{2}i\Pi_\rho \pm \Pi_F
\end{eqnarray}

We will use equation (\ref{eq:49}) and its Hermitian conjugate as the starting point for deriving the equations of motion for the neutrino density matrices.

\section{derivation of quantum kinetic equations}
\subsection{Outline of the Derivation and Some Preliminaries}
Equation (\ref{eq:49}) has a complicated structure, containing the kinetic equations as well as algebraic constraints relating various components of $F$ to each other.  To derive the quantum kinetic equations, we systematically expand equation (\ref{eq:49}) in the separation of scales, using the power counting defined in equation (\ref{eq:48}).

We expect the statistical function $F$ to have an $O\left(1\right)$ piece of the form given by equations (\ref{eq:39})-(\ref{eq:40}) and (\ref{eq:44})-(\ref{eq:45}), plus a small correction due to nonzero interactions and neutrino masses.  This correction will be $O\left(\epsilon\right)$, while our kinetic equations will be constructed to $O\left(\epsilon^2\right)$.  Thus, the $O\left(\epsilon\right)$ correction to $F$ will enter into the kinetic equations, and must be calculated.

Our strategy is to first expand equation (\ref{eq:49}) to $O\left(\epsilon\right)$, and use this to find the first-order shift in $F$ due to the mass and interactions.  Then, we will insert the $O\left(\epsilon\right)$ expression for $F$ back into equation (\ref{eq:49}), expand to $O\left(\epsilon^2\right)$, and extract the equations of motion for the density matrices and spin coherence densities.

We will show, in a subsequent section, that $\Sigma$ corresponds to the matter and neutrino self-interaction potential arising from coherent forward scattering, and has the form
\begin{eqnarray}
\label{eq:51}
\Sigma=\left(\begin{array}{cc}\delta\Sigma_S & \Sigma_L\cdot\sigma \\ \Sigma_R\cdot\bar{\sigma} & \delta\Sigma_S^\dagger\end{array}\right)
\end{eqnarray}
where $\Sigma_L$ and $\Sigma_R$ are Hermitian, and, for Majorana fermions, trivially related to each other.  $\Sigma_{L/R}=O\left(\epsilon\right)$ and $\delta\Sigma_S=O\left(\epsilon^2\right)$.

To $O\left(\epsilon^2\right)$, the equations of motion for the statistical function can be written as follows:
\begin{eqnarray}
\label{eq:52}
\Omega F = -\frac{1}{2}i\left(\Pi^+ G^- - \Pi^- G^+\right)
\end{eqnarray}
and the Hermitian conjugate.  The operator $\Omega$ has the following structure:
\begin{eqnarray}
\label{eq:53}
\Omega=\left(\begin{array}{cc}-m-\delta\Sigma_S & \left(k+\frac{1}{2}i\partial-\tilde{\Sigma}_L\right)\cdot\sigma
\\
\left(k+\frac{1}{2}i\partial-\tilde{\Sigma}_R\right)\cdot\bar{\sigma} & -m^\dagger-\delta\Sigma_S^\dagger\end{array}\right)
\nonumber\\
\equiv \left(\not k+\frac{1}{2}i\not \partial-\tilde{\Sigma}-M\right)\ \ \ \ 
\end{eqnarray}
Here, $\tilde{\Sigma}=\Sigma+\delta\Sigma-\frac{1}{2}i\left(\partial^\mu\Sigma\right)\partial_\mu^k$, where $\Sigma$ is the $O\left(\epsilon\right)$ quantity, $\delta\Sigma$ is an $O\left(\epsilon^2\right)$ correction resulting from the $O\left(\epsilon\right)$ shift in the argument of $\Sigma\left[F\right]$, and the $O\left(\epsilon^2\right)$ derivative term comes from the Wigner transform.  The collisional gain-loss potentials $\Pi^{\pm}$ can, in general, have all possible components: 
\begin{eqnarray}
\label{eq:54}
\Pi^\pm=\left(\begin{array}{cc} \Pi_S+\frac{1}{2}i\Pi_T^{L,\mu\nu}S^L_{\mu\nu} & \Pi_L \cdot\sigma \\ \Pi_R \cdot\bar{\sigma} & \Pi_S^\dagger+\frac{1}{2}i\Pi_T^{R,\mu\nu}S^R_{\mu\nu}\end{array}\right)^\pm
\end{eqnarray}
where all quantities are $O\left(\epsilon^2\right)$.  We will see that if the spin coherence density is zero, the gain-loss potentials take on a simpler form, where $\Pi_S$ and $\Pi_T$ are zero to $O\left(\epsilon^2\right)$.

For Majorana neutrinos, we will find that $\Sigma_L$ is related to $\Sigma_R$ and $\Pi_L$ is related to $\Pi_R$.  This is because $\Sigma$ and $\Pi$ are functionals of the two-point function $G$, and mirror the relations between $G_L$ and $G_R$.  For now, however, we will treat all components of $\Sigma$, $\Pi$ and $G$ as independent, and make use of the Majorana conditions when we derive the final kinetic equations.\\

Regardless of whether the fermions are Majorana or Dirac, the components of $\Sigma$, $\Pi^\pm$ and $F$ have certain properties which follow from CPT invariance, which requires that these quantities be invariant under simultaneous Hermitian conjugation in spinor and flavor space.  We can write $F$, in the most general possible form, as
\begin{eqnarray}
\label{eq:55}
F=\left(\begin{array}{cc}F_S^L+\frac{1}{2}iF_T^L S_L & F_V^L\cdot\sigma \\ F_V^R\cdot\bar{\sigma} & F_S^R+\frac{1}{2}iF_T^R S_R\end{array}\right)
\end{eqnarray}
where the notation is $F_T^{L/R}S_{L/R}\equiv \left(F_T^{L/R}\right)^{\mu\nu}S^{L/R}_{\mu\nu}$.  The components of $F$ must satisfy $F_V^{L\dagger} = F_V^L$, $F_V^{R\dagger}=F_V^R$, $F_S^{L\dagger}=F_S^R$ and $F_T^{L\dagger}=F_T^R$.  The corresponding components of $\Sigma$ and $\Pi^\pm$ satisfy similar Hermiticity conditions.

\subsection{QKEs to $O\left(1\right)$:  Large and Small Components}
To $O\left(1\right)$, Equation (\ref{eq:52}) and its Hermitian conjugate simply give
\begin{eqnarray}
\label{eq:56}
\not k F = O\left(\epsilon\right)\ \ \ \ \ \ F\not k = O\left(\epsilon\right)
\end{eqnarray}
\noindent
This gives the approximate dispersion relation $k^2=0$ to $O\left(\epsilon\right)$.  Thus, we can choose the $z$-axis to be along $k$ and write down $k = \left|\vec{k}\right|\hat{\kappa}+O\left(\epsilon\right)$, where the components of $\hat{\kappa}$ are $\hat{\kappa} = \left({\rm sign}\left(k^0\right),0,0,1\right)$.  Note that since $\hat{\kappa}\approx \frac{k}{\left|\vec{k}\right|}$, the first component of $\hat{\kappa}$ is $\pm 1$, depending on whether we are dealing with a positive or negative value of $k^0$.

We introduce additional basis vectors, as follows:
\begin{eqnarray}
\label{eq:57}
\hat{\kappa}'&=&\left({\rm sign}\left(k^0\right),0,0,-1\right)
\nonumber\\
\hat{x}^1&=&\left(0, 1, 0, 0\right)
\nonumber\\
\hat{x}^2&=&\left(0,0,1,0\right)
\end{eqnarray}
\noindent
These basis vectors satisfy the relations
\begin{eqnarray}
\label{eq:58}
\hat{\kappa}^2=\hat{\kappa}'^2 &=& 0
\nonumber\\
\hat{\kappa}\cdot\hat{\kappa}' &=& 2
\nonumber\\
\hat{\kappa}\cdot\hat{x}^i=\hat{\kappa}'\cdot\hat{x}^i &=& 0
\nonumber\\
\hat{x}^i\cdot\hat{x}^j &=& -\delta^{ij}
\end{eqnarray}
Note that we have imposed the condition that $\hat{x}^1$, $\hat{x}^2$ and $\hat{z}=\left(0,0,0,1\right)$ form a right-handed set of basis vectors.  The momentum 4-vector $k$ can receive $O\left(\epsilon\right)$ corrections due to a shift in the dispersion relation induced by interactions.  However, the basis vectors remain the same, regardless of any such shifts.

In addition to the $O\left(1\right)$ dispersion relation, substituting the general form for $F$ in equation (\ref{eq:55}) into equation (\ref{eq:56}) gives the following constraints on the components of $F$:
\begin{eqnarray}
\label{eq:59}
F_S &=& O\left(\epsilon\right)
\nonumber\\
F_V^{L/R,\mu} &=& \hat{\kappa}^\mu F_{L/R}+O\left(\epsilon\right)
\nonumber\\
F_T^{L\mu\nu} &=& \frac{1}{2}F_T^i\left(\delta^{ij}-i\epsilon^{ij}\right)\left(\hat{\kappa}\wedge\hat{x}^j\right)^{\mu\nu}+O\left(\epsilon\right)
\nonumber\\
F_T^{R\mu\nu} &=& \frac{1}{2}F_T^i\left(\delta^{ij}+i\epsilon^{ij}\right)\left(\hat{\kappa}\wedge\hat{x}^j\right)^{\mu\nu}+O\left(\epsilon\right)
\end{eqnarray}
The wedge product notation is defined in the usual way, $\left(U\wedge V\right)^{\mu\nu}\equiv U^\mu V^\nu-U^\nu V^\mu$.  Note that we use the names $F_{L/R}$ and $F_T$ to denote both the full four-vector or tensor quantities and their components.  Since we will often use notation where the Lorentz indices are not explicitly shown, it is important to note whether an expression refers to the full quantity or the component.  This will be clear from context.
  
The expressions for $F_T^{L}$ and $F_T^R$ can be rewritten as follows:
\begin{eqnarray}
\label{eq:60}
F_T^{L\mu\nu}=\frac{1}{2}\left(F_T^1+iF_T^2\right)\left(\hat{\kappa}\wedge\left(\hat{x}^1-i\hat{x}^2\right)\right)^{\mu\nu}
\nonumber\\\equiv
 \left(\hat{\kappa}\wedge\left(\hat{x}^1-i\hat{x}^2\right)\right)^{\mu\nu}\Phi
\nonumber\\
F_T^{R\mu\nu}=\frac{1}{2}\left(F_T^1-iF_T^2\right)\left(\hat{\kappa}\wedge\left(\hat{x}^1+i\hat{x}^2\right)\right)^{\mu\nu}
\nonumber\\
\equiv
\left(\hat{\kappa}\wedge\left(\hat{x}^1+i\hat{x}^2\right)\right)^{\mu\nu}\Phi^\dagger
\end{eqnarray}

where we have defined $\Phi\equiv \frac{1}{2}\left(F_T^1+iF_T^2\right)$.

Since we have $k^2=0$ to $O\left(\epsilon\right)$, the components of $F$ have the form
\begin{eqnarray}
\label{eq:61}
F_{L/R}=2\pi\delta\left(k^2+O\left(\epsilon\right)\right)\left|\vec{k}\right| g_{L/R}
\nonumber\\
F_T^i = 2\pi\delta\left(k^2+O\left(\epsilon\right)\right)\left|\vec{k}\right| g_T^i
\end{eqnarray}

For a multi-flavor system, the notation $\delta\left(k^2+O\left(\epsilon\right)\right)$ is symbolic, since each component of the flavor matrices $g_{L/R}$ and $g_T^i$ will in general carry different corrections to the argument of the delta function.

To $O\left(\epsilon\right)$, we write $F$ as follows:
\begin{eqnarray}
\label{eq:62}
F\rightarrow F^{\left(1\right)}+\Delta
\end{eqnarray}
Here, $F^{\left(1\right)}$ incorporates the $O\left(\epsilon\right)$ correction to the dispersion relation, and has the form
\begin{eqnarray}
\label{eq:62-1}
F^{\left(1\right)}=\left(\begin{array}{cc}\frac{1}{2}i\Phi\left(\hat{\kappa}\wedge\hat{x}^-\right)\cdot S_L & F_L\left(\hat{\kappa}\cdot \sigma\right) \\ F_R\left(\hat{\kappa}\cdot\bar{\sigma}\right) & \frac{1}{2}i\Phi^\dagger\left(\hat{\kappa}\wedge\hat{x}^+\right)\cdot S_R\end{array}\right)
\end{eqnarray}
where $\hat{x}^\pm = \left(\hat{x}^1\pm i\hat{x}^2\right)$.  $\Delta$ is the set of $O\left(\epsilon\right)$ small components.  In general,
\begin{eqnarray}
\label{eq:63}
\Delta=\left(\begin{array}{cc}\Delta_S+\frac{1}{2}i\Delta_T^L S_L & \Delta_L\cdot\sigma \\ \Delta_R\cdot\bar{\sigma} & \Delta_S^\dagger+\frac{1}{2}i\Delta_T^R S_R\end{array}\right)
\end{eqnarray}

Note that the form of $F$ given by equations (\ref{eq:59})-(\ref{eq:61}) is consistent with equations (\ref{eq:39})-(\ref{eq:40}) and (\ref{eq:44})-(\ref{eq:45}), which are derived from free, massless field theory.  All correlation functions that we have found in Section IV.B. are included in the $O\left(1\right)$ expression for $F$.  Specifically,
\begin{eqnarray}
\label{eq:64}
F_L=2\pi\delta\left(k^2+O\left(\epsilon\right)\right)\left|\vec{k}\right|
\times\nonumber\\
\left(\frac{1}{2}-\theta\left(k^0\right)f\left(\vec{k}\right)-\theta\left(-k^0\right)\bar{f}\left(-\vec{k}\right)\right)
\nonumber\\\nonumber\\
F_R=2\pi\delta\left(k^2+O\left(\epsilon\right)\right)\left|\vec{k}\right|
\times\nonumber\\
\left(\frac{1}{2}-\theta\left(k^0\right)\bar{f}^T\left(\vec{k}\right)-\theta\left(-k^0\right)f^T\left(-\vec{k}\right)\right)
\nonumber\\\nonumber\\
\Phi = -2\pi\delta\left(k^2+O\left(\epsilon\right)\right)\left|\vec{k}\right|\times\nonumber\\
\left(\theta\left(k^0\right)\phi\left(\vec{k}\right)+\theta\left(-k^0\right)\phi^T\left(-\vec{k}\right)\right)
\end{eqnarray}

Note that the results of Section IV.B. place additional constraints on the form of $F$.  These constraints relate $F_L\left(k\right)$ to $F_R\left(-k\right)$ and $F_T\left(k\right)$ to $F_T\left(-k\right)$, and do not follow from Equation (\ref{eq:52}).  These constraints follow from the Majorana nature of the fermions, which was assumed in Section IV.B. but not in the derivation of Equation (\ref{eq:52}).  As mentioned above, we will use the more general formalism of Equation (\ref{eq:52}) and treat $F_L$ and $F_R$ as independent quantities, until we are ready to extract the equations of motion for the density matrices.

\subsection{QKEs to $O\left(\epsilon\right)$:  Small Components and the Dispersion Relation}
We next expand equation (\ref{eq:52}) order-by-order, first using the $O\left(\epsilon\right)$ expansion to find the small components $\Delta$ and the $O\left(\epsilon\right)$ shift in the dispersion relation, and then inserting the results into the $O\left(\epsilon^2\right)$ equations to obtain the kinetic equations.  To $O\left(\epsilon\right)$, equation (\ref{eq:52}) is
\begin{eqnarray}
\label{eq:65}
\not k\Delta+\left(\not k+\frac{1}{2}i\not \partial\right)F - \Sigma F - M F = O\left(\epsilon^2\right)
\end{eqnarray}

Decomposing this into irreducible representations of the Lorentz group gives the following set of equations:

Scalar:
\begin{eqnarray}
\label{eq:66}
k\cdot\Delta_R+\left(k+\frac{1}{2}i\partial\right)\cdot F_V^R - \Sigma_L\cdot F_V^R = O\left(\epsilon^2\right)
\\
\label{eq:67}
k\cdot\Delta_L+\left(k+\frac{1}{2}i\partial\right)\cdot F_V^L-\Sigma_R\cdot F_V^L = O\left(\epsilon^2\right)
\end{eqnarray}

Vector:
\begin{eqnarray}
\label{eq:68}
k\Delta_S^\dagger-k\cdot\Delta_T^R-\left(k+\frac{1}{2}i\partial\right)\cdot F_T^R+\Sigma_L\cdot F_T^R
\nonumber\\
-mF_V^L=O\left(\epsilon^2\right)
\\
\label{eq:69}
k\Delta_S+k\cdot\Delta_T^L+\left(k+\frac{1}{2}i\partial\right)\cdot F_T^L-\Sigma_R\cdot F_T^L
\nonumber\\
-m^\dagger F_V^R = O\left(\epsilon^2\right)
\end{eqnarray}
For the vector equations, the notation is $V\cdot T \equiv V^\mu T_{\mu\nu}$ and $T\cdot V\equiv T_{\nu\mu}V^\mu$.

Tensor:
\begin{eqnarray}
\label{eq:70}
\left(k\wedge\Delta_R+\left(k+\frac{1}{2}i\partial\right)\wedge F_V^R-\Sigma_L\wedge F_V^R\right)^L
\nonumber\\-\frac{1}{2}mF_T^L=O\left(\epsilon^2\right)
\\
\label{eq:71}
\left(k\wedge\Delta_L+\left(k+\frac{1}{2}i\partial\right)\wedge F_V^L-\Sigma_R\wedge F_V^L\right)^R
\nonumber\\+\frac{1}{2}m^\dagger F_T^R=O\left(\epsilon^2\right)
\end{eqnarray}
where the superscripts $L$ and $R$ denote anti-self-dual and self-dual projections, respectively; that is, for an antisymmetric tensor $T$, $T^L\equiv\frac{1}{2}\left(T-iT^\star\right)$ and $T^R\equiv\frac{1}{2}\left(T+iT^\star\right)$.\\

These equations, and their Hermitian conjugates, determine the form of the small components $\Delta$ and the dispersion relations for $F_{L/R}$ and $F_T$.  To solve the equations, it is useful to decompose all our quantities into components along the basis vectors in equation (\ref{eq:57}).  The decomposition for $F_{L/R}$ and $F_T^{L/R}$ is given by equations (\ref{eq:59})-(\ref{eq:61}).  For the other four-vector quantities we use
\begin{eqnarray}
\label{eq:72}
\partial &=& \frac{1}{2}\partial^{\kappa'}\hat{\kappa}+\frac{1}{2}\partial^{\kappa}\hat{\kappa}'-\partial^i\hat{x}^i
\\
\label{eq:73}
\Sigma_{L/R}&=&\frac{1}{2}\Sigma_{L/R}^{\kappa'}\hat{\kappa}+\frac{1}{2}\Sigma_{L/R}^{\kappa}\hat{\kappa}'-\Sigma_{L/R}^i\hat{x}^i
\\
\label{eq:74}
\Delta_{L/R}&=&\frac{1}{2}\Delta_{L/R}^{\kappa}\hat{\kappa}'-\Delta_{L/R}^i\hat{x}^i
\\
\label{eq:75}
k&=&\frac{1}{2}\left(k\cdot\hat{\kappa}'\right)\hat{\kappa}+\frac{1}{2}\left(k\cdot\hat{\kappa}\right)\hat{\kappa}'
\nonumber\\
&=&\frac{1}{2}\left(\left|\vec{k}\right|+E\right)\hat{\kappa}+\frac{1}{2}\left(k\cdot\hat{\kappa}\right)\hat{\kappa}'
\end{eqnarray}

Note that the $\Delta_{L/R}^{\kappa'}$ component does not appear, since this kind of first-order shift would be along the same direction as $F_{L/R}$ and can therefore be absorbed into the $O\left(1\right)$ quantity.  For a four-vector quantity $V$, we have labeled its component along any basis vector $\hat{w}$ as $V^w\equiv V\cdot\hat{w}$.  This choice of notation determines the particular signs and factors of $1/2$ in equations (\ref{eq:72})-(\ref{eq:75}).  For example, $\hat{\kappa}\cdot \Sigma_L=\Sigma^\kappa$.  Since, from the $O\left(1\right)$ dispersion relation, $E = \left|\vec{k}\right|+O\left(\epsilon\right)$, the $\hat{\kappa}$ component of $k$ is $\left|\vec{k}\right|+O\left(\epsilon\right)$.

The tensor small component is decomposed as follows:
\begin{eqnarray}
\label{eq:76}
\frac{1}{2}\left(\hat{\kappa}\wedge\hat{\kappa'}\right)\Delta_T^{\kappa\kappa'}+\left(\hat{x}^1\wedge\hat{x}^2\right)\Delta_T^{xx}
+\left(\hat{\kappa}'\wedge\hat{x}^i\right)\Delta_T^i
\end{eqnarray}
Again, the component proportional to $\hat{\kappa}\wedge\hat{x}^i$ does not appear, as this component can be absorbed into $F_T$.  The anti-self-dual and self-dual projections of  equation (\ref{eq:76}) are
\begin{eqnarray}
\label{eq:77}
\Delta_T^L=\frac{1}{2}\left(\hat{\kappa}'\wedge\hat{x}^i\right)\left(\delta^{ij}-i\epsilon^{ij}\right)\Delta_T^j+
\nonumber\\
\left(\frac{1}{2}\left(\hat{\kappa}\wedge\hat{\kappa}'\right)-i\left(\hat{x}^1\wedge\hat{x}^2\right)\right)\Delta_T
\nonumber\\\nonumber\\
\Delta_T^R=\frac{1}{2}\left(\hat{\kappa}'\wedge\hat{x}^i\right)\left(\delta^{ij}+i\epsilon^{ij}\right)\Delta_T^j+
\nonumber\\
\left(\frac{1}{2}\left(\hat{\kappa}\wedge\hat{\kappa}'\right)+i\left(\hat{x}^1\wedge\hat{x}^2\right)\right)\Delta_T^\dagger
\end{eqnarray}
where $\Delta_T \equiv \frac{1}{2}\left(\Delta_T^{\kappa\kappa'}+i\Delta_T^{xx}\right)$

We next use equations (\ref{eq:72})-(\ref{eq:76}) to decompose equations (\ref{eq:66})-(\ref{eq:71}) into components.  For the scalar equations, (\ref{eq:66})-(\ref{eq:67}), this gives
\begin{eqnarray}
\label{eq:78}
\left|\vec{k}\right|\Delta_R^\kappa+\left(k\cdot\hat{\kappa}\right)F_R+\frac{1}{2}i\partial^\kappa F_R-\Sigma_L^\kappa F_R &=& O\left(\epsilon^2\right)
\\
\label{eq:79}
\left|\vec{k}\right|\Delta_L^\kappa+\left(k\cdot\hat{\kappa}\right)F_L+\frac{1}{2}i\partial^\kappa F_L-\Sigma_R^\kappa F_L &=& O\left(\epsilon^2\right)
\end{eqnarray}
\noindent
The Hermitian portions of these equations are:
\begin{eqnarray}
\label{eq:80}
\left|\vec{k}\right|\Delta_R^\kappa+\left(k\cdot\hat{\kappa}\right)F_R-\frac{1}{2}\left\{\Sigma_L^\kappa, F_R\right\}&=&O\left(\epsilon^2\right)
\\
\label{eq:81}
\left|\vec{k}\right|\Delta_L^\kappa+\left(k\cdot\hat{\kappa}\right)F_L-\frac{1}{2}\left\{\Sigma_R^\kappa, F_L\right\}&=&O\left(\epsilon^2\right)
\end{eqnarray}

The anti-Hermitian portions of the scalar equations involve derivatives of $F$ along $\hat{\kappa}$, and are therefore kinetic equations, giving the evolution of the neutrino density matrices along the neutrino world line.  We will return to the kinetic equations when we expand to $O\left(\epsilon^2\right)$.\\

The vector equations (\ref{eq:68})-(\ref{eq:69}) include components along $\hat{\kappa}$ and $\hat{x}^i$ (the component along $\hat{\kappa}'$ is trivial to $O\left(\epsilon\right)$).  Before extracting these components, it is useful to separate the vector equations into those involving $\Delta_S$ and those involving $\Delta_T$.  Taking the Hermitian conjugate of equation (\ref{eq:68}) and adding this to equation (\ref{eq:69}) gives
\begin{eqnarray}
\label{eq:82}
2k\Delta_S+i\partial\cdot F_T^L-\left(\Sigma_R\cdot F_T^L + F_T^L\cdot\Sigma_L\right)
\nonumber\\-\left(m^\dagger F_R+F_L m^\dagger\right)=O\left(\epsilon^2\right)
\end{eqnarray}
Subtracting equation (\ref{eq:69}) from the Hermitian conjugate of equation (\ref{eq:68}) gives
\begin{eqnarray}
\label{eq:83}
2k\cdot\left(\Delta_T^L+F_T^L\right)-\left(\Sigma_R\cdot F_T^L-F_T^L\cdot\Sigma_L\right)
\nonumber\\
-\left(m^\dagger F_R-F_Lm^\dagger\right)=O\left(\epsilon^2\right)
\end{eqnarray}

The components of equations (\ref{eq:82}) and (\ref{eq:83}) along $\hat{\kappa}$ give $\Delta_S$ and $\Delta_T$ as functions of $F_L$, $F_R$ and $F_T^i$:
\begin{eqnarray}
\label{eq:84}
2\left|\vec{k}\right|\Delta_S-i\partial^i P_+^{ij}F_T^j+\left(\Sigma_R^i P_+^{ij}F_T^j-P_+^{ij}F_T^j\Sigma_L^i\right)\nonumber\\
-\left(m^\dagger F_R+F_L m^\dagger\right)=O\left(\epsilon^2\right)
\\\nonumber\\
\label{eq:85}
-2\left|\vec{k}\right|\Delta_T+\left(\Sigma_R^i P_+^{ij}F_T^j+P_+^{ij}F_T^j\Sigma_L^i\right)\nonumber\\
-\left(m^\dagger F_R-F_L m^\dagger\right)=O\left(\epsilon^2\right)
\end{eqnarray}
Here, $P_\pm^{ij}$ are projection operators on the $\hat{x}^1, \hat{x}^2$ plane, given by  $P_\pm^{ij}\equiv\frac{1}{2}\left(\delta^{ij}\pm i\epsilon^{ij}\right)$\\

The components of equation (\ref{eq:82}) along $\hat{x}^i$ give kinetic equations for $F_T^i$; we will return to these equations when we consider the $O\left(\epsilon^2\right)$ expansion.  The components of equation (\ref{eq:83}) along $\hat{x}^i$ are:
\begin{eqnarray}
\label{eq:86}
4\left|\vec{k}\right| P_-^{ij}\Delta_T^j+2\left(k\cdot\hat{\kappa}\right)P_+^{ij}F_T^j
\nonumber\\
-\left(\Sigma_R^\kappa P_+^{ij}F_T^j+P_+^{ij}F_T^j\Sigma_L^\kappa\right)=O\left(\epsilon^2\right)
\end{eqnarray}

Acting on this with $P_-$ and using $P_+ P_- = 0$ and $P_-P_-=P_-$ gives $P_-^{ij}\Delta_T^j = O\left(\epsilon^2\right)$.  The Hermitian conjugate is $P_+^{ij}\Delta_T^j=O\left(\epsilon^2\right)$; adding these equations together gives $\Delta_T^j = O\left(\epsilon^2\right)$.  The remainder of the equation, with its Hermitian conjugate, is
\begin{eqnarray}
\label{eq:87}
\left(k\cdot\hat{\kappa}\right)P_+^{ij}F_T^j-\frac{1}{2}\left(\Sigma_R^\kappa P_+^{ij}F_T^j+P_+^{ij}F_T^j\Sigma_L^\kappa\right)=O\left(\epsilon^2\right)
\\
\label{eq:88}
\left(k\cdot\hat{\kappa}\right)P_-^{ij}F_T^j-\frac{1}{2}\left(\Sigma_L^\kappa P_-^{ij}F_T^j+P_-^{ij}F_T^j\Sigma_R^\kappa\right)=O\left(\epsilon^2\right)
\end{eqnarray}
This is a set of dispersion relations for $F_T$; we will return to these equations later.

We next consider the tensor equations, (\ref{eq:70})-(\ref{eq:71}).  The components proportional to $\hat{\kappa}'\wedge\hat{x}^i$ are trivial to $O\left(\epsilon\right)$.  The components proportional to $\hat{\kappa}\wedge\hat{\kappa}'$ are
\begin{eqnarray}
\label{eq:89}
\left|\vec{k}\right|\Delta_R^{\kappa}-\left(k\cdot\hat{\kappa}\right) F_R-\frac{1}{2}i\partial^\kappa F_R+\Sigma_L^\kappa F_R=O\left(\epsilon^2\right)
\\
\label{eq:90}
\left|\vec{k}\right|\Delta_L^\kappa-\left(k\cdot\hat{\kappa}\right) F_L-\frac{1}{2}i\partial^\kappa F_L+\Sigma_R^\kappa F_L=O\left(\epsilon^2\right)
\end{eqnarray}
The Hermitian parts of these equations, together with equations (\ref{eq:78})-(\ref{eq:79}), give $\Delta_R^\kappa=O\left(\epsilon^2\right)$ and the dispersion relations for $F_L$ and $F_R$:
\begin{eqnarray}
\label{eq:91}
\left(k\cdot\hat{\kappa}\right)F_R-\frac{1}{2}\left\{\Sigma_L^\kappa, F_R\right\}=O\left(\epsilon^2\right)
\\
\label{eq:92}
\left(k\cdot\hat{\kappa}\right)F_L-\frac{1}{2}\left\{\Sigma_R^\kappa, F_L\right\}=O\left(\epsilon^2\right)
\end{eqnarray}
The anti-Hermitian part simply replicates the $O\left(\epsilon\right)$ kinetic equation obtained from the scalar equations.  The components along $\hat{x}^1\wedge\hat{x}^2$ are trivially related to those along $\hat{\kappa}\wedge\hat{\kappa}'$.\\  

The components of equations (\ref{eq:70})-(\ref{eq:71}) along $\hat{\kappa}\wedge\hat{x}^i$ are
\begin{eqnarray}
\label{eq:93}
P_+^{ij}\left(\left|\vec{k}\right|\Delta_R^j-\frac{1}{2}i \partial^j F_R+\Sigma_L^j F_R+\frac{1}{2}mF_T^j\right)
=O\left(\epsilon^2\right)\ 
\\
\label{eq:94}
P_-^{ij}\left(\left|\vec{k}\right|\Delta_L^j-\frac{1}{2}i\partial^j F_L+\Sigma_R^j F_L-\frac{1}{2}m^\dagger F_T^j\right)
=O\left(\epsilon^2\right)\ 
\end{eqnarray}

The Hermitian parts of equations (\ref{eq:93})-(\ref{eq:94}) give expressions for $\Delta_{L/R}^i$:
\begin{eqnarray}
\label{eq:95}
\left|\vec{k}\right|\Delta_R^i+\frac{1}{2}\epsilon^{ij}\partial^j F_R+\left(P_+^{ij}\Sigma_L^j F_R+F_RP_-^{ij}\Sigma_L^j\right)
\nonumber\\
+\frac{1}{2}\left(mP_+^{ij}F_T^j+P_-^{ij}F_T^j m^\dagger\right)=O\left(\epsilon^2\right)
\\\nonumber\\
\label{eq:96}
\left|\vec{k}\right|\Delta_L^i-\frac{1}{2}\epsilon^{ij}\partial^j F_L+\left(P_-^{ij}\Sigma_R^j F_L+F_LP_+^{ij}\Sigma_R^j\right)
\nonumber\\
-\frac{1}{2}\left(m^\dagger P_-^{ij} F_T^j + P_+^{ij}F_T^j m\right)=O\left(\epsilon^2\right)
\end{eqnarray}
The anti-Hermitian parts are trivially related to the Hermitian parts.

In summary, the equations to $O\left(\epsilon\right)$ give the following expressions for the small components:
\begin{eqnarray}
\label{eq:97}
\Delta_{L/R}^\kappa = O\left(\epsilon^2\right)\ \ \ \ \ \ \ \ \ \ \Delta_T^i = O\left(\epsilon^2\right)
\\\nonumber\\
\label{eq:98}
\Delta_S=\frac{1}{2\left|\vec{k}\right|}\left(m^\dagger F_R + F_L m^\dagger\right)
\nonumber\\
+\frac{P_+^{ij}}{2\left|\vec{k}\right|}\left(i\partial^i F_T^j-\left(\Sigma_R^i F_T^j-F_T^j\Sigma_L^i\right)\right)
\\\nonumber\\
\label{eq:99}
\Delta_T=-\frac{1}{2\left|\vec{k}\right|}\left(m^\dagger F_R-F_Lm^\dagger\right)
\nonumber\\
+\frac{P_+^{ij}}{2\left|\vec{k}\right|}\left(\Sigma_R^i F_T^j + F_T^j \Sigma_L^i\right)
\\\nonumber\\
\label{eq:100}
\Delta_L^i = \frac{1}{2\left|\vec{k}\right|}\left(m^\dagger P_-^{ij}F_T^j +P_+^{ij}F_T^j m\right)
\nonumber\\
+\frac{1}{\left|\vec{k}\right|}\left(\frac{1}{2}\epsilon^{ij}\partial^j F_L-\left(P_-^{ij}\Sigma_R^j F_L+F_LP_+^{ij}\Sigma_R^j\right)\right)
\\\nonumber\\
\label{eq:101}
\Delta_R^i=-\frac{1}{2\left|\vec{k}\right|}\left(mP_+^{ij}F_T^j + P_-^{ij}F_T^j m^\dagger\right)
\nonumber\\
-\frac{1}{\left|\vec{k}\right|}\left(\frac{1}{2}\epsilon^{ij}\partial^j F_R+\left(P_+^{ij}\Sigma_L^j F_R+F_RP_-^{ij}\Sigma_L^j\right)\right)
\end{eqnarray}

We also obtain dispersion relations for $F_T$ and $F_{L/R}$, given by equations (\ref{eq:87})-(\ref{eq:88}) and (\ref{eq:91})-(\ref{eq:92}).

\subsection{Kinetic Equations for $F_{L/R}$}
We now construct equations for the evolution of $F_L$ and $F_R$, which encode the particle densities, to $O\left(\epsilon^2\right)$.  These equations are derived from the scalar components of equation (\ref{eq:52}).  To $O\left(\epsilon^2\right)$, the scalar equations are
\begin{eqnarray}
\label{eq:102}
k\cdot\left(F_R+\Delta_R\right)+\frac{1}{2}i\partial\cdot F_R-\tilde{\Sigma}_L\cdot F_R
\nonumber\\
+\frac{1}{2}i\partial\cdot\Delta_R - \Sigma_L\cdot\Delta_R - m\Delta_S
\nonumber\\
=-\frac{1}{2}i\left(\Pi^+_L \cdot F_R^- - \Pi^-_L\cdot F_R^+\right)
\nonumber\\
+\frac{1}{8}i\left(\Pi^{L+}_T G^{L-}_T-\Pi^{L-}_T G^{L+}_T\right)
\\\nonumber\\
\label{eq:103}
k\cdot\left(F_L+\Delta_L\right)+\frac{1}{2}i\partial\cdot F_L-\tilde{\Sigma}_R\cdot F_L
\nonumber\\
+\frac{1}{2}i\partial\cdot\Delta_L-\Sigma_R\cdot\Delta_L-m^\dagger\Delta_S^\dagger
\nonumber\\
=-\frac{1}{2}i\left(\Pi^+_R\cdot F_L^- - \Pi^-_R\cdot F_L^+\right)
\nonumber\\
+\frac{1}{8}i\left(\Pi^{R+}_T G^{R-}_T-\Pi^{R-}_T G^{R+}_T\right)
\end{eqnarray}
where we have used the notation $\Pi_T G_T\equiv\left(\Pi_T\right)_{\mu\nu}G_T^{\mu\nu}$.
Taking the anti-Hermitian parts of these equations and decomposing the four-vector quantities into components gives
\begin{eqnarray}
\label{eq:104}
i\partial^\kappa F_R
\,   - \frac{i}{ 2| \vec{k}|} \{ \partial^i \Sigma_L^i,   F_R \} 
-\left(\tilde{\Sigma}_L^\kappa F_R - F_R\tilde{\Sigma}_L^{\kappa\dagger}\right)
\nonumber\\
-i\partial^i\Delta_R^i + \left[\Sigma_L^i, \Delta_R^i\right]-\left(m\Delta_S-\Delta_S^\dagger m^\dagger\right)
=iC_R  
\\
\label{eq:105} 
i\partial^\kappa F_L
- \frac{i}{ 2| \vec{k}|} \{ \partial^i \Sigma_R^i,   F_L \}  
-\left(\tilde{\Sigma}_R^\kappa F_L - F_L\tilde{\Sigma}_R^{\kappa\dagger}\right)
\nonumber\\
-i\partial^i\Delta_L^i+\left[\Sigma_R^i, \Delta_L^i\right]-\left(m^\dagger\Delta_S^\dagger-\Delta_S m\right)
=iC_L 
\end{eqnarray}
where
\begin{eqnarray}
\label{eq:106}
C_R=-\frac{1}{2}\left(\left\{\Pi_L^{\kappa+},G_R^-\right\}-\left\{\Pi_L^{\kappa-},G_R^+\right\}\right)+C_T^R
\\
\label{eq:107}
C_L=-\frac{1}{2}\left(\left\{\Pi_R^{\kappa+},G_L^-\right\}-\left\{\Pi_R^{\kappa-},G_L^+\right\}\right)+C_T^L
\end{eqnarray}
The quantities $G^\pm$ are defined in equation (\ref{eq:50}).  The terms $C_T^L$ and $C_T^R$ involve the tensor components of $\Pi$, and are given by
\begin{eqnarray}
\label{eq:108}
C_T^L =\nonumber\\
 \frac{1}{8}\left(\Pi_T^{R+}G_T^{R-}+G_T^{L-}\Pi_T^{L+}-\Pi_T^{R-}G_T^{R+}-G_T^{L+}\Pi_T^{L-}\right)
 \\
 \label{eq:109}
 C_T^R=\nonumber\\
 \frac{1}{8}\left(\Pi_T^{L+}G_T^{L-}+G_T^{R-}\Pi_T^{R+}-\Pi_T^{L-}G_T^{L+}-G_T^{R+}\Pi_T^{R-}\right)
\end{eqnarray}

Next, we break this expression down into components along the basis vectors.  Since $G_T^\pm$ contains only components proportional to $\hat{\kappa}\wedge\hat{x}^i$, the contraction $G_{T\mu\nu}^{\pm}\Pi_T^{\mu\nu\mp}$ will only have nonzero contributions from components of $\Pi_T^\mp$ that are proportional to $\hat{\kappa}'\wedge\hat{x}^i$.  Thus, we can write
\begin{eqnarray}
\label{eq:112}
\Pi_T^{L\pm} = \Pi_T^{i\pm}P_+^{ij}\left(\hat{\kappa}'\wedge\hat{x}^j\right)
\\
\label{eq:113}
\Pi_T^{R\pm} = \Pi_T^{i\pm} P_-^{ij}\left(\hat{\kappa}'\wedge\hat{x}^j\right)
\end{eqnarray}

We now use $G_T^\pm=\pm F_T$, switch to the notation $\Phi \equiv \frac{1}{2}\left(F_T^1+iF_T^2\right)$ and similarly define $P_T^{\pm} \equiv \frac{1}{2}\left(\Pi_T^1+i\Pi_T^2\right)^\pm$.   With this notation, the terms appearing in equations (\ref{eq:106}) and (\ref{eq:107}) are
\begin{eqnarray}
\label{eq:116}
C_T^R=\left(P_T^++P_T^-\right)^\dagger\Phi+\Phi^\dagger\left(P_T^++P_T^-\right)
\\
\label{eq:117}
C_T^L=\left(P_T^++P_T^-\right)\Phi^\dagger+\Phi \left(P_T^++P_T^-\right)^\dagger
\end{eqnarray}

Next, we use equations (\ref{eq:98}) and (\ref{eq:100})-(\ref{eq:101}) to express the small components $\Delta_S$ and $\Delta_{L/R}^i$ in terms of $F_{L/R}$ and $F_T^i$.  Equation (\ref{eq:105}) contains the following combination of small components:  $U_L\equiv-i\partial^i\Delta_L^i+\left[\Sigma_R^i,\Delta_L^i\right]-\left(m^\dagger\Delta_S^\dagger-\Delta_S m\right)$, and equation (\ref{eq:104}) contains a similar combination, which we denote $U_R$.  We separate this into parts that depend on $F_{L/R}$ and $F_T$:
\begin{eqnarray}
\label{eq:118}
U_L\left[F_{L/R}\right]=-i\partial^i\Delta_L^i\left[F_{L/R}\right]+\left[\Sigma_R^i,\Delta_L^i\left[F_{L/R}\right]\right]
\nonumber\\
-\left(m^\dagger\Delta_S^\dagger\left[F_{L/R}\right]-\Delta_S\left[F_{L/R}\right]m\right)
\\\nonumber\\
\label{eq:119}
U_L\left[F_T\right]=-i\partial^i\Delta_L^i\left[F_T\right]+\left[\Sigma_R^i,\Delta_L^i\left[F_T\right]\right]
\nonumber\\
-\left(m^\dagger\Delta_S^\dagger\left[F_T\right]-\Delta_S\left[F_T\right]m\right)
\end{eqnarray}

Using equations (\ref{eq:98}) and (\ref{eq:100}) gives
\begin{eqnarray}
\label{eq:120}
U_L\left[F_L\right]=\frac{1}{2\left|\vec{k}\right|}i\partial^i\left\{\Sigma_R^i,F_L\right\}-
\nonumber\\
\frac{1}{2\left|\vec{k}\right|}\left[m^\dagger m-\epsilon^{ij}\partial^i\Sigma_R^j+\Sigma_R^i\Sigma_R^i-i\left[\Sigma_R^1,\Sigma_R^2\right],F_L\right]
\\\nonumber\\
\label{eq:121}
U_L\left[F_T\right]=\frac{1}{2\left|\vec{k}\right|}\left(\Sigma_R^i m^\dagger P_-^{ij}F_T^j-P_+^{ij}F_T^j m\Sigma_R^i\right)
\nonumber\\
-\frac{1}{2\left|\vec{k}\right|}\left(m^\dagger\Sigma_L^i P_-^{ij}F_T^j-P_+^{ij}F_T^j\Sigma_L^i m\right)\ \ 
\end{eqnarray}
Similarly, we calculate $U_R\left[F_R\right]$ and $U_R\left[F_T\right]$:
\begin{eqnarray}
\label{eq:122}
U_R\left[F_R\right]=\frac{1}{2\left|\vec{k}\right|}i\partial^i\left\{\Sigma_L^i,F_R\right\}-
\nonumber\\
\frac{1}{2\left|\vec{k}\right|}\left[mm^\dagger+\epsilon^{ij}\partial^i\Sigma_L^j+\Sigma_L^i\Sigma_L^i+i\left[\Sigma_L^1,\Sigma_L^2\right],F_R\right]
\\\nonumber\\
\label{eq:123}
U_R\left[F_T\right]=-\frac{1}{2\left|\vec{k}\right|}\left(\Sigma_L^i m P_+^{ij}F_T^j-P_-^{ij}F_T^j m^\dagger\Sigma_L^i\right)
\nonumber\\
+\frac{1}{2\left|\vec{k}\right|}\left(m\Sigma_R^i P_+^{ij}F_T^j-P_-^{ij}F_T^j\Sigma_R^i m^\dagger\right)
\end{eqnarray}
The equations for $F_{L/R}$ are coupled to $F_T^i$ via the $U_{L/R}\left[F_T\right]$ terms as well as terms contained in $C_L$ and $C_R$.  Therefore, in addition to the kinetic equations for $F_{L/R}$, which are related to the usual neutrino density matrices, we will need to derive the kinetic equations for $F_T^i$, which encode coherence between left-handed and right-handed neutrinos.  Note that the coupling of $F_T^i$ to $F_{L/R}$ vanishes in the limit of isotropy.  This is as expected, since in the isotropic limit, conservation of angular momentum prohibits the interconversion of left-handed and right-handed states.

Using the notation $\Phi=\frac{1}{2}\left(F_T^1+iF_T^2\right)$, we write the kinetic equations for $F_L$ and $F_R$ as follows:
\begin{eqnarray}
\label{eq:124}
i\partial^\kappa F_R+\frac{1}{2\left|\vec{k}\right|}i  \left\{\Sigma_L^i, \partial^i  F_R\right\}+\frac{1}{2}i\left\{\partial_\mu\Sigma_L^\kappa, \partial_k^\mu F_R\right\}
\nonumber\\
-\left[H_L, F_R\right]+U_R\left[\Phi\right]=iC_R\left[F_L,F_R,\Phi\right]
\\
\label{eq:125}
i\partial^\kappa F_L+\frac{1}{2\left|\vec{k}\right|} \left\{\Sigma_R^i, \partial^i  F_L\right\}+\frac{1}{2}i\left\{\partial_\mu \Sigma_R^\kappa, \partial_k^\mu F_L\right\}
\nonumber\\
-\left[H_R,F_L\right]+U_L\left[\Phi\right]=iC_L\left[F_L,F_R,\Phi\right]
\end{eqnarray}
\noindent
where the Hamiltonian-like operators are
\begin{eqnarray}
\label{eq:126}
H_L=\Sigma_L^\kappa+\delta\Sigma_L^\kappa
\nonumber\\
+\frac{1}{2\left|\vec{k}\right|}\left(mm^\dagger+\epsilon^{ij}\partial^i\Sigma_L^j+4\Sigma_L^-\Sigma_L^+\right)
\\
\label{eq:127}
H_R=\Sigma_R^\kappa+\delta\Sigma_R^\kappa
\nonumber\\
+\frac{1}{2\left|\vec{k}\right|}\left(m^\dagger m-\epsilon^{ij}\partial^i\Sigma_R^j+4\Sigma_R^+\Sigma_R^-\right)
\end{eqnarray}
and the couplings to the spin coherence density are
\begin{eqnarray}
\label{eq:128}
U_R\left[\Phi\right]=
\nonumber\\
\frac{1}{\left|\vec{k}\right|}\left(\left(m\Sigma_R^--\Sigma_L^-m\right)\Phi
+\Phi^\dagger\left(m^\dagger\Sigma_L^+-\Sigma_R^+m^\dagger\right)\right)
\\
\label{eq:129}
U_L\left[\Phi\right]=
\nonumber\\
-\frac{1}{\left|\vec{k}\right|}\left(\left(m^\dagger\Sigma_L^+-\Sigma_R^+m^\dagger\right)\Phi^\dagger+\Phi\left(m\Sigma_R^--\Sigma_L^-m\right)\right)
\end{eqnarray}
Here, $\Sigma^\pm\equiv\frac{1}{2}\left(\Sigma^1\pm i\Sigma^2\right)$; while
$C_L$ and $C_R$ correspond to Boltzmann collision terms, as will be shown below.  These are given by equations (\ref{eq:106})-(\ref{eq:107}) and (\ref{eq:116})-(\ref{eq:117}).

\subsection{Kinetic Equations for Spin Coherence}
We see that the equations of motion for $F_L$ and $F_R$, which encode the density matrices for the particles, are coupled to the spin coherence density $\Phi$.  We will see below that this spin coherence can mediate oscillations between particles of opposite helicity.  We now derive the equations of motion for $\Phi$.\\

We begin with kinetic equations for $F_T$, which can be derived from the vector components of equation (\ref{eq:52}).  To $O\left(\epsilon^2\right)$, the vector equations are
\begin{eqnarray}
\label{eq:130}
\left(k+\frac{1}{2}i\partial\right)\Delta_S^\dagger-\Sigma_L\Delta_S^\dagger-m\left(F_L+\Delta_L\right)
\nonumber\\
-\left(k+\frac{1}{2}i\partial\right)\cdot\left(F_T^R+\Delta_T^R\right)+\tilde{\Sigma}_L\cdot\left(F_T^R+\Delta_T^R\right)
\nonumber\\
=\frac{1}{2}i\left(\Pi_L^+\cdot F_T^{R-}-\Pi_L^-\cdot F_T^{R+}\right)
\nonumber\\
-\frac{1}{2}i\left(\Pi_S^+ F_L^- - \Pi_S^- F_L^+ +\Pi_L^{T+}\cdot F_L^--\Pi_L^{T-}\cdot F_L^+\right)
\\\nonumber\\
\label{eq:131}
\left(k+\frac{1}{2}i\partial\right)\Delta_S-\Sigma_R\Delta_S-m^\dagger\left(F_R+\Delta_R\right)
\nonumber\\
+\left(k+\frac{1}{2}i\partial\right)\cdot\left(F_T^L+\Delta_T^L\right)-\tilde{\Sigma}_R\cdot\left(F_T^L+\Delta_T^L\right)
\nonumber\\
=-\frac{1}{2}i\left(\Pi_R^+\cdot F_T^{L-}-\Pi_R^-\cdot F_T^{L+}\right)
\nonumber\\
-\frac{1}{2}i\left(\Pi_S^{\dagger+}F_R^--\Pi_S^{\dagger-}F_R^+-\Pi_R^{T+}\cdot F_R^-+\Pi_R^{T-}\cdot F_R^+\right)
\end{eqnarray}

We take the Hermitian conjugate of the equation (\ref{eq:130}), add to equation (\ref{eq:131}), and then choose the $\hat{x}^i$ components and act with $P_+^{ij}$.  This gives
\begin{eqnarray}
\label{eq:132}
i\partial^\kappa P_+^{ij}F_T^j-\left(\tilde{\Sigma}_R^\kappa P_+^{ij}F_T^j-P_+^{ij}F_T^j\tilde{\Sigma}_L^{\dagger\kappa}\right)
\nonumber\\
-\frac{i}{2 |\vec{k}|}   \left(  \partial^n \Sigma_R^n    \  P_+^{ij}F_T^j  +   \  P_+^{ij}F_T^j \  \partial^n \Sigma_L^n  \right)
\nonumber \\
+\frac{i}{2 |\vec{k}|}   P_+^{ij}  \left(\partial^j \Sigma_R^n    \  P_+^{nm}F_T^m +   \  P_+^{nm}F_T^m \  \partial^j \Sigma_L^n   
\right)
\nonumber\\
+P_+^{ij}\left(\left(m^\dagger\Delta_R^j+\Delta_L^jm^\dagger\right)+\left(\Sigma_R^j\Delta_S+\Delta_S\Sigma_L^j\right)\right)
\nonumber\\
+P_+^{ij}\left(i\partial^j\Delta_T-\left(\Sigma_R^j\Delta_T-\Delta_T\Sigma_L^j\right)\right)=iC_T^i
\end{eqnarray}
where
\begin{eqnarray}
\label{eq:133}
C_T^i=\frac{1}{2}\left(\Pi_R^{+\kappa} P_+^{ij}F_T^{j}+P_+^{ij}F_T^{j}\Pi_L^{+\kappa}\right)
\nonumber\\
+\frac{1}{2}\left(\Pi_R^{-\kappa} P_+^{ij}F_T^{j}+P_+^{ij}F_T^{j}\Pi_L^{-\kappa}\right)
\nonumber\\
-P_+^{ij}\left(\Pi_T^{j+}G_R^-+G_L^-\Pi_T^{j+}-\Pi_T^{j-}G_R^+-G_L^+\Pi_T^{j-}\right)
\end{eqnarray}

Writing this in terms of the complex matrix $\Phi$, defined above:
\begin{eqnarray}
\label{eq:134}
i\partial^\kappa\Phi-\left(\tilde{\Sigma}_R^\kappa\Phi-\Phi\tilde{\Sigma}_L^{\dagger\kappa}\right)
+i\partial^+\Delta_T-\left(\Sigma_R^+\Delta_T-\Delta_T\Sigma_L^+\right)
\nonumber\\
- \frac{i}{2 |\vec{k}|}  \ \left( \partial^i \Sigma_R^i   \,  \Phi + \Phi \,   \partial^i \Sigma_L^i \right) 
+ \frac{i}{ |\vec{k}|}  \ \left(  \partial^+ \Sigma_R^-    \,  \Phi + \Phi  \,  \partial^+ \Sigma_L^-  \right) 
\nonumber \\
+\left(m^\dagger\Delta_R^++\Delta_L^+m^\dagger\right)+\left(\Sigma_R^+\Delta_S+\Delta_S\Sigma_L^+\right)=iC_\Phi \qquad 
\end{eqnarray}
where, using $P_T^\pm=\frac{1}{2}\left(\Pi_T^1+i\Pi_T^2\right)^\pm$,
\begin{eqnarray}
\label{eq:135}
C_\Phi=
\frac{1}{2}\left(\left(\Pi_R^{+\kappa}+\Pi_R^{-\kappa}\right)\Phi+\Phi\left(\Pi_L^{+\kappa}+\Pi_L^{-\kappa}\right)\right)
\nonumber\\
-P_T^+ G_R^--G_L^-P_T^++P_T^- G_R^++G_L^+P_T^-
\end{eqnarray}

We separate the combination of small components in equation (\ref{eq:132}) into a part dependent on $\Phi$ and one dependent on $F_{L/R}$:
\begin{eqnarray}
\label{eq:136}
V\left[\Phi\right]+V\left[F_{L/R}\right]=i\partial^+\Delta_T-\left(\Sigma_R^+\Delta_T-\Delta_T\Sigma_L^+\right)
\nonumber\\
+\left(m^\dagger\Delta_R^++\Delta_L^+m^\dagger\right)+\left(\Sigma_R^+\Delta_S+\Delta_S\Sigma_L^+\right)
\end{eqnarray}

Using equations (\ref{eq:98})-(\ref{eq:101}) for the small components, we obtain
\begin{eqnarray}
\label{eq:137}
V\left[\Phi\right]=\frac{1}{2\left|\vec{k}\right|}i\partial^i\left(\Sigma_R^i\Phi+\Phi\Sigma_L^i\right)
\nonumber\\
-\frac{1}{2\left|\vec{k}\right|}\left(m^\dagger m +2i\partial^-\Sigma_R^++4\Sigma_R^+\Sigma_R^-\right)\Phi
\nonumber\\
+\frac{1}{2\left|\vec{k}\right|}\Phi\left(mm^\dagger-2i\partial^-\Sigma_L^++4\Sigma_L^-\Sigma_L^+\right)
\\
\label{eq:138}
V\left[F_{L/R}\right]=-\frac{1}{\left|\vec{k}\right|}\left(m^\dagger\Sigma_L^+F_R-F_Lm^\dagger\Sigma_L^+\right)
\nonumber\\
+\frac{1}{\left|\vec{k}\right|}\left(\Sigma_R^+m^\dagger F_R-F_L\Sigma_R^+ m^\dagger\right)
\end{eqnarray}

We arrange the kinetic equation for $\Phi$ as follows:
\begin{eqnarray}
\label{eq:139}
i\partial^\kappa\Phi+\frac{1}{2\left|\vec{k}\right|}i  \left(\Sigma_R^i   \,  \partial^i  \Phi+ \partial^i \Phi \, \Sigma_L^i\right)
\nonumber\\
+\frac{1}{2}i\left(\partial_\mu\Sigma_R^\kappa\partial_k^\mu\Phi+\partial_k^\mu\Phi\partial_\mu\Sigma_L^\kappa\right)
\nonumber\\
-\left(H_\Phi\Phi-\Phi \bar{H}_\Phi\right)+V\left[F_{L/R}\right]=iC_\Phi
\end{eqnarray}
where $V\left[F_{L/R}\right]$ is given by equation (\ref{eq:138}), and the operators $H_\phi$ and $\bar{H}_\phi$ are given by
\begin{eqnarray}
\label{eq:140}
H_\Phi&=&  H_R 
\\
\label{eq:141}
\bar{H}_\Phi&= &H_L ~.
\end{eqnarray}

\subsection{The Majorana Conditions and Dispersion Relation}
We now extract the kinetic equations for particle and antiparticle density matrices.  These equations can be obtained by integrating the equations of motion for $F_L$ and $F_R$ over positive or negative energies.  

For Majorana neutrinos, the equations of motion for $F_L$ and $F_R$ must be redundant; that is, the positive-energy component of $F_L$ contains the same information as the negative-energy component of $F_R$. 
Specifically, $F_L\left(k\right)=F_R^T\left(-k\right)$ and $\Phi\left(k\right)=\Phi^T\left(-k\right)$.  The redundancy of the equations of motion requires
\begin{eqnarray}
\label{eq:142}
m=m^T
\nonumber\\
\Sigma_R=-\Sigma_L^T\equiv\Sigma
\end{eqnarray}  
The condition $m=m^T$ follows from the form of the Majorana mass term.  When we calculate the matter potential and the gain-loss potentials below, we will see that the other conditions are also satisfied.  This follows simply from the fact that the potentials $\Sigma$ and $\Pi$ are functionals of the two-point function, and the Majorana constraints on the form of the two-point function lead to the appropriate constraints on $\Sigma$ and $\Pi$.

In addition to imposing the Majorana constraints, we must solve the dispersion relations for $F_L$, $F_R$ and $F_T$, given by equations (\ref{eq:91})-(\ref{eq:92}) and (\ref{eq:87})-(\ref{eq:88}), to $O\left(\epsilon\right)$.  We solve equation (\ref{eq:91}), by transforming to the basis in flavor space that diagonalizes $\Sigma_L^\kappa$.  In this basis, $F_R$ satisfies equation (\ref{eq:91}) if it has the form
\begin{eqnarray}
\label{eq:143}
F_R=\left(\begin{array}{ccc} \delta\left(1,1\right)g_R^{11} & \delta\left(1,2\right)g_R^{12} & ... \\ \delta\left(2,1\right)g_R^{21} & \delta\left(2,2\right)g_R^{22} & ... \\ ... & ... & ...\end{array}\right)
\end{eqnarray}
Here, $\delta\left(I,J\right)$ is an expression containing a delta function that enforces the condition $k\cdot\hat{\kappa}-\frac{1}{2}\left(\Sigma_L^I+\Sigma_L^J\right)=O\left(\epsilon^2\right)$, where $\Sigma_L^I$ is the $I$th eigenvalue of $\Sigma_L^\kappa$.  We wish to write this as $2\pi\delta\left(k^2+O\left(\epsilon\right)\right)\left|\vec{k}\right|$, to match the $O\left(1\right)$ expression $F_L=2\pi\delta\left(k^2\right)\left|\vec{k}\right| g\left(k\right)$.  Therefore, the appropriate form of the delta function is $\delta\left(I,J\right)=2\pi\delta\left(k^2-\left|\vec{k}\right|\left(\Sigma_L^I+\Sigma_L^J\right)+O\left(\epsilon^2\right)\right)\left|\vec{k}\right|$.\\

Using flavor projection operators $P_I$, where $P_1=\left(\begin{array}{ccc} 1 & 0 & ... \\ 0 & 0 & ... \\ ... & ... & ...\end{array}\right)$, $P_2=\left(\begin{array}{ccc} 0 & 0 & ... \\ 0 & 1 & ... \\ ... & ... & ...\end{array}\right)$, etc, we can write
\begin{eqnarray}
\label{eq:144}
F_R = \sum_{IJ}2\pi\delta\left(k^2-\left|\vec{k}\right|\left(\Sigma_L^I+\Sigma_L^J\right)\right)\left|\vec{k}\right| P_I g_R P_J
\end{eqnarray}
We can now transform to an arbitrary basis (such as the flavor basis) by using the unitary matrix $U_L$, which transforms from the desired basis to one in which $\Sigma_L^\kappa$ is diagonal, and use equation (\ref{eq:40}) to express $F_R$ in terms of $f$ and $\bar{f}$.  The result is
\begin{eqnarray}
\label{eq:145}
F_R=2\pi\left|\vec{k}\right|\sum_{IJ}\delta\left(k^2-\left|\vec{k}\right|\left(\Sigma_L^I+\Sigma_L^J\right)\right)\left(U_L^\dagger P_I U_L\right)\times
\nonumber\\
\left(\frac{1}{2}-\theta\left(k^0\right)\bar{f}^T\left(\vec{k}\right)-\theta\left(-k^0\right)f^T\left(-\vec{k}\right)\right)\left(U_L^\dagger P_J U_L\right)\ \ \ 
\end{eqnarray}
Similarly,
\begin{eqnarray}
\label{eq:146}
F_L=2\pi\left|\vec{k}\right|\sum_{IJ}\delta\left(k^2-\left|\vec{k}\right|\left(\Sigma_R^I+\Sigma_R^J\right)\right)\left(U_R^\dagger P_I U_R\right)\times
\nonumber\\
\left(\frac{1}{2}-\theta\left(k^0\right)f\left(\vec{k}\right)-\theta\left(-k^0\right)\bar{f}\left(-\vec{k}\right)\right)\left(U_R^\dagger P_J U_R\right)\ \ \ 
\end{eqnarray}
where the density matrices $f$ and $\bar{f}$ are expressed in the original flavor basis.  For spin coherence, the dispersion relation is given by equations (\ref{eq:87})-(\ref{eq:88}).  In terms of the quantity $\Phi$, these equations give
 \begin{eqnarray}
 \label{eq:147}
\left(k\cdot\hat{\kappa}\right)\Phi-\frac{1}{2}\left(\Sigma_R^\kappa\Phi+\Phi\Sigma_L^\kappa\right)=O\left(\epsilon^2\right).
\end{eqnarray}
\noindent
Note that $\Phi$ satisfies the dispersion relation if it has the form
\begin{eqnarray}
\label{eq:148}
\Phi=-2\pi\left|\vec{k}\right|\sum_{IJ}\delta\left(k^2-\left|\vec{k}\right|\left(\Sigma_R^I+\Sigma_L^J\right)\right)\left(U^\dagger_R P_I U_R\right)\ 
\nonumber\\
\times \left(\theta\left(k^0\right)\phi\left(\vec{k}\right)+\theta\left(-k^0\right)\phi^T\left(-\vec{k}\right)\right)\left(U_L^\dagger P_J U_L\right)\ \ 
\end{eqnarray}

\subsection{Equations of Motion for Density Matrices and Spin Coherence Densities}

We can now find the equations of motion for the density matrices of Majorana neutrinos.  These equations can be obtained by integrating the equation of motion for $F_L$, equation (\ref{eq:125}), over positive energies, and similarly integrating equation (\ref{eq:124}) for $F_R$ over positive energies and taking the transpose.  We also integrate equation (\ref{eq:139}) over positive energies to obtain the equations of motion for the spin coherence density.   Due to the Majorana nature of the fermions, these equations are redundant with those obtained by integrating over negative energies; the redundancy is satisfied if the Majorana conditions on the mass and the matter potentials, equation (\ref{eq:142}), hold.  Performing the integration and imposing the Majorana conditions gives
\begin{eqnarray}
\label{eq:149}
i\partial^\kappa f^{\left(1\right)}+\frac{1}{2\left|\vec{k}\right|}i      \left\{\Sigma^i,   \partial^i    f\right\}
-  \frac{1}{2}i\left\{ \frac{\partial\Sigma^\kappa}{\partial{\vec{x}}}   , \frac{\partial f}{\partial \vec{k}}\right\}
\nonumber\\
-\left[H, f\right]^{\left(1\right)}+U\left[\phi\right]=iC\left[f,\bar{f},\phi\right]
\\\nonumber\\
\label{eq:150}
i\partial^\kappa\bar{f}^{\left(1\right)}-\frac{1}{2\left|\vec{k}\right|}i   \left\{\Sigma^i,   \partial^i  \bar{f}\right\}
+ \frac{1}{2}i\left\{ \frac{\partial\Sigma^\kappa}{\partial{\vec{x}}}   , \frac{\partial \bar{f}}{\partial \vec{k}}\right\}
\nonumber\\
-\left[\bar{H},\bar{f}\right]^{\left(1\right)}+\bar{U}\left[\phi\right]=i\bar{C}\left[f,\bar{f},\phi\right]
\end{eqnarray}
\begin{eqnarray}
\label{eq:151}
i\partial^\kappa\phi^{\left(1\right)}+\frac{1}{2\left|\vec{k}\right|}  i  \left(\Sigma^i \,  \partial^i \phi-  \partial^i \phi \,   \Sigma^{iT}\right)
\nonumber\\
-  \frac{1}{2}i   \left(
  \frac{\partial\Sigma^\kappa}{\partial{\vec{x}}}   \cdot  \frac{\partial \phi}{\partial \vec{k}}  
- 
\frac{\partial \phi}{\partial \vec{k}}  \cdot      \frac{\partial\Sigma^{\kappa T}}{\partial{\vec{x}}}  
\right) 
\nonumber \\
-\left(H_\Phi\phi-\phi \bar{H}_\Phi\right)^{\left(1\right)}+V\left[f, \bar{f}\right]=iC_\phi\left[\phi,f,\bar{f} \right]
\end{eqnarray}

Since $\Sigma_L$ and $\Sigma_R$ are related by the Majorana condition, we use the notation $\Sigma\equiv\Sigma_R=-\Sigma_L^T$.  The terms immediately following the first derivative term, {\it i.e.,} those involving anticommutators and derivatives of the matter potential, give trajectory deviation and a shift in energy of the particles in response to a changing matter potential.

The Hamiltonian operators for neutrinos and anti-neutrinos, $H$ and $\bar{H}$, are:
\begin{eqnarray}
\label{eq:152}
H=\Sigma^{\kappa}+\delta\Sigma^\kappa
\nonumber\\
+\frac{1}{2\left|\vec{k}\right|}\left(m^\dagger m-\epsilon^{ij}\partial^i\Sigma^j+\Sigma^i\Sigma^i-i\left[\Sigma^1,\Sigma^2\right]\right)
\\
\label{eq:153}
\bar{H}=\Sigma^{\kappa}+\delta\Sigma^\kappa
\nonumber\\
-\frac{1}{2\left|\vec{k}\right|}\left(m^\dagger m-\epsilon^{ij}\partial^i\Sigma^j+\Sigma^i\Sigma^i-i\left[\Sigma^1,\Sigma^2\right]\right)
\end{eqnarray}

The terms coupling the kinetic equations to the spin coherence are:
\begin{eqnarray}
\label{eq:154}
U=\frac{1}{\left|\vec{k}\right|}\left(\Sigma^+m^\star\phi^\dagger-\phi m\Sigma^-\right)
\nonumber\\
+\frac{1}{\left|\vec{k}\right|}\left(m^\star\Sigma^{+T}\phi^\dagger-\phi\Sigma^{-T}m\right)
\\
\label{eq:155}
\bar{U}=-\frac{1}{\left|\vec{k}\right|}\left(\Sigma^+m^\star\phi^\star-\phi^Tm\Sigma^-\right)
\nonumber\\
-\frac{1}{\left|\vec{k}\right|}\left(m^\star\Sigma^{+T}\phi^\star-\phi^T\Sigma^{-T}m\right)
\end{eqnarray}

The collision terms on the right-hand side are
\begin{eqnarray}
\label{eq:156}
C=
\frac{1}{2}\left(
\left\{ \tilde{\Pi}_R^{\kappa+}, f \right\} 
- \left\{ \tilde{\Pi}_R^{\kappa-}, 1 - f\right\}
\right)
\nonumber\\
+ \left(\tilde{P}_T^++\tilde{P}_T^-\right)   \phi^\dagger   + \phi  \left(\tilde{P}_T^++\tilde{P}_T^-\right)^\dagger 
\end{eqnarray}
\begin{eqnarray}
\label{eq:157}
\bar{C}=
\frac{1}{2}\left(
\left\{  \left[\tilde{\Pi}_L^{\kappa+}\right]^T , \bar{f} \right\} 
- \left\{\left[\tilde{\Pi}_L^{\kappa-}\right]^T , 1 - \bar{f}\right\}
\right)
\nonumber\\
+ \left(\tilde{P}_T^++\tilde{P}_T^- \right)^T    \phi^\star  
+ \phi^T  \left(\tilde{P}_T^++\tilde{P}_T^-\right)^\star 
\end{eqnarray}
\begin{eqnarray}
\label{eq:157-1}
C_\phi=
\frac{1}{2}\left[ \left(\tilde{\Pi}_R^{\kappa+}+\tilde{\Pi}_R^{\kappa -}\right)\phi+\phi\left(\tilde{\Pi}_L^{\kappa+}+\tilde{\Pi}_L^{\kappa-}\right)\right]
\nonumber \\
+ f \, \tilde{P}_T^+ - (1 - f)  \tilde{P}_T^-
+ \tilde{P}_T^+ \bar{f}^T -  \tilde{P}_T^-  \left(1 -  \bar{f}^T \right)
\end{eqnarray}
where
\begin{eqnarray}
\label{eq:157-2}
\tilde{\Pi}_{L,R}^{\kappa \pm}    \left(\vec{k}\right)=\int_0^{\infty}dk^0\ \Pi_{L,R}^{\kappa \pm}  \left(k\right) \ \delta (k^0 - |\vec{k}|)
\nonumber\\
 \tilde{P}_T^{\pm} \left(\vec{k}\right)=\int_0^\infty dk^0\ 
P_T^\pm     \left(k\right)  \ \delta (k^0 - |\vec{k}|)
\nonumber 
\end{eqnarray}

The first two terms in $C$ and $\bar{C}$ correspond to the gain-loss terms in the Boltzmann equation, including Fermi blocking.  The remainder represent coupling to the spin coherence $\phi$ via collisional processes.

The superscript \lq\lq $\ ^{\left(1\right)}$\rq\rq\  we take to indicate terms that include corrections stemming from a shift in the dispersion relation, up to $O\left(\epsilon^2\right)$. 
Specifically,
\begin{eqnarray}
\label{eq:159}
f^{\left(1\right)}=\int_0^\infty\frac{dk^0}{2\pi}\left(-2F_L\right)=
\nonumber\\
f-\sum_{IJ}\frac{\Sigma^I+\Sigma^J}{2\left|\vec{k}\right|}\left(U^\dagger P_I U\right)f\left(U^\dagger P_J U\right)
\\
\label{eq:160}
\bar{f}^{\left(1\right)}=\int_0^{\infty}\frac{dk^0}{2\pi}\left(-2F_R\right)^T=
\nonumber\\
\bar{f}+\sum_{IJ}\frac{\Sigma^I+\Sigma^J}{2\left|\vec{k}\right|}\left(U^\dagger P_J U\right)\bar{f}\left(U^\dagger P_I U\right)
\end{eqnarray}
and
\begin{eqnarray}
\label{eq:161}
\left[H,f\right]^{\left(1\right)}=\left[H\left(\epsilon\right),f^{\left(1\right)}\right]+\left[H\left(\epsilon^2\right),f\right]
\end{eqnarray}
where $H\left(\epsilon\right)$ and $H\left(\epsilon^2\right)$ are the $O\left(\epsilon\right)$ and
$O\left(\epsilon^2\right)$ contributions to $H$.

The quantities appearing in the equation of motion for spin coherence are Hamiltonian-like quantities acting on $\phi$ itself
\begin{eqnarray}
\label{eq:162}
H_\Phi&=& H 
\\
\label{eq:163}
\bar{H}_\Phi&=& - \bar{H}^T  
\end{eqnarray}
as well as a term coupling $\phi$ to $f$ and $\bar{f}$:
\begin{eqnarray}
\label{eq:164}
V\left[f,\bar{f}\right]=\frac{1}{\left|\vec{k}\right|}\left(m^\star\Sigma^{+T}\bar{f}^T-fm^\star\Sigma^{+T}\right)
\nonumber\\
+\frac{1}{\left|\vec{k}\right|}\left(\Sigma^+m^\star \bar{f}^T-f\Sigma^+m^\star\right)~.
\end{eqnarray}

The quantity $\phi^{\left(1\right)}$ incorporates corrections due to the dispersion relation:
\begin{eqnarray}
\label{eq:165}
\phi^{\left(1\right)}=\phi-\sum_{IJ}\frac{\Sigma_I-\Sigma_J}{2\left|\vec{k}\right|}\left(U^\dagger P_I U\right)\phi\left(U^T P_J U^\star\right)
\end{eqnarray}\\

\subsection{$2N_f \times 2N_f$ Notation}
Equations (\ref{eq:149})-(\ref{eq:151}), the quantum kinetic equations, can be written more compactly as follows:
\begin{eqnarray}
\label{eq:165-1}
iD\left[{\cal F}\right]-\left[{\cal H},{\cal F}\right]  =i{\cal C}\left[\cal F\right]
\end{eqnarray}

Here, for 3 neutrino flavors, ${\cal F}$ and ${\cal H}$ are $6 \times 6$ matrices having the following block structure:
\begin{eqnarray}
\label{eq:165-2}
{\cal F}\equiv\left(\begin{array}{cc}f & \phi \\ \phi^\dagger & \bar{f}^T\end{array}\right)\ \ \ \ \ \ {\cal H}\equiv\left(\begin{array}{cc}H & H_{\nu\bar{\nu}} \\ H_{\nu\bar{\nu}}^\dagger & -\bar{H}^T\end{array}\right)
\end{eqnarray}
The quantities $H$ and $\bar{H}$ are the neutrino and anti-neutrino Hamiltonians, given by equations (\ref{eq:152}) and (\ref{eq:153}), while $H_{\nu\bar{\nu}}$ is given by
\begin{eqnarray}
\label{eq:165-3}
H_{\nu\bar{\nu}} = -\frac{1}{\left|\vec{k}\right|}\left(\Sigma^+m^\star+m^\star\Sigma^{+T}\right)
\end{eqnarray}
The derivative term is
\begin{eqnarray}
\label{eq:165-4}
iD\left[{\cal F}\right] = i\partial^\kappa {\cal F}^{\left(1\right)} + \frac{i}{2\left|\vec{k}\right|}\left\{\left(\begin{array}{cc}\Sigma^i & 0 \\ 0 & -\Sigma^{i T}\end{array}\right), \partial^i {\cal F}\right\}
\nonumber\\
- \frac{1}{2}i\left\{     
\frac{\partial}{\partial \vec{x}}
\left(\begin{array}{cc}\Sigma^\kappa & 0 \\ 0 & -\Sigma^{\kappa T}\end{array}\right), \frac{\partial{\cal F}}{\partial \vec{k}}\right\}
\end{eqnarray}
and the collision term is
\begin{eqnarray}
\label{eq:165-5}
C=\left(\begin{array}{cc} C & C_\phi \\ C^\dagger_\phi & \bar{C}^T\end{array}\right)
\end{eqnarray}
where $C$, $\bar{C}$ and $C_\phi$ are given by equations ($\ref{eq:156}$), ($\ref{eq:157}$) and ($\ref{eq:157-1}$).

\section{neutrino interactions with matter}
In this section, we compute the matter potential $\Sigma$ for neutrinos.  We also show how the gain-loss potentials $\Pi^\pm$ are calculated, and explicitly compute some of the terms in $\Pi^\pm$ to show that these quantities can be identified with the gain-loss terms in the Boltzmann equation.

\subsection{Matter Potential}
The matter potential corresponds to the local piece of the neutrino self-energy, as given by Equation (\ref{eq:25}).  Since, in the low-energy limit, the $W$ and $Z$ boson propagators are local (proportional to $\delta\left(x-y\right)$), to leading order the matter potential is given by the one-loop diagrams shown in Figure 1.  We note that in general, the leading-order form of the weak boson propagator receives small corrections, which may be physically important in some environments \cite{Fuller87,Notzold:1988fv,Blennow:2008lr,Esteban-Pretel:2008fk,Esteban-Pretel:2008qy,Esteban-Pretel:2010uq}.  For simplicity, we do not include these corrections here; however, incorporating them would be relatively straightforward.

\begin{figure}
\includegraphics[width=2.5in]{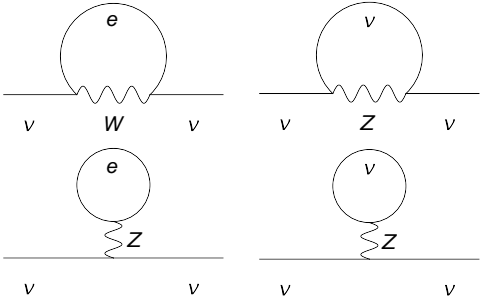}
\caption{Feynman graphs for neutral and charged current one-loop contributions to neutrino self-energy. }
\label{feynman1}
\end{figure}

Note that the one-loop diagrams involving only neutrino propagators include all corrections to the neutrino two-point function, since the neutrino two-point function is treated as a dynamical quantity.  As a consequence, the diagrams already include all "bubble" diagrams with bubbles branching off an internal neutrino line.  However, since we are not treating charged leptons as dynamical, there are additional contributions corresponding to corrections to the charged lepton two-point function.  Examples of such contribution are given in Figure 2.  Diagrams such as this generate a neutrino magnetic moment, thus giving neutrinos a small effective interaction with the electromagnetic field.  These diagrams also give a small effective mass splitting between muon and tau neutrinos, due to the different mass of the virtual charged lepton on the internal lines.  Since the sub-diagram involves the electromagnetic, rather than the weak interaction, even higher-order diagrams like this can give a larger contribution to $\Sigma$ than two-loop diagrams involving only the weak interaction. Nevertheless, for simplicity, we will not include such diagrams here, and simply use the leading-order expressions for the charged lepton two-point function.  However, it should be kept in mind that the charged lepton corrections, though small, nevertheless may prove important in neutrino flavor evolution in supernovae, as demonstrated in Ref.s \cite{Fuller87,Botella:1987lr,Mirizzi:2009lr}.\\

\begin{figure}
\includegraphics[width=2.5in]{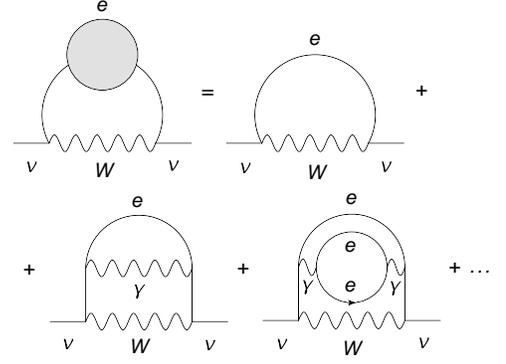}
\caption{Examples of diagrams that incorporate corrections to the charged lepton two-point function.  For simplicity, we neglect all but the leading-order diagram in this section.}
\label{feynman15}
\end{figure}

Having made these simplifications, we compute the first diagram in Fig. 1.  Note that this diagram cannot involve an arrow-clashing charged lepton propagator (involving either an odd number of mass insertions, or any kind of charged lepton spin coherence) because the arrow-clashing propagator always connects the charged lepton field to its Dirac counterpart, which does not interact via the charged current interaction.  Therefore, the only contributions to $\Sigma$ from this diagram are those given in Fig. 3.

\begin{figure}
\includegraphics[width=2.5in]{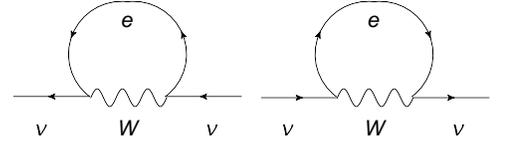}
\caption{Contributions to the charged current one-loop diagram}
\label{feynman16}
\end{figure}

In position space, these diagrams give
\begin{eqnarray}
\label{eq:166}
\Sigma_{IJ,\alpha\dot{\alpha}}^{W,e}\left(x,y\right)=
\nonumber\\
i\delta^4\left(x-y\right)\left(-i2\sqrt{2}G_F\right)\sigma^\mu_{\alpha\dot{\beta}}G_{IJ}^{e, \dot{\beta}\beta}\left(x,y\right)\sigma_{\mu\beta\dot{\alpha}}
\\
\label{eq:167}
\Sigma_{IJ}^{W,e,\dot{\alpha}\alpha}\left(x,y\right)=
\nonumber\\
i\delta^4\left(x-y\right)\left(-i2\sqrt{2}G_F\right)\bar{\sigma}^{\mu\dot{\alpha}\beta}G^e_{IJ,\beta\dot{\beta}}\left(x,y\right)\bar{\sigma}_{\mu\dot{\beta}\alpha}
\end{eqnarray}
The superscript $W$ indicates that this is the contribution to the matter potential stemming from the charged current interaction.

Upon Wigner transformation, this is
\begin{eqnarray}
\label{eq:168}
\Sigma^{W,e}_{IJ,\alpha\dot{\alpha}}\left(x\right)=2\sqrt{2}G_F\int\frac{d^4q}{\left(2\pi\right)^4} \sigma^\mu_{\alpha\dot{\beta}} F_{IJ}^{e, \dot{\beta}\beta}\left(x,q\right)\sigma_{\mu, \beta\dot{\alpha}}\ \ \ 
\\
\label{eq:169}
\Sigma^{W,e,\dot{\alpha}\alpha}_{IJ}\left(x\right)=2\sqrt{2}G_F\int\frac{d^4q}{\left(2\pi\right)^4}\ \bar{\sigma}^{\mu,\dot{\alpha}\beta} F^e_{IJ,\beta\dot{\beta}}\left(x,q\right)\bar{\sigma}_\mu^{\dot{\beta}\alpha}\ \ \ 
\end{eqnarray}

In the flavor basis, neglecting corrections from interactions with the plasma, the statistical function for charged fermions is
\begin{eqnarray}
\label{eq:170}
F_{IJ}^{e, \dot{\alpha}\alpha}\left(x,q\right)=
\nonumber\\
2\pi\sum_{K}\delta\left(q^2-m_K^2\right)q\cdot\bar{\sigma}^{\dot{\alpha}\alpha}\left(P_K\right)_{JI}
\times\nonumber\\
\left(\frac{1}{2}-\theta\left(q^0\right)f^e_{R,K}\left(x,\vec{q}\right)-\theta\left(-q^0\right)\bar{f}^e_{L,K}\left(x,-\vec{q}\right)\right)
\\
\label{eq:171}
F_{IJ,\alpha\dot{\alpha}}^{e}\left(x,q\right)=
\nonumber\\
2\pi\sum_{K}\delta\left(q^2-m_K^2\right)q\cdot\sigma_{\alpha\dot{\alpha}}\left(P_K\right)_{IJ}
\times\nonumber\\
\left(\frac{1}{2}-\theta\left(q^0\right)f_{L,K}^{e}\left(x,\vec{q}\right)-\theta\left(-q^0\right)\bar{f}_{R,K}^{e}\left(x,-\vec{q}\right)\right)
\end{eqnarray}
Here, the flavor index $K$ denotes electrons, muons and tauons.   $m_K$ is the charged lepton mass corresponding to flavor $K$, $\left(P_K\right)_{IJ}$ are flavor projection matrices, $f^e_{L,K}$ is the density of left-handed charged leptons of flavor $K$, and $\bar{f}^e_{R,K}$ is the density of right-handed charged anti-leptons of flavor $K$.  Note that this expression assumes that there is no coherence between charged leptons of different flavor.  This assumption is motivated by two arguments.  First, mass-squared splittings between charged leptons are large, so at low energies, flavor coherence would be difficult to generate.  Second, charged leptons interact much more strongly than neutrinos.  Scattering is expected to cause decoherence, so that even if charged lepton flavor coherence could be generated, it would be quickly destroyed by interactions.

In supernovae, and in certain epochs in the early Universe, the temperature is too low for a substantial number of muons or tauons to be present in the plasma.  In this case, we can set $f_K,\bar{f}_K\approx 0$ for $K\not=1$.

Performing the integrals in equations (\ref{eq:168})-(\ref{eq:169}) over $q^0$ and using the definition of $\Sigma_{L/R}$ in equation (\ref{eq:51})  gives

\begin{eqnarray}
\label{eq:172}
\Sigma_L^{W,e}\left(x\right)=
\nonumber\\
-4\sqrt{2}G_F\sum_K P_K^T\int\tilde{dq}_K\ q^\mu_K\left(f_{L,K}^e\left(x,\vec{q}\right)-\bar{f}_{R,K}^e\left(x,\vec{q}\right)\right)
\nonumber\\
=-4\sqrt{2}G_F\sum_K P_K^T J^\mu_{L,K}\left(x\right)\ \ \ 
\\
\label{eq:173}
\Sigma_R^{W,e}\left(x\right)=
\nonumber\\
4\sqrt{2}G_F\sum_K P_K\int\tilde{dq}_K\ q^\mu_K \left(f_{L,K}^e\left(x,\vec{q}\right)-\bar{f}_{R,K}^e\left(x,\vec{q}\right)\right)
\nonumber\\
=4\sqrt{2}G_F\sum_K P_K J^\mu_{L,K}\left(x\right)\ \ \ 
\end{eqnarray}

Here, $\tilde{dq}_K\equiv\frac{d^3\vec{q}}{\left(2\pi\right)^3 2E_{q,K}}$ and $q^\mu_K=\left(E_{q,K}, \vec{q}\right)$, with $E_{q,K}=\sqrt{\vec{q}^2+m_K^2}$.  $J^\mu_{L,K}$ is the current associated with left-handed charged leptons of flavor $K$.

The second diagram in Fig. 1 has a similar structure, and gives the following contribution to $\Sigma$:
\begin{eqnarray}
\label{eq:174}
\Sigma_L^\nu\left(x\right)&=&-2\sqrt{2}G_F\left(J^\mu_{\left(\nu\right)}\left(x\right)\right)^T
\nonumber\\
\Sigma_R^\nu\left(x\right)&=&2\sqrt{2} G_F J^\mu_{\left(\nu\right)}\left(x\right)
\end{eqnarray}

where $J_{\left(\nu\right)}^\mu\left(x\right)$ is the neutrino current, given by
\begin{eqnarray}
\label{eq:175}
J_{\left(\nu\right)}^\mu\left(x\right)=\int\tilde{dq}\ q^\mu\left(f\left(x,\vec{q}\right)-\bar{f}\left(x,\vec{q}\right)\right)
\end{eqnarray}

For neutrinos, we also obtain contributions to $\Sigma_{\alpha}^{\ \beta}$ and $\Sigma_{\ \dot{\beta}}^{\dot{\alpha}}$ by including the arrow-clashing propagator in the loop.  These components of $\Sigma$ can in general have a tensor component and a scalar component.  However, the tensor component is proportional to $\bar{\sigma}^\mu S_L^{\rho\sigma} \sigma_\mu$ or $\sigma^\mu S_R^{\rho\sigma} \bar{\sigma}_\mu$, which vanishes in four spacetime dimensions, so there is no tensor contribution to $\Sigma$.  The scalar component, on the other hand, is proportional to the scalar component of the neutrino two-point function, which is an $O\left(\epsilon\right)$ quantity.  Consequently, the scalar component of $\Sigma$ is $O\left(\epsilon^2\right)$.  Since this appears in the kinetic equations as a correction to the mass, and the mass always enters as a part of an $O\left(\epsilon^2\right)$ term, the shift in the mass due to the scalar component of $\Sigma$ produces an $O\left(\epsilon^3\right)$ term, which can be neglected.

Note that the neutrino current contains an $O\left(\epsilon\right)$ correction due to a shift in the dispersion relation.  Another correction comes from the $O\left(\epsilon\right)$ contribution to $F$ from the small components $\Delta_{L/R}$.  These corrections result in an $O\left(\epsilon^2\right)$ shift in $\Sigma_{L/R}$, which is denoted in the quantum kinetic equations as $\delta\Sigma_{L/R}$.  Thus, we define $\Sigma$ as the quantity that is calculated by using the massless, free-field, $O\left(1\right)$ expression for the current, while $\delta\Sigma$ contains the $O\left(\epsilon\right)$ corrections from the masses and interactions.

Similarly, we calculate the two lower diagrams in Fig. 1 to obtain the following contributions to $\Sigma_R^\mu$:
\begin{eqnarray}
\label{eq:176}
4\sqrt{2}G_F{\rm \bf 1}
\times\nonumber\\\sum_K\left(\left(\sin^2\theta_W-\frac{1}{2}\right)J^\mu_{L,K}+\sin^2\theta_W J^\mu_{R,K}\right)
\end{eqnarray}
and
\begin{eqnarray}
\label{eq:177}
2\sqrt{2}G_F\left({\rm tr}\ J^\mu_{\left(\nu\right)}\right){\rm \bf 1}\ \sigma_{\mu\alpha\dot{\beta}}
\end{eqnarray}
and similarly for the $\bar{\sigma}$ component of $\Sigma$.  Here, ${\rm \bf 1}$ is the flavor unit matrix, and the trace is over flavor.  The complete expression for the matter potential $\Sigma$ to $O\left(\epsilon\right)$ is, therefore,
\begin{eqnarray}
\label{eq:178}
\Sigma_R^\mu = \Sigma_R^{\left(e\right)\mu} + \Sigma_R^{\left(\nu\right)\mu}=
4\sqrt{2} G_F\sum_K\times
\nonumber\\
\left(\left(P_K+{\rm\bf 1}\left(\sin^2\theta_W-\frac{1}{2}\right)\right)J_{L,K}^\mu+{\rm \bf 1}\sin^2\theta_W J_{R,K}\right)
\nonumber\\
+2\sqrt{2} G_F\left(J_{\left(\nu\right)}^\mu+{\rm \bf 1}\left({\rm tr}\ J_{\left(\nu\right)}^\mu\right)\right)\ \ 
\end{eqnarray}

\subsection{Collision Terms}
In this section, we consider the quantities  $\Pi^\pm$ that appear on the right-hand side of the quantum kinetic equations.  We will see that these terms have the gain-loss structure of a Boltzmann collision term.  We will refer to them as the gain-loss potentials.

$\Pi^\pm$ are linear combinations of $\Pi_\rho$ and $\Pi_F$ given by equation (\ref{eq:50}).  In position space, $\Pi_\rho$ and $\Pi_F$ are nonlocal components of the self-energy.  In our model, all nonlocal contributions correspond to two-loop (or higher-order) diagrams involving the exchange of at least two $W$ or $Z$ bosons.  To two-loop order, the diagrams that contribute to $\Pi_{\rho,F}$ are shown in Figure 4.
\begin{figure}
\includegraphics[width=3.5in]{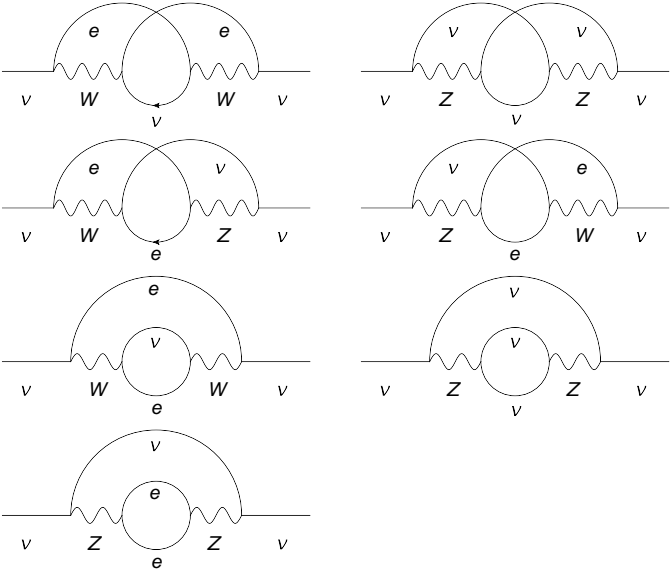}
\caption{Feynman graphs showing two-loop contributions to neutrino self-energy. }
\label{feynman2}
\end{figure}

These diagrams give a large number of terms corresponding to various scattering processes, which must all be included in a complete treatment of inelastic scattering of neutrinos off charged leptons and other neutrinos.  Since we do not present numerical computations of neutrino scattering in this paper, we will not calculate every term in detail.  We will show that the $\Pi^{\pm}$ produce Boltzmann-like gain-loss terms 
and for the purpose of illustration we will  compute only one of the terms in detail. 
A calculation of the  full  collision term will be presented in upcoming work. 

\subsubsection{Example: $\nu \nu$ scattering  neglecting spin coherence}

As an illustration, we consider inelastic processes involving only neutrinos and anti-neutrinos, ignoring the presence of electrons and other particles in the 
thermal bath.  This means we consider only the contribution from the upper-right  and lower-right diagrams in Fig. 4, which involve only neutrino lines.  
First, consider the upper-right diagram:
placing arrows on the fermion lines produces 16 arrangements that contribute to this diagram.  There are four possible combinations of external arrow directions, which pick out the particular component of $\Pi^\pm$ that is being calculated.  For each combination of external arrows, there are four possible combinations of internal arrow directions, which determine the components of $G$ that the given contribution to $\Pi^\pm$ depends on.  For example, the contributions to $\Pi^{\dot{\alpha}\alpha}$ from this diagram are given in Figure 5; there are similar contributions to $\Pi_{\alpha\dot{\alpha}}$, $\Pi_\alpha^{\ \beta}$ and $\Pi^{\dot{\alpha}}_{\dot{\beta}}$, which correspond to different directions for the external arrows.
\begin{figure}
\includegraphics[width=3.5in]{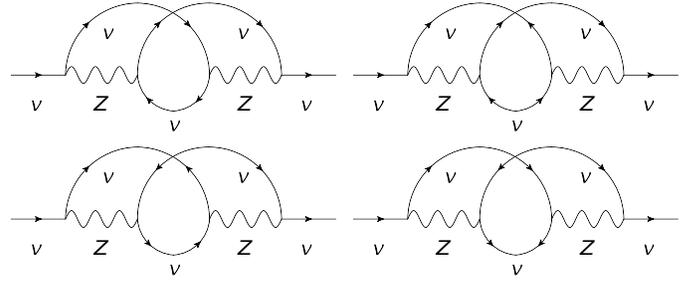}
\caption{Contributions to $\Pi^{\dot{\alpha}\alpha}$ corresponding to the upper-right diagram in Fig. 4}
\label{feynman2-1}
\end{figure}

All diagrams in Fig. 5 except the upper-left include two factors of arrow-clashing two-point functions for neutrinos.  The arrow-clashing two-point functions contain a scalar and a tensor component; the scalar component is $O\left(\epsilon\right)$, while the tensor component can in general be $O\left(1\right)$ if there is spin coherence.  The $O\left(\epsilon\right)$ terms can be dropped, since the two-loop diagrams are already $O\left(\epsilon^2\right)$.  Then, if spin coherence is present, so that $\phi=O\left(1\right)$, all four diagrams contribute.  However, in the absence of spin coherence, only the first diagram is $O\left(\epsilon^2\right)$; the remaining three are $O\left(\epsilon^4\right)$ and can be dropped.  Moreover, any contribution to $\Pi_{\alpha}^{\ \beta}$ or $\Pi^{\dot{\alpha}}_{\ \dot{\beta}}$ must contain at least one arrow-clashing internal line, and therefore these quantities are at least $O\left(\epsilon^3\right)$ and can be dropped in the absence of spin coherence.

For the sake of brevity, here we consider only terms that do not depend on spin coherence.  The procedure for calculating the other terms will be similar.

In position space, in terms of two-point functions, the upper-left diagram in Fig. 5 gives  
\begin{eqnarray}
\label{eq:179}
\Pi^{\dot{\alpha}\alpha}  \left(x,y\right)=
- 2 \, 
\delta^4\left(x-w\right)\delta^4\left(z-y\right)G_F^2\times\nonumber\\
\bar{\sigma}^{\mu\dot{\alpha}\beta} G^{\left(\nu\right)}_{\beta\dot{\beta}}\left(x,z\right)\bar{\sigma}^{\nu\dot{\beta}\gamma}G^{\left(\nu\right)}_{\gamma\dot{\gamma}}\left(z,w\right)\bar{\sigma}_\mu^{\dot{\gamma}\delta}G^{\left(\nu\right)}_{\delta\dot{\delta}}\left(w,y\right)\bar{\sigma}_\nu^{\dot{\delta}\alpha}
\end{eqnarray}
To proceed further, first, we calculate the appropriate combinations of spectral and statistical components, $\Pi^+$ and $\Pi^-$, defined by Equation (\ref{eq:50}).  
When performing this calculation we do not need to keep track of the details of the spin and flavor structure of two-point function products, since the decomposition into spectral and statistical components is the same regardless of these details.  As a result, we can write, symbolically,
\begin{eqnarray}
\label{eq:180}
\Pi\left(x,y\right) \sim G_1\left(x,y\right) G_2\left(y,x\right) G_3\left(x,y\right)
\end{eqnarray}
This notation simply indicates that $\Pi$ is composed of three distinct two-point functions, which are then contracted in some way and multiplied by the appropriate couplings  and electroweak boson propagators.  Note that the delta functions in equation  (\ref{eq:179}) 
allow us to write all two-point functions as functions of only $x$ and $y$.

We can write $G\left(x,y\right)=\theta\left(x^0-y^0\right)G^+\left(x,y\right)-\theta\left(y^0-x^0\right)G^-\left(x,y\right)$, and similarly $\Pi\left(x,y\right)=\theta\left(x^0-y^0\right)\Pi^+\left(x,y\right)-\theta\left(y^0-x^0\right)\Pi^-\left(x,y\right)$.  Then, setting $x^0 > y^0$, we obtain
\begin{eqnarray}
\label{eq:181}
\Pi^+\left(x,y\right)\sim    -  G_1^+\left(x,y\right) G_2^-\left(y,x\right) G_3^+\left(x,y\right)
\end{eqnarray}
Similarly, for $x^0 < y^0$, we obtain
\begin{eqnarray}
\label{eq:182}
\Pi^-\left(x,y\right)\sim
-  G_1^-\left(x,y\right) G_2^+\left(y,x\right) G_3^-\left(x,y\right)
\end{eqnarray}
Next, we Wigner transform equation (\ref{eq:179}), and use equations (\ref{eq:181})-(\ref{eq:182}) to obtain $\Pi^\pm$.  This gives the following expression:
\begin{eqnarray}
\label{eq:183}
\Pi^\pm\left(k\right)= \int \prod_{i=1}^3\frac{d^4q_i}{\left(2\pi\right)^4}\left(2\pi\right)^4\delta^4\left(k-q_1-q_2-q_3\right)
\nonumber\\
2G_F^2\bar{\sigma}^\mu G^\pm\left(q_1\right)\bar{\sigma}^\nu G^\mp\left(-q_2\right)\bar{\sigma}_\mu G^\pm\left(q_3\right)\bar{\sigma}_\nu\ \ 
\end{eqnarray}
The dependence of $\Pi^\pm$ and the two-point functions on the position $x$ is implied. 
We can change $-q_2\rightarrow q_2$ to obtain
\begin{eqnarray}
\label{eq:184}
\Pi^\pm\left(k\right)= \int \prod_{i=1}^3\frac{d^4q_i}{\left(2\pi\right)^4}\left(2\pi\right)^4\delta^4\left(k+q_2-q_1-q_3\right)
\nonumber\\
2G_F^2\bar{\sigma}^\mu G^\pm\left(q_1\right)\bar{\sigma}^\nu G^\mp\left(q_2\right)\bar{\sigma}_\mu G^\pm\left(q_3\right)\bar{\sigma}_\nu\ \ 
\end{eqnarray}
Every two-point function $G^\pm$ contains a positive- and a negative-energy piece, and is proportional to an on-shell delta function, which to leading order is $2\pi\delta\left(q_i^2\right)$.  This, together with the overall momentum-conserving delta function, implies that the only terms giving a nonzero contribution to the integral are those where all four of $\left(k^0,q^0_i\right)$ are positive (corresponding to the neutrino-neutrino scattering process), those where all four are negative (corresponding to antineutrino-antineutrino scattering), and those where two are positive and two are negative (describing neutrino-antineutrino scattering).

We consider the term in which all energies are positive, which describes neutrino-neutrino scattering.  Using $G^\pm=-\frac{1}{2}i\rho\pm F$, using the $O\left(1\right)$ expressions for $F$ and $\rho$ given by equations (\ref{eq:39})-(\ref{eq:40}), (\ref{eq:44})-(\ref{eq:45}) and (\ref{eq:46})-(\ref{eq:47}), and omitting spin coherence, we obtain
\begin{eqnarray}
\label{eq:185}
\Pi^{+,\dot{\alpha}\alpha} \left(k\right)=
\int\prod_{i=1}^3\tilde{dq}_i\left(2\pi\right)^4\delta^4\left(k+q_2-q_1-q_3\right)
\nonumber\\
2 G_F^2\left(\bar{\sigma}^\mu\sigma_\rho\bar{\sigma}^\nu\sigma_\sigma\bar{\sigma}_\mu\sigma_\tau\bar{\sigma}_\nu\right)^{\dot{\alpha}\alpha}q_1^\rho q_2^\sigma q_3^\tau
\times\nonumber\\
\left(1 - f\left(\vec{q}_1\right) \right)  f\left(\vec{q}_2\right)  \left(1 -   f\left(\vec{q}_3\right) \right)
\nonumber\\
=-16 G_F^2\int\prod_{i=1}^3\tilde{dq}_i\left(2\pi\right)^4\delta^4\left(k+q_2-q_1-q_3\right)
\nonumber\\
\left(q_2\cdot\bar{\sigma}^{\dot{\alpha}\alpha}\right)\left(q_1\cdot q_3\right)
\left(1 - f\left(\vec{q}_1\right) \right)  f\left(\vec{q}_2\right)  \left(1 -   f\left(\vec{q}_3\right) \right)
\end{eqnarray}
Similarly, the contribution to $\Pi^-$ is:
\begin{eqnarray}
\label{eq:186}
\Pi^{-,\dot{\alpha}\alpha} \left(k\right)=
\nonumber\\
-16  G_F^2\int\prod_{i=1}^3\tilde{dq}_i\left(2\pi\right)^4\delta^4\left(k+q_2-q_1-q_3\right)
\nonumber\\
\left(q_2\cdot\bar{\sigma}^{\dot{\alpha}\alpha}\right)\left(q_1\cdot q_3\right)
f\left(\vec{q}_1\right) \left(1-f\left(\vec{q}_2\right)\right)f\left(\vec{q}_3\right)
\end{eqnarray}
\noindent
Since we have chosen the term for which $k^0$ is positive, this expression enters into the collision term for neutrinos.  The corresponding contribution to the collision term in Equation 
(\ref{eq:156})  is
\begin{eqnarray}
\label{eq:187}
8G_F^2
\frac{1}{\big| \vec{k} \big|}
\int \prod_{i=1}^3\tilde{dq}_i\left(2\pi\right)^4\delta^4\left(k+q_2-q_1-q_3\right)
\times\nonumber\\
\left(k\cdot q_2\right)\left(q_1\cdot q_3\right)
\times\nonumber\\
\left(\left\{1-f, f_1\left(1-f_2\right)f_3\right\}-\left\{f, \left(1-f_1\right)f_2\left(1-f_3\right)\right\}\right)
\end{eqnarray}
where $f = f (\vec{k})$ and $f_i = f\left(\vec{q}_i\right)$.

To obtain the complete piece of the collision term that describes neutrino-neutrino scattering, we also need to include the lower-right diagram in Fig. 4.  We also introduce $s\equiv\left(k+q_2\right)^2=\left(q_1+q_3\right)^2$.  For the approximately massless neutrinos, $s \approx 2k\cdot q_2 = 2q_1\cdot q_3$.  The collision term for neutrino-neutrino scattering is then given by
\begin{eqnarray}
\label{eq:188}
C_{\nu\nu\leftrightarrow\nu\nu}=
\frac{2 G_F^2}{\big| \vec{k} \big|}
\int \prod_i \tilde{dq}_i\delta^4\left(k+q_2-q_1-q_3\right)s^2
\nonumber\\
\left\{1-f, f_1\left[
{\rm tr}_F\left(\left(1-f_2\right)f_3\right)
+
\left(1-f_2\right)f_3\right]
\right\}
\nonumber\\
-\left\{f,\left(1-f_1\right)\left[ {\rm tr}_F\left(f_2\left(1-f_3\right)  \right)
+
f_2\left(1-f_3\right)\right]  \right\}
\end{eqnarray}

This  contribution to the collision term clearly has the gain-loss structure of the Boltzmann equation with Fermi blocking, describing $\nu\nu\leftrightarrow\nu\nu$ scattering.  However, unlike in the Boltzmann equation, the densities $f$ are flavor matrices, and the collision term has nontrivial flavor structure.

We can make the connection to the usual Boltzmann term by considering a case in which there is no coherence between neutrino flavors, so that the density matrices $f$ are all diagonal in the same basis.  Then, the anticommutators become products of the diagonal terms, which are just the neutrino densities, and the collision term for flavor $I$ reduces to the Boltzmann form:
\begin{eqnarray}
\label{eq:189}
C_{\nu\nu\leftrightarrow\nu\nu}^I=
\frac{4 G_F^2}{\big| \vec{k} \big|} 
\int \prod_i \tilde{dq}_i\delta^4\left(k+q_2-q_1-q_3\right)
s^2\times
\nonumber\\
 \Bigg\{\left(1-f^I\right)f_1^I    
 \Bigg[
  2  \left(1-f_2^I\right)f_3^I 
+  \sum_{J \not= I}\left(1-f_2^J\right)f_3^J
\Bigg]
\nonumber\\-
 f^I\left(1-f_1^I\right)  
 \Bigg[
2   f_2^I\left(1-f_3^I\right)
+ \sum_{J \not= I} f_2^J\left(1-f_3^J\right) 
\Bigg]\Bigg\}\ \ 
\end{eqnarray}
This corresponds to the usual Boltzmann term describing scattering of neutrinos off each other, with one incoming and outgoing neutrino described by $f\leftrightarrow f_1$ and the other by $f_2\leftrightarrow f_3$.  In the above expression, repeated indices are not summed over unless the sum is explicitly indicated.  
From the above formula we see that the total scattering rate for $\nu_I \nu_I$ is twice that for $\nu_I \nu_J$ with $J \neq I$,  
consistent with the discussion in Ref.~\cite{Flowers:1976fj} . 

\subsubsection{Generalizations}

So far, we have only considered diagrams for neutrino-neutrino scattering, and assumed that the spin coherence is zero.  When all processes are included, we obtain collision terms that have the following structure:
\begin{eqnarray}
\label{eq:190}
C=C_{\nu\nu\leftrightarrow\nu\nu}+C_{\nu\bar{\nu}\leftrightarrow\nu\bar{\nu}}+
\nonumber\\
+C_{\nu e\leftrightarrow \nu e}+C_{\nu \bar{e}\leftrightarrow \nu \bar{e}}+C_{\nu \bar{\nu}\leftrightarrow e\bar{e}}
+C'\left[f, \bar{f}, \phi\right]
\end{eqnarray}
where $C'$ is a set of additional terms dependent on spin coherence, which are zero when $\phi=0$.
These can be calculated in the same way as the rest of the collision terms, but with different arrangements of two-component spinor arrows within the Feynman diagrams. 
The other collision terms, $\bar{C}$ and $C_\phi$,  have a similar structure.

\section{properties of the quantum kinetic equations}
We now examine the quantum kinetic equations, equations (\ref{eq:149})-(\ref{eq:151}) (summarized in equation \ref{eq:165-1}), and consider some of their properties.  In the previous section, we have seen that the right-hand sides of equations (\ref{eq:149})-(\ref{eq:151}) correspond to the Boltzmann collision terms, with some additional flavor structure and dependence on coherence.  We now show that the quantum kinetic equations replicate the usual equations for coherent flavor evolution in the low-density limit.  We also discuss the spin coherence terms, and show that these terms can potentially lead to coherent transformation between neutrino and anti-neutrino states.

\subsection{Low-Density Limit}
The low-density limit is realized in certain situations in nature, for example, in the supernova envelope, or in the early Universe after weak decoupling.  In this limit, we neglect the collision term, since this is proportional to $G_F^2$, but retain the matter potential, which is proportional to $G_F$.  Furthermore, we assume that the matter potential $\Sigma$ is much smaller than the vacuum mass $m$, but comparable to $\frac{m^2}{E}$.  With these assumptions we can demote $\Sigma$ from $O\left(\epsilon\right)$ to $O\left(\epsilon^2\right)$, and drop higher-order terms involving $m\Sigma$, $\partial\Sigma$ and $\Sigma^2$.

In this regime, the quantum kinetic equations become
\begin{eqnarray}
\label{eq:192}
i\partial^{\kappa}f-\left[\Sigma^\kappa+\frac{m^\star m}{2\left|\vec{k}\right|},f\right]=0
\\
\label{eq:193}
i\partial^\kappa \bar{f}-\left[\Sigma^\kappa-\frac{m^\star m}{2\left|\vec{k}\right|},\bar{f}\right]=0
\end{eqnarray}

In the low-density limit, or in the isotropic limit, the spin coherence density $\phi$ is decoupled from the equations for $f$ and $\bar{f}$.  Therefore, in the low-density limit, there is no need to solve equation (\ref{eq:151}) for the spin coherence density.

Equations (\ref{eq:192}) and (\ref{eq:193}) are equivalent to the usual equations for coherent flavor evolution, for example those described in Ref.s~\cite{Fuller87,Notzold:1988fv,Sawyer:1990lr,Qian93,Samuel:1993sf,Qian95,Samuel:1996rm,Elze:2000vh,Pastor:2002zl,Pastor02,Balantekin05,Fuller06,Duan06a,Duan06b,Duan06c,Hannestad:2006qd,Duan07a,Duan07b,Balantekin:2007kx,Duan07c,Duan08,Kneller:2008rt,Lunardini08,Dasgupta:2008kx,Gava:2009yq,Dasgupta:2011uq,Friedland:2010yq,Duan:2011lr,Mirizzi:2012qy}.  The equations describe phenomena such as coherent oscillations, the Mikheyev-Smirnov-Wolfenstein (MSW) effect \cite{Wolfenstein78,Mikheyev85}, and collective flavor transformation due to the neutrino self-coupling terms present in $\Sigma$.  These phenomena are described in detail in Ref.~\cite{Duan:2010fr}.

\subsection{Spin Coherence}
A feature that appears at high densities and in the presence of anisotropy in the neutrino field is the coupling of the quantum kinetic equations for $f$ and $\bar{f}$ to a new dynamical quantity, the spin coherence density $\phi$.  We now examine the possible consequences of this coupling.

It is clear from the form of equation (\ref{eq:165-1}) that $\phi$ represents coherence between neutrinos and anti-neutrinos, and the $H_{\nu\bar{\nu}}$ term gives mixing between neutrino and anti-neutrino states.  The effects of spin coherence conserve the total number of neutrinos plus antineutrinos for each momentum, but not the two separately.  This can be seen by taking the trace of equation (\ref{eq:165-1}) to obtain

\begin{eqnarray}
\label{eq:194}
{\rm tr} D\left[{\cal F}\right]= {\rm tr}\ {\cal C}\left[{\cal F}\right]
\end{eqnarray}

Since ${\rm tr}{\cal F}\left(\vec{k}\right) = {\rm tr}f\left(\vec{k}\right) + {\rm tr}\bar{f}\left(\vec{k}\right)$, ${\rm tr}{\cal F}\left(\vec{k}\right)$ corresponds to the total density of neutrinos plus anti-neutrinos of momentum $\vec{k}$.  The derivative combination ${\rm tr} D\left[{\cal F}\right]$ can be interpreted as simply a derivative of the neutrino plus antineutrino density along a light-like world line, which deviates slightly from the world line of an actual neutrino due to an index of refraction from the matter and neutrino potentials.  

As a consequence, along the particle world line the total neutrino plus antineutrino density for a given momentum can change in response to the collision term, but not in response to spin coherence.  However, in the presence of spin coherence, the quantities ${\rm tr}f$ and ${\rm tr}\bar{f}$ are not individually conserved, so the difference between neutrino and antineutrino densities can undergo coherent evolution.

Therefore, the coupling to the spin coherence can lead to a coherent process that converts neutrinos to antineutrinos, and vice versa.  The mixing term $H_{\nu\bar{\nu}}$ involves a combination of the neutrino mass $m$ and spacelike components of the matter and neutrino potential orthogonal to the momentum, $\Sigma^\pm=\frac{1}{2}\left(\Sigma^1\pm i\Sigma^2\right)$.  We see from this that three conditions are necessary for a coherent change of helicity:  (1) the particles must have a mass; (2) there must be an anisotropic matter or neutrino potential with a component orthogonal to the particle's momentum; and, (3) the spin coherence density $\phi$ must be present.

The anisotropy condition can be satisfied in the context of a supernova explosion or a compact object merger.  One source of anisotropy, which is present even in spherically symmetric models, is the outgoing flux of neutrinos.  A neutrino moving at a nonzero angle with respect to the radial direction will receive a contribution to $H_{\nu\bar{\nu}}$ from interactions with other outgoing neutrinos.

The mixing Hamiltonian, $H_{\nu\bar{\nu}}$, is $O\left(\epsilon^2\right)$ while the diagonal blocks, $H$ and $-\bar{H}^T$,  are $O\left(\epsilon\right)$.  Thus, under generic conditions, we expect the effects of mixing between neutrinos and anti-neutrinos to be small.  However, we can potentially obtain large effects  ``at resonance", when there is a degeneracy between eigenvalues of $H$ and $-\bar{H}$.  This is analogous to the MSW resonance effect, where a small neutrino mass can lead to large-scale flavor transformation at resonance.  Note that, unlike in the decoupled equations of motion for $f$ and $\bar{f}$, equations (\ref{eq:192})-(\ref{eq:193}), the flavor-independent components of $H$ and $\bar{H}$ that are proportional to the flavor unit matrix must be included. Therefore, to determine the conditions for neutrino-antineutrino resonance in a realistic model it is necessary to include the neutral current contributions to the matter potential, including contributions from coherent forward scattering of neutrinos on nuclei and nucleons.

The Hamiltonian ${\cal H}$ and the combined neutrino-antineutrino density matrix ${\cal F}$ in equation (\ref{eq:165-1}) bear some resemblance to the description of coherent evolution of neutrinos with a nonzero transition magnetic moment in the presence of a magnetic field \cite{Dvornikov:2011dv,de-Gouvea:2012fk,de-Gouvea:2013lr}.  However, a Standard Model neutrino magnetic moment arises from loop corrections, and is therefore quite small, requiring very large magnetic fields to obtain neutrino-antineutrino mixing.  Our effect comes from the weak interaction, which has a handle on neutrino helicity without the need to consider higher-order loop corrections, and does not require a large external magnetic field.

Whether large-scale neutrino-antineutrino transformation will actually take place in a supernova explosion is a difficult question, due to the neutrino-neutrino interaction terms in the Hamiltonian and the possibility for nonlinear feedback.  Resolving this question is likely to require sufficiently realistic numerical simulations.  The results from Ref.s~\cite{de-Gouvea:2012fk,de-Gouvea:2013lr} suggest that the presence of even a small neutrino-antineutrino mixing term in the Hamiltonian could potentially lead to large-scale neutrino-antineutrino transformation.

\section{Comparison with previous work} 
\label{sect:comparisons}

Our approach to neutrino quantum kinetics  heavily relies on previous  studies of transport
equations from quantum field theory (CTP and 2PI techniques)  for both
scalars and fermion fields (see~\cite{Calzetta:1988qy,Blaizot:2002uq,Berges:2004qy} and references therein),  
and  their non-trivial generalization to multi-flavor cases in the context of electroweak
baryogenesis~\cite{Prokopec:2004lr,Prokopec:2004fk,Konstandin:2005qy,Cirigliano:2010lr,Cirigliano:2011lr,Fidler:2012lr,Herranen:2010kx}
and leptogenesis~\cite{Garbrecht:2012fj,Beneke:2010dz,Beneke:2010wd,Garny:2009qn,Garny:2009rv}. 

Compared to previous field-theoretical analyses, our work contains the following new elements:
(i) we clearly spell out a power counting in ratio of scales that is
specific to neutrinos (ultra-relativistic  weakly  interacting particles
in an environment that is nearly homogenous on the scale of a de Broglie
wavelength) and expand the kinematics around light-like four-momenta.
(ii) We make no assumptions of isotropy and treat spin degrees of freedom
in full generality, which leads us to discover spin-coherence correlations
that have been neglected in the past.  

We are not aware of any other work that derives quantum kinetic equations for neutrinos in a fully
anisotropic environment, or provides a description of the evolution of
neutrino spin degrees of freedom.  Since the neutrino fields in the astrophysical environments (supernovae, compact object mergers) of interest for application of the QKEs are {\it inherently} anisotropic, the features of our QKEs that arise from a non-isotropic neutrino field are potentially very important.  Anisotropy, spin coherence, and the
interplay between spin and flavor degrees of freedom may play an important
role in these environments.

Neutrino QKEs have been derived in the past using different
first-principles approaches and approximation schemes.   Our approach is
very closely related to the one of Raffelt and Sigl~\cite{Sigl:1993fr}. 
 In fact, the ``matrix of densities"  introduced in  \cite{Sigl:1993fr} can be related to certain Lorentz
components of the Wigner transformed neutrino two-point function used in
our work.  Moreover, as in  \cite{Sigl:1993fr}  we do rely on perturbation theory and there
is a one-to-one correspondence between the assumptions made in these two
works.  The end-results of our analysis match the one  of Ref.~\cite{Sigl:1993fr}  up to
the inclusion of spin-coherence densities (which is new in our work).

More recently,  a new approach to neutrino quantum kinetics has been
proposed Ref.~\cite{Volpe:2013lr}, based on many-body techniques and the BBGKY hierarchy.
Again, there is a correspondence between Ref.~\cite{Volpe:2013lr}  and the field-theoretic
treatment. In general, in field theory the non-equilibrium system is
described by the set of all $n$-point Green's functions. These obey  coupled
integro-differential equations, equivalent to the BBGKY equations~\cite{Calzetta:1988qy}.  
We truncate this hierarchy by writing down dynamical equations only for the
two point functions and expressing all higher order Green's functions as a
perturbative series in terms of the two-point functions.   Here we assume
that higher order correlations are absent in the initial state and  we make
essential use of our power counting in terms of weak interactions:  the
methods used here do not generalize to strongly interacting / correlated
systems.
Furthermore, when considering the dynamics of two-point functions,  we
neglect particle-antiparticle  pairing  correlations  (see discussion
following  Eq.~(\ref{eq:36})).  This is consistent with our power counting
assumption that physical quantities vary slowly on the scale of the
neutrino de Broglie wavelength.  Nonetheless, these correlations that pair
particles and antiparticles of opposite momenta 
(first discussed in the context of neutrino kinetics in Ref.~\cite{Volpe:2013lr}) 
could  be included in our formalism.  In fact, evolution equations that couple these
particle-antiparticle  densities to the standard particle-particle and
antiparticle-antiparticle densities can be derived  in the field theory
framework~\cite{Fidler:2012lr,Herranen:2010kx}. 
%
In the context of time-dependent multi-flavor mass matrices in the Early
Universe (at the electroweak phase transition), it was shown in  \cite{Fidler:2012lr}  that
particle-antiparticle correlations can dynamically arise from a vanishing
(equilibrium) initial condition and  can play an important role in
baryogenesis. We are not aware of any numerical exploration  of the role of
these correlations in a non-homogeneous supernova environment.

Finally, let us discuss the structure of  our collision terms
(Eqs.~\ref{eq:156}, \ref{eq:157}, \ref{eq:157-1}, \ref{eq:165-5}), in comparison to other work. 
Even though  here  we do not calculate explicitly all the vector and tensor componenst of the self-energies  $\Pi_{L,R}^\pm$, 
it is clear that our collision term is non-diagonal both in flavor 
and spin, thus producing decoherence of any linear superposition of flavor or spin states. 
Neglecting spin coherence, the structure of our result 
matches the ``non-abelian" matrix structure in flavor space discussed in Ref.~\cite{Sigl:1993fr}. 
We note, however,  that many {\it ad hoc} treatments of the QKEs, 
including recent ones~\cite{Zhang:2013lka}, 
completely miss  the off-diagonal  entries of the collision term,  
which  are  required  by quantum mechanical considerations.

\section{Conclusion}
\label{sect:conclusion}
We have produced a self-consistent derivation of the quantum kinetic equations (QKEs) that govern how neutrino flavor evolves in medium.  This derivation started from first principles relying only on quantum field theory and assumed standard model interactions for neutrinos.  To our knowledge, this is the first such self-consistent first-principles derivation of QKEs for flavored fermions in an anisotropic environment. 
Our result,  Eq.~(\ref{eq:165-1}), captures the correct structure of the QKEs in anisotropic environments, but is somewhat formal 
because the self-energies on the right-hand-side are not fully calculated.      
In a future paper we will present  a detailed analysis of the inelastic collision term, including spin coherence, 
thus making our results   amenable to implementation in  numerical simulations. 

Specializing to ultra relativistic Majorana neutrinos and making expansions in small parameters, equation (\ref{eq:48}), our QKEs assume the usual form which describes coherent neutrino flavor evolution in low density media.  Likewise, at high density, where neutrino scattering is dominant, the collision terms in our QKEs assume Boltzmann-like forms.  
This is consistent with studies that have shown that the Boltzmann equation could be derived directly from quantum field theory \cite{Berges:2005lr}.

In the low density, coherent regime our QKEs are broadly similar to those derived from previous treatments, for example those of Ref.s~ \cite{Sigl:1993fr,Raffelt:1993kx}.  In the scattering-dominated Boltzmann limit and between these two limiting cases, however, there are differences. Unlike previous studies, we follow in detail neutrino spin degrees of freedom, and in this sector there are surprises.

We have found a new dynamical quantity associated with spin coherence.  At low density we find that the equation of motion for this quantity decouples from the rest of the QKEs describing neutrino flavor evolution.  This equation describes Majorana neutrino spin (helicity) evolution in a matter and neutrino background.  An obvious feature we find is that spin coherence can only arise in conditions where neutrino fluxes and/or matter potentials are not isotropic.  Such conditions never arise in a standard Friedman-LaMaitre-Robertson-Walker early Universe expansion, but might occur in out of equilibrium environments like those associated with phase transition-induced nucleation of topological defects like bubbles or domain walls \cite{Kolb:1992lr,Shi:1999fk}.  By contrast, the region above the proto-neutron star in core collapse supernovae and the neutron star merger environment are both characterized by gross anisotropy in matter and neutrino fields.

The terms driving coherent spin flip in our QKEs stem from products of neutrino absolute mass and spacelike projections of the matter potentials (hence the requirement for anisotropy).  Unlike coherent flavor transformation, which is sensitive only to the mass-squared differences between different neutrino flavors, coherent spin flip is sensitive to the neutrino absolute mass.

Also, unlike coherent flavor transformation, coherent spin flip is sensitive to the Majorana or Dirac nature of neutrinos.  In this paper, we have specialized to Majorana neutrinos, but extending our treatment to Dirac neutrinos is straightforward.  The simplest way to introduce Dirac neutrinos in our model is to add an additional field describing sterile neutrinos, $\nu_s$.  For pure Dirac neutrinos, the mass term always connects the active neutrino field, $\nu$, with the sterile field, $\nu_s$.  Because the spin flip term carries a single power of the mass, for Dirac neutrinos it will result in transformation between active and sterile states.   However, for Majorana neutrinos, coherent spin flip generates transformation between active neutrinos and active antineutrinos.

It is not known at present whether coherent spin flip can result in large-scale transformation between right-handed and left-handed neutrino states in supernovae.  Due to nonlinearity and complexity of the QKEs, the resolution of this question likely requires sufficiently detailed and realistic numerical modeling.  If numerical simulations do show that effects from coherent spin flip are large enough to produce a detectable signature in the supernova neutrino spectrum, then measurement of a supernova neutrino signal could in principle be used to constrain the absolute neutrino mass and determine the Majorana vs. Dirac nature of neutrinos.

Additionally, both neutrino production ({\it e.g.,} Ref.s~\cite{Fryer:2009qy}) and neutrino energy deposition in the core collapse supernova shock re-heating (accretion) phase and the neutron-to-proton ratio ({\it e.g.,} Ref.~\cite{Qian93})  in any neutrino-heated outflow nucleosynthesis can be very sensitive to the relative fluxes and energy spectra of $\nu_e$ and $\bar\nu_e$. Consequently, for these processes, any large-scale inter-conversion of neutrinos and antineutrinos could be significant.

Simulations of the core collapse supernova and neutron star merger environments are some of the most sophisticated numerical calculations being done at present with, in some cases, state-of-the-art multi-dimensional radiation hydrodynamics coupled with detailed equation of state and other microphysics,  {\it e.g.,} Ref.s~\cite{Herant:1994uq,Mezzacappa:1998fr,Mezzacappa:2001fk,Mezzacappa:2005lr,Bruenn:2010qy,Keil:2003qy,Kotake:2006fk,Murphy:2008zr,Ott:2009fj,Muller:2010kx,Brandt:2011mz,Muller:2012ly,Cardall:2012rt,Hanke:2012ys,Ellinger:2013gf,Murphy:2013yq,Fryer:2013kx,Tamborra:2013fj,Bruenn:2013vn,Suwa:2013lr}. A key conclusion that can be drawn from these studies is that neutrinos and their interactions are important in many aspects of compact object evolution and nucleosynthesis. However, experiment has now caught up with theory in a sense. It is an experimental fact that neutrinos have nonzero rest masses and that neutrino flavors mix in vacuum.  
This physics is, for the most part, not in these otherwise very sophisticated simulations. 
The work presented here, a self-consistent approach to treating this physics, suggests that there are unresolved issues in the neutrino-supernova story.\\
\\

\begin{acknowledgments}
This work was supported in part by
NSF grant PHY-09-70064 at UCSD and by the DOE Office of Science and the LDRD Program
at LANL, and by the University of California Office of the President and the UC HIPACC
collaboration.
We would also like to acknowledge support from the DOE/LANL Topical Collaboration.
We thank J.~Carlson, J.~F.~Cherry, A.~Friedland, K.~Intriligator, B.~Keister, C.~Lee, A.~Manohar, M.~J.~Ramsey-Musolf, S.~Reddy, M.~Roberts, and S.~Tulin for useful discussions.

\end{acknowledgments}

\bibliographystyle{h-physrev}
\bibliography{allref}

\begin{thebibliography}{100}

\bibitem{Schwinger:1961uq}
J.~{Schwinger},
\newblock Journal of Mathematical Physics {\bf 2}, 407 (1961).

\bibitem{Harris:1981qy}
R.~A. {Harris} and L.~{Stodolsky},
\newblock \jcp {\bf 74}, 2145 (1981).

\bibitem{Harris:1982fk}
R.~A. {Harris} and L.~{Stodolsky},
\newblock Physics Letters B {\bf 116}, 464 (1982).

\bibitem{Stodolsky:1987lr}
L.~{Stodolsky},
\newblock \prd {\bf 36}, 2273 (1987).

\bibitem{Manohar:1987ys}
A.~{Manohar},
\newblock Physics Letters B {\bf 186}, 370 (1987).

\bibitem{Habib:1996yq}
S.~{Habib}, Y.~{Kluger}, E.~{Mottola}, and J.~P. {Paz},
\newblock Physical Review Letters {\bf 76}, 4660 (1996), arXiv:hep-ph/9509413.

\bibitem{Cooper:1997rt}
F.~{Cooper}, S.~{Habib}, Y.~{Kluger}, and E.~{Mottola},
\newblock \prd {\bf 55}, 6471 (1997), arXiv:hep-ph/9610345.

\bibitem{Berges:2002vn}
J.~{Berges},
\newblock Nuclear Physics A {\bf 699}, 847 (2002), arXiv:hep-ph/0105311.

\bibitem{Berges:2003fj}
J.~{Berges}, S.~{Bors{\'a}nyi}, and J.~{Serreau},
\newblock Nuclear Physics B {\bf 660}, 51 (2003), arXiv:hep-ph/0212404.

\bibitem{Berges:2005lr}
J.~{Berges} and I.-O. {Stamatescu},
\newblock Physical Review Letters {\bf 95}, 202003 (2005),
  arXiv:hep-lat/0508030.

\bibitem{Muller:2006kx}
B.~{M{\"u}ller} and A.~{Sch{\"a}fer},
\newblock \prc {\bf 73}, 054905 (2006), arXiv:hep-ph/0512100.

\bibitem{Giraud:2010lr}
A.~{Giraud} and J.~{Serreau},
\newblock Physical Review Letters {\bf 104}, 230405 (2010), 0910.2570.

\bibitem{Raffelt:1993fj}
G.~{Raffelt} and G.~{Sigl},
\newblock Astroparticle Physics {\bf 1}, 165 (1993), arXiv:astro-ph/9209005.

\bibitem{Sigl:1993fr}
G.~{Sigl} and G.~{Raffelt},
\newblock Nuclear Physics B {\bf 406}, 423 (1993).

\bibitem{Raffelt:1993kx}
G.~{Raffelt}, G.~{Sigl}, and L.~{Stodolsky},
\newblock Physical Review Letters {\bf 70}, 2363 (1993), arXiv:hep-ph/9209276.

\bibitem{McKellar:1994uq}
B.~H.~J. {McKellar} and M.~J. {Thomson},
\newblock \prd {\bf 49}, 2710 (1994).

\bibitem{Sawyer:2005yg}
R.~F. {Sawyer},
\newblock \prd {\bf 72}, 045003 (2005), arXiv:hep-ph/0503013.

\bibitem{Strack:2005fk}
P.~{Strack} and A.~{Burrows},
\newblock \prd {\bf 71}, 093004 (2005), arXiv:hep-ph/0504035.

\bibitem{Cardall:2008lr}
C.~Y. {Cardall},
\newblock \prd {\bf 78}, 085017 (2008), 0712.1188.

\bibitem{Herranen:2008qy}
M.~{Herranen}, K.~{Kainulainen}, and P.~{Matti Rahkila},
\newblock Journal of High Energy Physics {\bf 9}, 32 (2008), 0807.1435.

\bibitem{Herranen:2009fk}
M.~{Herranen}, K.~{Kainulainen}, and P.~M. {Rahkila},
\newblock Nuclear Physics B {\bf 810}, 389 (2009), 0807.1415.

\bibitem{Gava:2009yq}
J.~{Gava}, J.~{Kneller}, C.~{Volpe}, and G.~C. {McLaughlin},
\newblock Physical Review Letters {\bf 103}, 071101 (2009), 0902.0317.

\bibitem{Volpe:2013lr}
C.~{Volpe}, D.~{V{\"a}{\"a}n{\"a}nen}, and C.~{Espinoza},
\newblock ArXiv e-prints  (2013), 1302.2374.

\bibitem{Enqvist:1991yq}
K.~{Enqvist}, K.~{Kainulainen}, and J.~{Maalampi},
\newblock Nuclear Physics B {\bf 349}, 754 (1991).

\bibitem{Barbieri:1991fj}
R.~{Barbieri} and A.~{Dolgov},
\newblock Nuclear Physics B {\bf 349}, 743 (1991).

\bibitem{Dodelson:1994rt}
S.~{Dodelson} and L.~M. {Widrow},
\newblock Physical Review Letters {\bf 72}, 17 (1994), arXiv:hep-ph/9303287.

\bibitem{Shi:1996vn}
X.~{Shi},
\newblock \prd {\bf 54}, 2753 (1996), arXiv:astro-ph/9602135.

\bibitem{Foot:1997rt}
R.~{Foot} and R.~R. {Volkas},
\newblock \prd {\bf 55}, 5147 (1997), arXiv:hep-ph/9610229.

\bibitem{Bell:1999fk}
N.~F. {Bell}, R.~R. {Volkas}, and Y.~Y.~Y. {Wong},
\newblock \prd {\bf 59}, 113001 (1999), arXiv:hep-ph/9809363.

\bibitem{Dolgov:2000fr}
A.~D. {Dolgov}, S.~H. {Hansen}, G.~{Raffelt}, and D.~V. {Semikoz},
\newblock Nuclear Physics B {\bf 590}, 562 (2000), arXiv:hep-ph/0008138.

\bibitem{Volkas:2000uq}
R.~R. {Volkas} and Y.~Y.~Y. {Wong},
\newblock \prd {\bf 62}, 093024 (2000), arXiv:hep-ph/0007185.

\bibitem{Abazajian:2001lr}
K.~{Abazajian}, G.~M. {Fuller}, and M.~{Patel},
\newblock \prd {\bf 64}, 023501 (2001), arXiv:astro-ph/0101524.

\bibitem{Dolgov:2002ve}
A.~D. {Dolgov} and S.~H. {Hansen},
\newblock Astroparticle Physics {\bf 16}, 339 (2002), arXiv:hep-ph/0009083.

\bibitem{Kusenko:2005qy}
A.~{Kusenko}, S.~{Pascoli}, and D.~{Semikoz},
\newblock Journal of High Energy Physics {\bf 11}, 28 (2005),
  arXiv:hep-ph/0405198.

\bibitem{Boyanovsky:2007fk}
D.~{Boyanovsky},
\newblock \prd {\bf 76}, 103514 (2007), 0706.3167.

\bibitem{Boyanovsky:2007kx}
D.~{Boyanovsky} and C.~M. {Ho},
\newblock \prd {\bf 76}, 085011 (2007), 0705.0703.

\bibitem{Boyanovsky:2007lr}
D.~{Boyanovsky} and C.-M. {Ho},
\newblock Journal of High Energy Physics {\bf 7}, 30 (2007),
  arXiv:hep-ph/0612092.

\bibitem{Kishimoto:2008pd}
C.~T. {Kishimoto} and G.~M. {Fuller},
\newblock \prd {\bf 78}, 023524 (2008), 0802.3377.

\bibitem{Kusenko:2009lr}
A.~{Kusenko},
\newblock Physics Reports {\bf 481}, 1 (2009), 0906.2968.

\bibitem{Cirigliano:2010lr}
V.~{Cirigliano}, C.~{Lee}, M.~J. {Ramsey-Musolf}, and S.~{Tulin},
\newblock \prd {\bf 81}, 103503 (2010), 0912.3523.

\bibitem{Cirigliano:2011lr}
V.~{Cirigliano}, C.~{Lee}, and S.~{Tulin},
\newblock \prd {\bf 84}, 056006 (2011), 1106.0747.

\bibitem{Qian93}
Y.~{Qian} {\em et~al.},
\newblock Physical Review Letters {\bf 71}, 1965 (1993).

\bibitem{Bethe:1980zr}
H.~A. {Bethe}, J.~H. {Applegate}, and G.~E. {Brown},
\newblock \apj {\bf 241}, 343 (1980).

\bibitem{Bethe:1985lr}
H.~A. {Bethe} and J.~R. {Wilson},
\newblock \apj {\bf 295}, 14 (1985).

\bibitem{Fuller:1992eu}
G.~M. {Fuller}, R.~{Mayle}, B.~S. {Meyer}, and J.~R. {Wilson},
\newblock \apj {\bf 389}, 517 (1992).

\bibitem{Dasgupta:2011uq}
B.~{Dasgupta}, E.~P. {O'Connor}, and C.~D. {Ott},
\newblock ArXiv e-prints  (2011), 1106.1167.

\bibitem{Fuller:2009lr}
G.~M. {Fuller} and C.~T. {Kishimoto},
\newblock Physical Review Letters {\bf 102}, 201303 (2009), 0811.4370.

\bibitem{Dodelson:2009qy}
S.~{Dodelson} and M.~{Vesterinen},
\newblock Physical Review Letters {\bf 103}, 171301 (2009), 0907.2887.

\bibitem{Duan:2010fr}
H.~{Duan}, G.~M. {Fuller}, and Y.-Z. {Qian},
\newblock Annual Review of Nuclear and Particle Science {\bf 60}, 569 (2010),
  1001.2799.

\bibitem{Sawyer:1990lr}
R.~F. {Sawyer},
\newblock \prd {\bf 42}, 3908 (1990).

\bibitem{Loreti:1994qy}
F.~N. {Loreti} and A.~B. {Balantekin},
\newblock \prd {\bf 50}, 4762 (1994), arXiv:nucl-th/9406003.

\bibitem{Loreti:1995fk}
F.~N. {Loreti}, Y.-Z. {Qian}, G.~M. {Fuller}, and A.~B. {Balantekin},
\newblock \prd {\bf 52}, 6664 (1995), arXiv:astro-ph/9508106.

\bibitem{Kneller:2008rt}
J.~P. {Kneller}, G.~C. {McLaughlin}, and J.~{Brockman},
\newblock \prd {\bf 77}, 045023 (2008), 0705.3835.

\bibitem{Kneller:2010ys}
J.~{Kneller} and C.~{Volpe},
\newblock \prd {\bf 82}, 123004 (2010), 1006.0913.

\bibitem{Raffelt:2013qy}
G.~{Raffelt}, S.~{Sarikas}, and D.~{de Sousa Seixas},
\newblock ArXiv e-prints  (2013), 1305.7140.

\bibitem{Mirizzi:2013uq}
A.~{Mirizzi},
\newblock ArXiv e-prints  (2013), 1308.1402.

\bibitem{Mirizzi:2013fj}
A.~{Mirizzi},
\newblock ArXiv e-prints  (2013), 1308.5255.

\bibitem{Cherry:2012lr}
J.~F. {Cherry}, J.~{Carlson}, A.~{Friedland}, G.~M. {Fuller}, and
  A.~{Vlasenko},
\newblock Physical Review Letters {\bf 108}, 261104 (2012), 1203.1607.

\bibitem{Sarikas:2012fk}
S.~{Sarikas}, I.~{Tamborra}, G.~{Raffelt}, L.~{H{\"u}depohl}, and H.-T.
  {Janka},
\newblock \prd {\bf 85}, 113007 (2012), 1204.0971.

\bibitem{Mirizzi:2012qy}
A.~{Mirizzi} and P.~D. {Serpico},
\newblock \prd {\bf 86}, 085010 (2012), 1208.0157.

\bibitem{Cherry:2013lr}
J.~F. {Cherry}, J.~{Carlson}, A.~{Friedland}, G.~M. {Fuller}, and
  A.~{Vlasenko},
\newblock \prd {\bf 87}, 085037 (2013), 1302.1159.

\bibitem{Martin:2011v6}
S.~P. {Martin},
\newblock {A Supersymmetry Primer},
\newblock {unpublished}, 2011.

\bibitem{Dreiner:2010lr}
H.~K. {Dreiner}, H.~E. {Haber}, and S.~P. {Martin},
\newblock Physics Reports {\bf 494}, 1 (2010), 0812.1594.

\bibitem{Berges:2004qy}
J.~{Berges},
\newblock {Introduction to Nonequilibrium Quantum Field Theory},
\newblock in {\em American Institute of Physics Conference Series}, edited by
  M.~E. {Bracco}, M.~{Chiapparini}, E.~{Ferreira}, and T.~{Kodama}, , American
  Institute of Physics Conference Series Vol. 739, pp. 3--62, 2004,
  arXiv:hep-ph/0409233.

\bibitem{Prokopec:2004lr}
T.~{Prokopec}, M.~G. {Schmidt}, and S.~{Weinstock},
\newblock Annals of Physics {\bf 314}, 208 (2004), arXiv:hep-ph/0312110.

\bibitem{Prokopec:2004fk}
T.~{Prokopec}, M.~G. {Schmidt}, and S.~{Weinstock},
\newblock Annals of Physics {\bf 314}, 267 (2004), arXiv:hep-ph/0406140.

\bibitem{Konstandin:2005qy}
T.~{Konstandin}, T.~{Prokopec}, and M.~G. {Schmidt},
\newblock Nuclear Physics B {\bf 716}, 373 (2005), arXiv:hep-ph/0410135.

\bibitem{Konstandin:2006uq}
T.~{Konstandin} and T.~{Ohlsson},
\newblock Physics Letters B {\bf 634}, 267 (2006), arXiv:hep-ph/0511010.

\bibitem{Cirigliano:2010yq}
V.~{Cirigliano}, Y.~{Li}, S.~{Profumo}, and M.~J. {Ramsey-Musolf},
\newblock Journal of High Energy Physics {\bf 1}, 2 (2010), 0910.4589.

\bibitem{Calzetta:1988qy}
E.~{Calzetta} and B.~L. {Hu},
\newblock \prd {\bf 37}, 2878 (1988).

\bibitem{Fuller87}
G.~M. {Fuller}, R.~W. {Mayle}, J.~R. {Wilson}, and D.~N. {Schramm},
\newblock \apj {\bf 322}, 795 (1987).

\bibitem{Notzold:1988fv}
D.~{N{\"o}tzold} and G.~{Raffelt},
\newblock Nuclear Physics B {\bf 307}, 924 (1988).

\bibitem{Blennow:2008lr}
M.~{Blennow}, A.~{Mirizzi}, and P.~D. {Serpico},
\newblock \prd {\bf 78}, 113004 (2008), 0810.2297.

\bibitem{Esteban-Pretel:2008fk}
A.~{Esteban-Pretel} {\em et~al.},
\newblock \prd {\bf 78}, 085012 (2008), 0807.0659.

\bibitem{Esteban-Pretel:2008qy}
A.~{Esteban-Pretel}, S.~{Pastor}, R.~{Tom{\`a}s}, G.~G. {Raffelt}, and
  G.~{Sigl},
\newblock \prd {\bf 77}, 065024 (2008), 0712.1137.

\bibitem{Esteban-Pretel:2010uq}
A.~{Esteban-Pretel}, R.~{Tom{\`a}s}, and J.~W.~F. {Valle},
\newblock \prd {\bf 81}, 063003 (2010), 0909.2196.

\bibitem{Botella:1987lr}
F.~J. {Botella}, C.-S. {Lim}, and W.~J. {Marciano},
\newblock \prd {\bf 35}, 896 (1987).

\bibitem{Mirizzi:2009lr}
A.~{Mirizzi}, S.~{Pozzorini}, G.~G. {Raffelt}, and P.~D. {Serpico},
\newblock Journal of High Energy Physics {\bf 10}, 20 (2009), 0907.3674.

\bibitem{Flowers:1976fj}
E.~G. {Flowers} and P.~G. {Sutherland},
\newblock Astrophys. J. Letters {\bf 208}, L19 (1976).

\bibitem{Samuel:1993sf}
S.~{Samuel},
\newblock \prd {\bf 48}, 1462 (1993).

\bibitem{Qian95}
Y.~{Qian} and G.~M. {Fuller},
\newblock \prd {\bf 51}, 1479 (1995), arXiv:astro-ph/9406073.

\bibitem{Samuel:1996rm}
S.~{Samuel},
\newblock \prd {\bf 53}, 5382 (1996), arXiv:hep-ph/9604341.

\bibitem{Elze:2000vh}
H.-T. Elze, T.~Kodama, and R.~Opher,
\newblock Phys.Rev. {\bf D63}, 013008 (2001), astro-ph/0007024.

\bibitem{Pastor:2002zl}
S.~{Pastor}, G.~{Raffelt}, and D.~V. {Semikoz},
\newblock \prd {\bf 65}, 053011 (2002), arXiv:hep-ph/0109035.

\bibitem{Pastor02}
S.~{Pastor} and G.~{Raffelt},
\newblock Physical Review Letters {\bf 89}, 191101 (2002),
  arXiv:astro-ph/0207281.

\bibitem{Balantekin05}
A.~B. {Balantekin} and H.~{Y{\"u}ksel},
\newblock New Journal of Physics {\bf 7}, 51 (2005), arXiv:astro-ph/0411159.

\bibitem{Fuller06}
G.~M. {Fuller} and Y.~{Qian},
\newblock \prd {\bf 73}, 023004 (2006), arXiv:astro-ph/0505240.

\bibitem{Duan06a}
H.~{Duan}, G.~M. {Fuller}, J.~{Carlson}, and Y.~{Qian},
\newblock \prd {\bf 74}, 105014 (2006), arXiv:astro-ph/0606616.

\bibitem{Duan06b}
H.~{Duan}, G.~M. {Fuller}, J.~{Carlson}, and Y.~{Qian},
\newblock Physical Review Letters {\bf 97}, 241101 (2006),
  arXiv:astro-ph/0608050.

\bibitem{Duan06c}
H.~{Duan}, G.~M. {Fuller}, and Y.~{Qian},
\newblock \prd {\bf 74}, 123004 (2006), arXiv:astro-ph/0511275.

\bibitem{Hannestad:2006qd}
S.~{Hannestad}, G.~G. {Raffelt}, G.~{Sigl}, and Y.~Y.~Y. {Wong},
\newblock \prd {\bf 74}, 105010 (2006), arXiv:astro-ph/0608695.

\bibitem{Duan07a}
H.~{Duan}, G.~M. {Fuller}, and Y.~{Qian},
\newblock \prd {\bf 76}, 085013 (2007), 0706.4293.

\bibitem{Duan07b}
H.~{Duan}, G.~M. {Fuller}, J.~{Carlson}, and Y.~{Qian},
\newblock \prd {\bf 75}, 125005 (2007), arXiv:astro-ph/0703776.

\bibitem{Balantekin:2007kx}
A.~B. {Balantekin} and Y.~{Pehlivan},
\newblock Journal of Physics G Nuclear Physics {\bf 34}, 47 (2007),
  arXiv:astro-ph/0607527.

\bibitem{Duan07c}
H.~{Duan}, G.~M. {Fuller}, J.~{Carlson}, and Y.~{Qian},
\newblock Physical Review Letters {\bf 99}, 241802 (2007), 0707.0290.

\bibitem{Duan08}
H.~{Duan}, G.~M. {Fuller}, J.~{Carlson}, and Y.~{Qian},
\newblock Physical Review Letters {\bf 100}, 021101 (2008), 0710.1271.

\bibitem{Lunardini08}
C.~{Lunardini}, B.~{M{\"u}ller}, and H.~{Janka},
\newblock \prd {\bf 78}, 023016 (2008), 0712.3000.

\bibitem{Dasgupta:2008kx}
B.~{Dasgupta}, A.~{Dighe}, A.~{Mirizzi}, and G.~{Raffelt},
\newblock \prd {\bf 78}, 033014 (2008), 0805.3300.

\bibitem{Friedland:2010yq}
A.~{Friedland},
\newblock Physical Review Letters {\bf 104}, 191102 (2010), 1001.0996.

\bibitem{Duan:2011lr}
H.~{Duan}, A.~{Friedland}, G.~C. {McLaughlin}, and R.~{Surman},
\newblock Journal of Physics G Nuclear Physics {\bf 38}, 035201 (2011),
  1012.0532.

\bibitem{Wolfenstein78}
L.~{Wolfenstein},
\newblock \prd {\bf 17}, 2369 (1978).

\bibitem{Mikheyev85}
S.~P. {Mikheyev} and A.~Y. {Smirnov},
\newblock Yad. Fiz. {\bf 42} (1985).

\bibitem{Dvornikov:2011dv}
M.~Dvornikov,
\newblock Nucl.Phys. {\bf B855}, 760 (2012), 1108.5043.

\bibitem{de-Gouvea:2012fk}
A.~{de Gouv{\^e}a} and S.~{Shalgar},
\newblock Journal of Cosmology and Astroparticle Physics {\bf 10}, 27 (2012),
  1207.0516.

\bibitem{de-Gouvea:2013lr}
A.~{de Gouv{\^e}a} and S.~{Shalgar},
\newblock Journal of Cosmology and Astroparticle Physics {\bf 4}, 18 (2013),
  1301.5637.

\bibitem{Blaizot:2002uq}
J.-P. {Blaizot} and E.~{Iancu},
\newblock Physics Reports {\bf 359}, 355 (2002), hep-ph/0101103.

\bibitem{Fidler:2012lr}
C.~{Fidler}, M.~{Herranen}, K.~{Kainulainen}, and P.~M. {Rahkila},
\newblock Journal of High Energy Physics {\bf 2}, 65 (2012), 1108.2309.

\bibitem{Herranen:2010kx}
M.~{Herranen}, K.~{Kainulainen}, and P.~{Matti Rahkila},
\newblock Journal of High Energy Physics {\bf 12}, 72 (2010), 1006.1929.

\bibitem{Garbrecht:2012fj}
B.~{Garbrecht} and M.~{Herranen},
\newblock Nuclear Physics B {\bf 861}, 17 (2012), 1112.5954.

\bibitem{Beneke:2010dz}
M.~Beneke, B.~Garbrecht, C.~Fidler, M.~Herranen, and P.~Schwaller,
\newblock Nucl.Phys. {\bf B843}, 177 (2011), 1007.4783.

\bibitem{Beneke:2010wd}
M.~Beneke, B.~Garbrecht, M.~Herranen, and P.~Schwaller,
\newblock Nucl.Phys. {\bf B838}, 1 (2010), 1002.1326.

\bibitem{Garny:2009qn}
M.~Garny, A.~Hohenegger, A.~Kartavtsev, and M.~Lindner,
\newblock Phys.Rev. {\bf D81}, 085027 (2010), 0911.4122.

\bibitem{Garny:2009rv}
M.~Garny, A.~Hohenegger, A.~Kartavtsev, and M.~Lindner,
\newblock Phys.Rev. {\bf D80}, 125027 (2009), 0909.1559.

\bibitem{Zhang:2013lka}
Y.~Zhang and A.~Burrows,
\newblock (2013), 1310.2164.

\bibitem{Kolb:1992lr}
E.~W. {Kolb} and Y.~{Wang},
\newblock \prd {\bf 45}, 4421 (1992).

\bibitem{Shi:1999fk}
X.~{Shi} and G.~M. {Fuller},
\newblock Physical Review Letters {\bf 83}, 3120 (1999),
  arXiv:astro-ph/9904041.

\bibitem{Fryer:2009qy}
C.~L. {Fryer},
\newblock \apj {\bf 699}, 409 (2009), 0711.0551.

\bibitem{Herant:1994uq}
M.~{Herant}, W.~{Benz}, W.~R. {Hix}, C.~L. {Fryer}, and S.~A. {Colgate},
\newblock \apj {\bf 435}, 339 (1994), arXiv:astro-ph/9404024.

\bibitem{Mezzacappa:1998fr}
A.~{Mezzacappa} {\em et~al.},
\newblock \apj {\bf 495}, 911 (1998), arXiv:astro-ph/9709188.

\bibitem{Mezzacappa:2001fk}
A.~{Mezzacappa} {\em et~al.},
\newblock Physical Review Letters {\bf 86}, 1935 (2001),
  arXiv:astro-ph/0005366.

\bibitem{Mezzacappa:2005lr}
A.~{Mezzacappa},
\newblock Annual Review of Nuclear and Particle Science {\bf 55}, 467 (2005).

\bibitem{Bruenn:2010qy}
S.~W. {Bruenn} {\em et~al.},
\newblock ArXiv e-prints  (2010), 1002.4914.

\bibitem{Keil:2003qy}
M.~T. {Keil}, G.~G. {Raffelt}, and H.-T. {Janka},
\newblock \apj {\bf 590}, 971 (2003), arXiv:astro-ph/0208035.

\bibitem{Kotake:2006fk}
K.~{Kotake}, K.~{Sato}, and K.~{Takahashi},
\newblock Reports on Progress in Physics {\bf 69}, 971 (2006),
  arXiv:astro-ph/0509456.

\bibitem{Murphy:2008zr}
J.~W. {Murphy} and A.~{Burrows},
\newblock \apj {\bf 688}, 1159 (2008), 0805.3345.

\bibitem{Ott:2009fj}
C.~D. {Ott} {\em et~al.},
\newblock Journal of Physics Conference Series {\bf 180}, 012022 (2009),
  0907.4043.

\bibitem{Muller:2010kx}
B.~{M{\"u}ller}, H.~{Janka}, and H.~{Dimmelmeier},
\newblock The Astrophysical Journal Supplement {\bf 189}, 104 (2010),
  1001.4841.

\bibitem{Brandt:2011mz}
T.~D. {Brandt}, A.~{Burrows}, C.~D. {Ott}, and E.~{Livne},
\newblock \apj {\bf 728}, 8 (2011), 1009.4654.

\bibitem{Muller:2012ly}
B.~{M{\"u}ller}, H.-T. {Janka}, and A.~{Heger},
\newblock \apj {\bf 761}, 72 (2012), 1205.7078.

\bibitem{Cardall:2012rt}
C.~Y. {Cardall},
\newblock Nuclear Physics B Proceedings Supplements {\bf 229}, 315 (2012).

\bibitem{Hanke:2012ys}
F.~{Hanke}, A.~{Marek}, B.~{M{\"u}ller}, and H.-T. {Janka},
\newblock \apj {\bf 755}, 138 (2012), 1108.4355.

\bibitem{Ellinger:2013gf}
C.~I. {Ellinger}, G.~{Rockefeller}, C.~L. {Fryer}, P.~A. {Young}, and
  S.~{Park},
\newblock ArXiv e-prints  (2013), 1305.4137.

\bibitem{Murphy:2013yq}
J.~W. {Murphy}, J.~C. {Dolence}, and A.~{Burrows},
\newblock \apj {\bf 771}, 52 (2013), 1205.3491.

\bibitem{Fryer:2013kx}
C.~L. {Fryer},
\newblock ArXiv e-prints  (2013), 1307.6141.

\bibitem{Tamborra:2013fj}
I.~{Tamborra}, F.~{Hanke}, B.~{Mueller}, H.-T. {Janka}, and G.~{Raffelt},
\newblock ArXiv e-prints  (2013), 1307.7936.

\bibitem{Bruenn:2013vn}
S.~W. {Bruenn} {\em et~al.},
\newblock Astrophys. J. Letters {\bf 767}, L6 (2013), 1212.1747.

\bibitem{Suwa:2013lr}
Y.~{Suwa} {\em et~al.},
\newblock \apj {\bf 764}, 99 (2013), 1206.6101.

\end{thebibliography}

\end{document}